\documentclass[12pt]{article}
\usepackage{epsfig}
\usepackage{amsmath}
\usepackage{hhline}
\usepackage{amssymb}
\usepackage{times}
\usepackage{cite}
\usepackage{lscape}
\usepackage{rotating}

\newlength{\dinwidth}
\newlength{\dinmargin}
\begin{document}
\newcommand{\pom}{{I\!\!P}}
\newcommand{\reg}{{I\!\!R}}
\def\gsim{\,\lower.25ex\hbox{$\scriptstyle\sim$}\kern-1.30ex%
\raise 0.55ex\hbox{$\scriptstyle >$}\,}
\def\lsim{\,\lower.25ex\hbox{$\scriptstyle\sim$}\kern-1.30ex%
\raise 0.55ex\hbox{$\scriptstyle <$}\,}
\newcommand{\trm}{m_{\perp}}
\newcommand{\trp}{p_{\perp}}
\newcommand{\trmm}{m_{\perp}^2}
\newcommand{\trpp}{p_{\perp}^2}
\newcommand{\alp}{\alpha_s}
\newcommand{\alps}{\alpha_s}
\newcommand{\sqrts}{$\sqrt{s}$}
\newcommand{\LO}{$O(\alpha_s^0)$}
\newcommand{\Oa}{$O(\alpha_s)$}
\newcommand{\Oaa}{$O(\alpha_s^2)$}
\newcommand{\PT}{p_{\perp}}
\newcommand{\JPSI}{J/\psi}
\newcommand{\PO}{I\!\!P}
\newcommand{\xbj}{x}
\newcommand{\xpom}{x_{\PO}}
\newcommand{\zpom}{z_{\PO}}
\newcommand{\dgr}{^\circ}
\newcommand{\gev}{\,\mbox{Ge}}
\newcommand{\GeV}{\rm GeV}
\newcommand{\xp}{x_p}
\newcommand{\xpi}{x_\pi}
\newcommand{\xg}{x_\gamma}
\newcommand{\xgj}{x_\gamma^{jet}}
\newcommand{\xpj}{x_p^{jet}}
\newcommand{\xpij}{x_\pi^{jet}}
\renewcommand{\deg}{^\circ}
\newcommand{\qsq}{\ensuremath{Q^2} }
\newcommand{\gevsq}{\ensuremath{\mathrm{GeV}^2} }
\newcommand{\et}{\ensuremath{E_t^*} }
\newcommand{\rap}{\ensuremath{\eta^*} }
\newcommand{\gp}{\ensuremath{\gamma^*}p }
\newcommand{\dsiget}{\ensuremath{{\rm d}\sigma_{ep}/{\rm d}E_t^*} }
\newcommand{\dsigrap}{\ensuremath{{\rm d}\sigma_{ep}/{\rm d}\eta^*} }
\newcommand {\gapprox}
   {\, \raisebox{-0.7ex}{$\stackrel {\textstyle>}{\sim} \,$}}
\newcommand {\lapprox}
   {\, \raisebox{-0.7ex}{$\stackrel {\textstyle<}{\sim} \,$}}

\def\Journal#1#2#3#4{{#1} {\bf #2}, #4 (#3)}
\def\NCA{\em Nuovo Cimento}
\def\NIM{\em Nucl. Instrum. Methods}
\def\NIMA{{\em Nucl. Instrum. Methods} {\bf A}}
\def\NPB{{\em Nucl. Phys.}   {\bf B}}
\def\PLB{{\em Phys. Lett.}   {\bf B}}
\def\PRL{\em Phys. Rev. Lett.}
\def\PRD{{\em Phys. Rev.}    {\bf D}}
\def\PR{{\em Phys. Rev.}    }
\def\PRP{{\em Phys. Rep.}    }
\def\ZPC{{\em Z. Phys.}      {\bf C}}
\def\ZP{{\em Z. Phys.}      }
\def\EJC{{\em Eur. Phys. J.} {\bf C}}
\def\EJA{{\em Eur. Phys. J.} {\bf A}}
\def\CPC{\em Comp. Phys. Commun.}
\def\SJNP{{\em Sov. J. Nucl. Phys.}}
\def\SPJETP{{\em Sov. Phys. JETP}}
\def\JETPL{{\em JETP Lett.}}


\begin{titlepage}


\noindent
 \begin{flushleft}
 DESY 11-166 \hfill ISSN 0418-9833\\
 September 2011
\end{flushleft}







\vspace*{2cm}

\begin{center}
\begin{Large}

{\boldmath \bf
  Measurement of Dijet Production in Diffractive Deep-Inelastic Scattering with a Leading Proton at HERA
}

\vspace{2cm}

H1 Collaboration

\end{Large}
\end{center}

\vspace{2cm}

\begin{abstract}
\noindent
 The cross section of diffractive deep-inelastic scattering 
 $ep \to e X p$ is measured,
 where the system $X$ contains at least two jets and
the leading final state proton is detected in the H1 Forward Proton
 Spectrometer. The measurement is performed for fractional proton longitudinal momentum loss $\xpom < 0.1$ and covers the range $0.1 < |t| < 0.7~{\rm  GeV}^2$ 
in squared four-momentum transfer at the proton vertex and $4<Q^2<110~{\rm GeV}^2$ in  photon virtuality.
 The differential cross sections extrapolated to $|t| < 1~\GeV^2$ 
 are in agreement with next-to-leading order QCD predictions based on
 diffractive parton distribution functions extracted from measurements of inclusive and dijet cross sections in diffractive deep-inelastic scattering. 
 The data are also compared with leading order Monte Carlo models.
\end{abstract}

\vspace{1cm}

\begin{center}
  Submitted to \EJC
\end{center}

\end{titlepage}

\newpage


F.D.~Aaron$^{5,48}$,           
C.~Alexa$^{5}$,                
V.~Andreev$^{25}$,             
S.~Backovic$^{30}$,            
A.~Baghdasaryan$^{38}$,        
S.~Baghdasaryan$^{38}$,        
E.~Barrelet$^{29}$,            
W.~Bartel$^{11}$,              
K.~Begzsuren$^{35}$,           
A.~Belousov$^{25}$
% ...

\bigskip{\noindent\it
% ...
$ ^{5}$ National Institute for Physics and Nuclear Engineering (NIPNE) ,
          Bucharest, Romania$^{ m}$ \\
% ...
$ ^{11}$ DESY, Hamburg, Germany \\
% ...
$ ^{25}$ Lebedev Physical Institute, Moscow, Russia$^{ e}$ \\
% ...
$ ^{29}$ LPNHE, Universit\'e Pierre et Marie Curie Paris 6,
          Universit\'e Denis Diderot Paris 7, CNRS/IN2P3, Paris, France \\
 $ ^{30}$ Faculty of Science, University of Montenegro,
          Podgorica, Montenegro$^{ n}$ \\
% ...
$ ^{35}$ Institute of Physics and Technology of the Mongolian
          Academy of Sciences, Ulaanbaatar, Mongolia \\
% ...
$ ^{38}$ Yerevan Physics Institute, Yerevan, Armenia \\
% ...
$ ^{48}$ Also at Faculty of Physics, University of Bucharest,
          Bucharest, Romania \\
% ...
$ ^e$ Supported by the Deutsche Forschungsgemeinschaft \\
% ...
$ ^m$ Supported by the Romanian National Authority for Scientific Research
      under the contract PN 09370101 \\
 $ ^n$ Partially Supported by Ministry of Science of Montenegro,
      no. 05-1/3-3352 \\
}
\newpage

 \section{Introduction}

Diffractive processes such as $ep \rightarrow eXY$, where the systems $X$ and $Y$ are separated in rapidity, have been
studied  extensively in deep-inelastic scattering (DIS) at the electron\footnote{In this paper ``electron'' is used to denote both electron 
and positron unless otherwise stated.}-proton collider HERA~\cite{H1Diff94,H1LRG,H1DiJets,H1DiJets2,H1Charm,ZEUSLPS2,ZEUSDiJets,ZEUSDPDF}. Diffractive DIS events can be viewed as resulting from processes in which
the photon probes a net colour singlet
combination of exchanged partons. The photon virtuality $Q^2$, the high transverse momentum of jets or a heavy quark mass
can provide a hard scale for perturbative QCD calculations.
For semi-inclusive DIS processes such as $ep \rightarrow eXp^{\prime}$ the hard scattering QCD collinear factorisation theorem \cite{Collins}
allows the definition of diffractive parton distribution functions (DPDFs).
The dependence of diffractive DIS on a hard scale 
can thus be treated in a manner similar to the treatment of inclusive DIS, for example through
the application of the
DGLAP parton evolution equations~\cite{DGLAP,DGLAP2,DGLAP3,DGLAP4}.
DPDFs have been determined from QCD fits to diffractive DIS measurements 
at HERA~\cite{H1LRG,H1DiJets,ZEUSDPDF}. The inclusive diffractive DIS cross section is 
directly proportional to the sum of the quark DPDFs and constrains the gluon DPDF via scaling violations. 
The production of diffractive hadronic final states containing heavy 
quarks or jets proceeds mainly via boson gluon fusion (BGF) and therefore 
directly constrains the diffractive gluon density~\cite{H1DiJets,ZEUSDPDF}.     

In previous analyses at HERA, diffractive DIS events 
have been selected on the basis
of the presence of
a large rapidity gap (LRG) between system $Y$, which consists of the outgoing proton or its dissociative excitations,
and the hadronic final state, system $X$~\cite{H1DiJets,H1DiJets2}. The main advantage of the LRG 
method is its high 
acceptance
for diffractive processes.
A complementary way to study diffraction is by direct
measurement of the outgoing proton, which remains intact in elastic interactions. This is achieved by the H1 experiment using the 
Forward Proton Spectrometer (FPS)~\cite{H1FPS,H1FPShera2}, which is a set of tracking detectors along the proton beam line. Despite the low geometrical acceptance of the 
FPS, 
this method of selecting diffractive events has several advantages.
The squared four-momentum transfer at the proton vertex, $t$, 
can be reconstructed with the FPS, while this is only possible in exclusive final states in the LRG case.  
The FPS method selects events in which the proton scatters
elastically, whereas the LRG method does not distinguish between the case where the scattered
proton remains intact or where it dissociates into a system of low mass $M_Y$.
The FPS method also allows measurements to be performed at higher values of fractional proton longitudinal momentum loss, $\xpom$,
than possible using the LRG method.

 This paper presents the first measurement of the cross section for the diffractive
 DIS process $ep \rightarrow e jjX^{\prime}p$, with two jets and a leading proton in the final state.
 The diffractive dijet cross sections are compared with next-to-leading order (NLO) QCD predictions based on DPDFs from 
 H1~\cite{H1LRG,H1DiJets} and with leading order (LO) Monte Carlo (MC) simulations based on different models.

 The dijet cross sections are measured for two event topologies: for a topology 
 where two jets are found in the central pseudorapidity range, labelled as \textquoteleft two central jets\textquoteright, 
 and for a topology where one jet is central and one jet is more forward\footnote{ The forward direction is 
 defined by the proton beam direction.}, labelled as \textquoteleft one central + one forward jet'. 
The universality of DPDFs is studied using events with two central jets.
 The distributions of the proton vertex variables
 $\xpom$ and $t$ are compared to those of the inclusive diffractive DIS case. 
This comparison tests the proton vertex factorisation hypothesis which assumes that the DIS variable factorise from the four-momentum of the final state proton.
 The data are also compared directly with the LRG measurement of the dijet cross section in diffractive DIS 
 \cite{H1DiJets} in order to test the compatibility of the two experimental techniques. 
Finally, events with one central and one forward jet 
are used to investigate diffractive DIS in a region of phase space where 
 effects beyond DGLAP parton evolution may be enhanced. This topology is not accessible with the LRG method since the rapidity gap requirement
limits the pseudorapidity of the reconstructed jets to the central region.

\begin{figure}[t] \unitlength 1mm
 \begin{center}
\vspace{-0.5cm}
 \epsfig{file=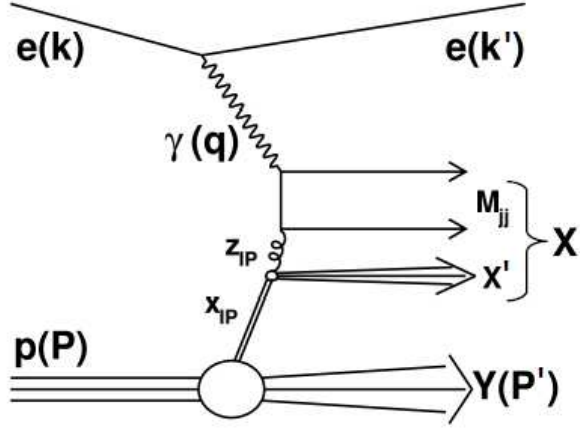,width=0.5\textwidth}
 \end{center}
\caption{The leading order boson gluon fusion diagram for dijet production in diffractive DIS.}
\label{kine}
\end{figure}

 \section{Kinematics}

 Figure~\ref{kine} illustrates the dominant process for diffractive dijet production in DIS. The incoming 
 electron with four-momentum $k$ interacts with the proton with four-momentum $P$ via the exchange of a virtual 
 photon with four-momentum $q$. The DIS kinematic variables are defined as:
\begin{equation} 
 Q^2 = -q^2 = (k-k')^2,~~~~~~~~~~x = \frac{-q^2}{2P \cdot q},~~~~~~~~~~y = \frac{P\cdot q}{P \cdot k},
\end{equation}
\noindent
 where $Q^2$ is the photon virtuality, $x$ is the longitudinal momentum fraction of the proton carried by the struck 
 quark and $y$ is the inelasticity of the process.
 These three variables are related via $Q^2 = xys$, where $s$ denotes the $ep$ 
centre-of-mass energy squared.

The hadronic final state of diffractive events consists of two systems $X$ and $Y$, separated by a gap in 
 rapidity. In general, the system $Y$ is 
 the outgoing proton or one of its low mass excitations. In events where 
 the outgoing proton remains intact, $M_Y = m_p$, the mass of the proton. The kinematics of diffractive DIS are described by:
\begin{equation}
 x_{I\!\!P} = \frac{q \cdot (P-P^\prime)}{q \cdot P},~~~~~~~~~~t = (P' - P)^2,~~~~~~~~~~\beta = \frac{-q^2}{2q \cdot (P-P^\prime)} = \frac{x}{x_{I\!\!P}},
\end{equation}
\noindent
 where $x_{I\!\!P}$ denotes the longitudinal momentum fraction of the proton carried by the colour singlet exchange, $t$ 
 is the squared four-momentum transfer at the proton vertex and $\beta$ is the fractional momentum of the 
 diffractive exchange carried by the struck parton. The longitudinal momentum fraction of the diffractive exchange 
carried by the parton entering the hard scatter is
\begin{equation}
 z_{I\!\!P} = \frac{q \cdot v}{q \cdot (P-P^\prime)},
\end{equation}
\noindent
where $v$ is the four-momentum of the parton.

\section{Theoretical Framework and Monte Carlo Models}
\label{models}

Within Regge phenomenology,
cross sections at high energies are described
by the exchange of Regge trajectories.
The diffractive cross section is dominated by a trajectory usually called the Pomeron ($\pom$).
In analyses of HERA data~\cite{H1LRG,H1DiJets,ZEUSDPDF},
diffractive DIS cross sections 
are interpreted assuming \lq proton vertex factorisation' which provides a description of diffractive DIS
in terms of a resolved Pomeron~\cite{res1,res2}.   
The QCD factorisation theorem and 
DGLAP parton evolution equations are applied to 
the dependence of the cross section on $Q^2$ and $\beta$, while
a Regge inspired approach is used to express the dependence on $\xpom$ and $t$.


The resolved Pomeron (RP) model~\cite{res1} is implemented in the RAPGAP event generator~\cite{RAPGAP}.
 RAPGAP implements both a leading Pomeron ($\pom$) trajectory
 and a sub-leading \lq Reggeon' ($\reg$). In this analysis the DPDF H1 2006 Fit B~\cite{H1LRG} is used, which employs the Owens pion PDFs~\cite{owens} for the partonic content of the Reggeon. 
 The Reggeon contribution is significant for $x_{I\!\!P} > 0.01$. 
Higher order QCD radiation is modelled by parton showers. 
 Processes with a resolved virtual photon are also included, with the photon structure function given by the SaS-G 2D 
 LO parameterisation~\cite{2gluphot}.

 In the two-gluon Pomeron (TGP) model~\cite{bart,bart2}, the 
 diffractive exchange is modelled at LO as the interaction of a colourless 
 pair of gluons with a $q\bar{q}$ or $q\bar{q}g$ configuration emerging from the photon. 
 The model is implemented in the RAPGAP generator.  Higher order effects are simulated using parton showers.
The unintegrated gluon PDF of set A0~\cite{a0ref} is used.

 In the soft colour interaction (SCI) model~\cite{lowx3,lowx4}, the diffractive exchange is modelled via 
 non-diffractive DIS scattering with subsequent colour rearrangement between the partons in the final state, which 
 can produce a colour singlet system separated by a large gap in pseudorapidity. A refined version of the SCI model 
 which uses a generalised area law (GAL) for the probability of having a soft colour interaction~\cite{gal} is 
 used in this analysis (SCI+GAL). Predictions for diffractive dijet production within the SCI+GAL model are obtained using the leading order 
 generator program LEPTO~\cite{lepto}. Higher order effects are simulated using parton showers~\cite{leptodglap1, leptodglap2}. The 
 calculations are based on the CTEQ6L~\cite{2gluprot} proton PDFs.
 The probability for a soft colour interaction, $R$, has been tuned to $0.3$ to describe the total diffractive 
 dijet cross section as measured using the \textquoteleft two central jets' topology.

In all three models hadronisation is simulated using the Lund string model~\cite{Lund} implemented within the
 PYTHIA program~\cite{Pythia}.

 In this analysis the dijet cross section is also compared to NLO QCD calculations.  
 Assuming proton vertex factorisation, NLO QCD predictions for the diffractive partonic dijet cross section are calculated 
 in bins of $x_{I\!\!P}$ using the 
 NLOJET++~\cite{nlojet} program and integrated over the full $x_{I\!\!P}$ range of the measurement. 
 The renormalisation and factorisation scales are set to $\mu_r = \mu_f = \sqrt{Q^2 + \langle P_T^*\rangle^2}$, where 
 $\langle P_T^*\rangle$ is 
 the mean of the transverse momenta of the two leading jets in the hadronic centre-of-mass frame. 
 In order to estimate the 
 uncertainties of the NLO QCD calculations due to missing higher orders, the factorisation scale $\mu_f$ and renormalisation 
 scale $\mu_r$ are varied simultaneously by factors of $0.5$ and $2$. The average uncertainty 
 arising from the variation of the scale is about $40\%$.
 The DPDFs used in the NLO QCD 
 calculations are 
H1 2006 Fit B \cite{H1LRG} and H1 2007 Jets \cite{H1DiJets}. The H1 2007 Jets fit is based on the diffractive inclusive and dijet 
 data while H1 2006 Fit B is based on inclusive diffractive data only.
In order to demonstrate the size of the NLO corrections,
 the QCD calculations are also performed at leading order.
 
 The NLO QCD partonic cross sections are corrected to the level of stable hadrons by evaluating effects due 
 to initial and final state parton showering, fragmentation and hadronisation. The hadronisation 
 corrections are defined in each bin as a ratio of the cross section obtained at the level of stable hadrons to the partonic cross 
 sections. Two sets of hadronisation corrections have been obtained using the RAPGAP generator using two 
 different parton shower models: parton showers based on leading logarithm DGLAP splitting 
 functions in leading order $\alpha_s$~\cite{DGLAP,DGLAP2,DGLAP3,DGLAP4} and parton showers based on the colour dipole model as implemented in  
  ARIADNE \cite{cdm}. The nominal set of corrections $(1+\delta_{had})$ is taken as the average of the two 
 sets,  while the 
 difference between them is considered as the hadronisation uncertainty. 
 The average hadronisation corrections are of about $0.9$ with an estimated uncertainty of about $7\%$. 
 Uncertainties arising due to scale variations and hadronisation corrections are added in 
 quadrature and quoted as a total uncertainty on NLO QCD predictions. 

 In order to compare with the results of the FPS measurements,
 NLO and LO QCD predictions as well as predictions of the RP model are scaled down by a 
 factor of $1.20$~\cite{H1FPShera2} due to the fact that the DPDF sets H1 2006 Fit B and H1 2007 Jets use LRG data 
 which contain a proton dissociation contribution.  
The $t$-dependence of the $\pom$ and $\reg$ fluxes implemented in the H1 DPDF sets 
 and the RP model are tuned to reproduce the $t$-dependence measured in inclusive diffractive 
 DIS with a leading proton in the final state~\cite{H1FPS}.

\section{Experimental Technique}

 The $e^{\pm}p$ data used in this analysis were collected with the H1 detector in the years 2005 to 2007 and correspond to an integrated
 luminosity of $156.6$~pb$^{-1}$.
 During this period the HERA collider was 
 operated at electron and proton beam energies of $E_e=27.6$~GeV and $E_p=920$~GeV
 respectively, corresponding to an $ep$ centre-of-mass energy of
 $\sqrt{s} = 319$~GeV.

\subsection{H1 detector}
\label{detector}

A detailed description of the H1 detector can be found elsewhere
\cite{h1detector,h1detector2,SPACAL}. Here, the components most
relevant for the presented measurement are described briefly.
A right-handed coordinate system is employed with the origin at the
nominal interaction point, where the $z$-axis pointing in the proton beam or forward
direction and the $x(y)$ axis points in the horizontal (vertical) direction.
The polar angle $\theta$ is measured with respect to the proton beam axis and the pseudorapidity is
defined as $\eta = -\ln \tan(\theta/2)$.

The Central Tracking Detector (CTD), with  a polar angle coverage of
$20\deg<\theta<160\deg$, is used
to reconstruct the interaction vertex and
to measure the momenta of
charged particles from the curvature of their trajectories
in the $1.16\,{\rm T}$ field provided by a superconducting solenoid. Scattered electrons with polar angles
in the range $154\deg\!\!<\!\!\theta_e^\prime\!\! <\!\!176\deg$ are measured in
a lead\,/\,scintillating-fibre calorimeter, the SpaCal \cite{SPACAL}.
The energy resolution is
$\sigma(E)/E\approx 7\%/\sqrt{E[\GeV]}\oplus 1\%$
as determined from the test beam measurement~\cite{SPACAL_TEST}.
A Backward Proportional Chamber (BPC) in front of the SpaCal
is used to measure the electron polar angle.
The finely segmented Liquid Argon (LAr) sampling calorimeter
surrounds the tracking system and covers the range in polar angle
$4\deg\!\!<\!\!\theta\!\!<\!\!154\deg$ corresponding to a pseudorapidity range $-1.5\!\!<\!\!\eta\!\!<\!\!3.4$. 
The LAr calorimeter consists of an electromagnetic section with
lead as the absorber and
a hadronic section with steel as the absorber.
The total depth varies with $\theta$ between $4.5$ and $8$ interaction lengths.
The energy resolution,
 determined from test beam  measurements~\cite{LAR,LAR2}, 
is $\sigma(E)/E\approx
11\%/\sqrt{E[\GeV]}\oplus 1\%$ for electrons and
$\sigma(E)/E\approx 50\%/\sqrt{E[\GeV]}\oplus 2\%$ for hadrons.
The hadronic final state is reconstructed using an
energy flow algorithm which
combines charged particles measured in the CTD with  information from
the SpaCal and LAr calorimeters~\cite{hadroo2}.

The luminosity is determined by
measuring the rate of the Bethe-Heitler process
$ep\rightarrow ep\gamma$ detected in a photon detector
located at $z=-103$~m.

The energy and scattering angle of the leading proton are obtained
from track measurements in the FPS \cite{FPS}.
Protons scattered at small angles
are deflected by the proton beam-line magnets into a system of
detectors placed within the proton beam pipe inside two movable
stations, known as Roman Pots. Both Roman Pot stations contain
four planes, where each plane consists of five layers of scintillating fibres, which together measure two
orthogonal coordinates in the $(x,y)$ plane. The fibre coordinate planes are
sandwiched between planes of 
scintillator tiles used for the trigger.
The stations approach the beam horizontally 
and are positioned at
$z = 61$~m and $z = 80$~m.    
The detectors are sensitive to scattered protons which
lose less than $10\%$ of their energy in the $ep$
interaction and are scattered through angles below $1$~mrad.

The energy resolution of the FPS is approximately $5$~GeV within the
measured range. The absolute energy scale uncertainty is $1$~GeV.
The effective resolution in the reconstruction of the transverse 
momentum components of the scattered proton with respect to the incident proton  
is determined to be
 \,$\sim\!50$~MeV for $P_x$
and \,$\sim\!150$~MeV for $P_y$, dominated by the intrinsic transverse momentum
spread of the proton beam at the interaction point.
The scale uncertainties in the
transverse momentum measurements are $10$~MeV for
$P_x$ and $30$~MeV for $P_y$. 
Further details of the analysis of the FPS resolution and scale uncertainties can be found elsewhere~\cite{H1FPShera2}.   
For a leading proton which passes through
both FPS stations, the track reconstruction efficiency is
$48\%$ on average.

\subsection{Kinematic reconstruction}

 The inclusive DIS variables $Q^2$, $x$ and the inelasticity $y$ are reconstructed by combining information  
 from the scattered electron and
 the hadronic final state using the following method~\cite{H1Diff94}:
\begin{eqnarray}
  y = y_e^2+y_d-y_d^2 \ \ \ , \ \ \
 Q^2=\frac{4E_e^2 (1-y)}{\tan ^2(\theta_e^\prime/2)} \ \ \ , \ \ \
 x=\frac{Q^2}{sy} \ .
\label{eq:y}
\end{eqnarray}
\noindent
 Here, $y_e$ and $y_d$ denote the values of $y$
 obtained from the scattered electron only (electron method) and
 from the angles
 of the electron and the hadronic final state (double angle
 method), respectively~\cite{dameth}.
 
 The observable $\xpom$ is reconstructed as: 
\begin{equation}
 \xpom = 1 - E'_p/E_p,
\label{eq:xpomfps}
\end{equation}
\noindent
where $E'_p$ is the measured energy of the leading proton in the FPS.
The quantity $\beta$ is reconstructed as $\beta = x/\xpom$.  The squared four-momentum transfer at the proton vertex 
is reconstructed using the transverse momentum $P_T$ of
 the leading proton measured with the FPS and $\xpom$
 as described above, such that:
\begin{eqnarray}
 t = t_{\rm min} - \frac{P_T^2}{1-\xpom} \ \ \ , \ \ \
t_{\rm min} = - \frac{\xpom^2 m_p^2}{1-\xpom} \ ,
\label{eq:tdef}
\end{eqnarray}
\noindent 
 where $|t_{\rm min}|$ is the minimum kinematically accessible value of $|t|$.
The absolute resolution in $t$ varies over the measured range from
$0.06$~GeV$^2$ at $|t| = 0.1$~GeV$^2$
to $0.17$~GeV$^2$ at $|t| = 0.7$~GeV$^2$.

An estimator for the momentum fraction $\zpom$ is defined at the level of stable hadrons as:
\begin{equation}
 \zpom = \frac{Q^2+M_{jj}^2}{\xpom y s},
\label{eq:zpomdef}
\end{equation}
\noindent
 where $M_{jj}$ denotes the invariant mass of the dijet system. The cross sections are studied in terms of the DIS variables $y, Q^2, \beta, \zpom$, the proton vertex variables $\xpom$ and 
 $t$, the jet variables $P_{T}^*$ and $\eta$, and
\begin{eqnarray}
 \langle P_T^*\rangle = \frac{1}{2}(P_{T,1}^* + P_{T,2}^*) \ \ , \ \ |\Delta\eta^*|=|\eta_1^*-\eta_2^*| \ \ , \ \ 
 |\Delta\phi^*|=|\phi_1^*-\phi_2^*| \ ,
\label{eq:detadef}
\end{eqnarray}
\noindent
 where $P_{T,1}^*, \eta_1^*,\phi_1^*$ and $P_{T,2}^*, \eta_2^*,\phi_2^*$ are transverse momenta, pseudorapidities and azimuthal angles of the axes of the
 leading and next-to-leading jets, respectively, 
reconstructed in the hadronic centre-of-mass frame. The indices $1$, $2$ stand for the two jets used in the specific 
analyses.

\subsection{Event selection}

 The events used in the \textquoteleft two central jets' and \textquoteleft one central + one forward jet' analyses are triggered on the basis of a coincidence of a signal in the FPS trigger scintillator tiles and in the electromagnetic SpaCal.
The trigger efficiency, calculated using events collected with independent triggers, is found to be $99\%$ 
 on average and is independent of kinematic variables.
 
\subsubsection{DIS selection}


 The selection of DIS events is based on the identification of the
 scattered electron as the most energetic electromagnetic cluster
 in the SpaCal calorimeter.
 The energy $E_e^\prime$ and polar angle $\theta_e^\prime$ of the scattered electron
 are determined from the
 SpaCal cluster 
and the interaction vertex reconstructed in the CTD.
 The electron candidate is required to be in range
 $154\deg\!<\!\theta^{\prime}_e\!<\!176\deg$ and $E_e^\prime > 10$~GeV. In 
 order to improve 
background rejection,
an additional requirement 
on the transverse cluster radius, estimated using square root energy 
 weighting~\cite{ELAN}, of less then $4$~cm is imposed.

 The reconstructed $z$ coordinate of the event vertex is required to
be within $\pm 35$~cm 
 of the mean position. At
 least one track originating from the interaction vertex and
 reconstructed in the CTD is required to have a transverse momentum
 above $0.1$~GeV. 

 The quantity $\sum(E-P_z)$, summed over the energies and longitudinal momenta of
 all reconstructed particles including the electron, is required to be
 between  $35$~GeV and $70$~GeV. For
 neutral current DIS events
 this quantity is expected to be twice the
 electron beam energy when neglecting detector effects
 and QED radiation. This requirement is applied to remove radiative DIS events and photoproduction background. 

In order to ensure a good detector acceptance
the measurement is restricted to the ranges  $4<Q^2<110$~GeV$^2$ and $0.05 < y <0.7$.

\subsubsection{Leading proton selection}



A high FPS acceptance is ensured by requiring 
the energy of the leading proton $E_p^\prime$ to be greater than $90\%$ of 
 the proton beam energy $E_p$ and the horizontal
 and vertical projections of the transverse momentum to 
be in the ranges $-0.63 < P_x < -0.27$~GeV and
 $|P_y| < 0.8$~GeV, respectively. 
 Additionally, $t$ is restricted to the range
 $0.1 <|t|< 0.7$~GeV$^2$.

The quantity $\sum(E+P_z)$, summed
 over all reconstructed
 particles including the leading proton, is required to be
 below $1880$~GeV. For neutral current DIS events this
 quantity is expected to be twice the
 proton beam energy. This requirement is applied to suppress
 cases where a DIS event reconstructed in the central detector coincides with background
 in the FPS, for example due to interactions between off-momentum protons from the beam halo with residual gas within the beampipe.

 Previous diffractive dijet DIS measurements~\cite{H1DiJets,H1DiJets2,ZEUSDiJets} and 
 DPDF fits~\cite{H1LRG,H1DiJets,ZEUSDPDF} have been performed for $|t_{\rm min}|<|t|<1$~GeV$^2$.
 To compare with these results, the cross sections are extrapolated to the range $|t_{\rm min}|<|t|<1$~GeV$^2$ using 
 the $t$ dependence measured in inclusive diffractive DIS with a leading proton in the final state~\cite{H1FPS}.

\subsubsection{Jet selection}
\label{jetselection}

 Reconstructed hadronic final state objects are used as input to the longitudinally 
 invariant $k_T$ jet 
 algorithm~\cite{ktalgorithm} using the $p_T$ recombination scheme with a jet radius of $1.0$ as implemented in 
 the FastJet package \cite{fast}. The jet finding algorithm is applied in the photon-proton 
 centre-of-mass system ($\gamma^*p$ frame). 
 The jet variables in the $\gamma^*p$ frame are denoted by a asterisk.  

 In the \textquoteleft two central jets' analysis, the requirements are
 $P_{T,1}^*>5$~GeV and 
 $P_{T,2}^*>4$~GeV for the leading 
 and next-to-leading jet, respectively. Asymmetric cuts are placed on the jet transverse momenta to 
restrict the phase space to 
a region where NLO calculations are reliable. 
The axes of the jets are required to lie within the pseudorapidity range
 $-1<\eta_{1,2}<2.5$ in 
 the laboratory frame.
The selected event topology is similar to that in the LRG dijet data used in the DPDF fits~\cite{H1DiJets,ZEUSDPDF}.  
 This data selection is used for testing the proton vertex factorisation hypothesis and the DPDFs in processes with 
 a leading proton in the final state.  

\renewcommand{\arraystretch}{1.35}
\begin{table}[ht]
\begin{center}
   \begin{tabular}{| c | c| c|}
\hline
   Selection   & two central jets                  & one central + one forward jet \\
\hline
  DIS   & \multicolumn{2}{|c|} {$4<Q^2<110~\GeV^2$} \\ 
& \multicolumn{2}{|c|}{$0.05 < y <0.7$} \\
\hline
Leading Proton  & \multicolumn{2}{|c|}{$x_{I\!\!P} < 0.1$}   \\
& \multicolumn{2}{|c|}{$|t|<1$~GeV$^2$} \\
\hline
&$P_{T,1}^* > 5~\GeV$               & $P_{T,c}^*,P_{T,f}^* > 3.5~\GeV$ \\
Jets &$P_{T,2}^* > 4~\GeV$               & $M_{jj} > 12~\GeV$ \\
&$-1 < \eta_{1,2} < 2.5$           & $-1 < \eta_{c} < 2.5$ \\
&                                  & $1 < \eta_{f} < 2.8 , \eta_f > \eta_{c}$ \\
\hline
\end{tabular}
\end{center}
\caption{Phase space of the diffractive dijet FPS measurements.}
\label{kinrange}
\end{table}

 The selection of the \textquoteleft one central + one forward jet' topology is motivated by the study of diffractive 
 DIS processes in a phase space 
where deviations from DGLAP parton evolution may be present. The requirement of a forward jet suppresses the parton $p_T$ ordering which is assumed by DGLAP evolution. At least one central jet with $-1<\eta_c<2.5$ and one forward jet with $1<\eta_f<2.8$, where $\eta_f > \eta_c$, are required with
 $P_{T}^*>3.5$~GeV. 
In addition, the invariant mass of the central-forward jet system is required to be
larger than  $12$~GeV to avoid the phase space region in which NLO QCD calculations are unreliable.

The selection criteria for the
 two analyses are summarised in table~\ref{kinrange}.
 The \textquoteleft two central jets' data sample contains $581$ events  and the \textquoteleft one central 
 + one forward jet' data sample contains $309$ events.

\section{Corrections to the Data and Cross Section Determination}

\subsection{Background subtraction}
\label{bgncor}

\begin{figure}[p] \unitlength 1mm
  \centering
    \vspace{-0.5cm}
    \epsfig{file=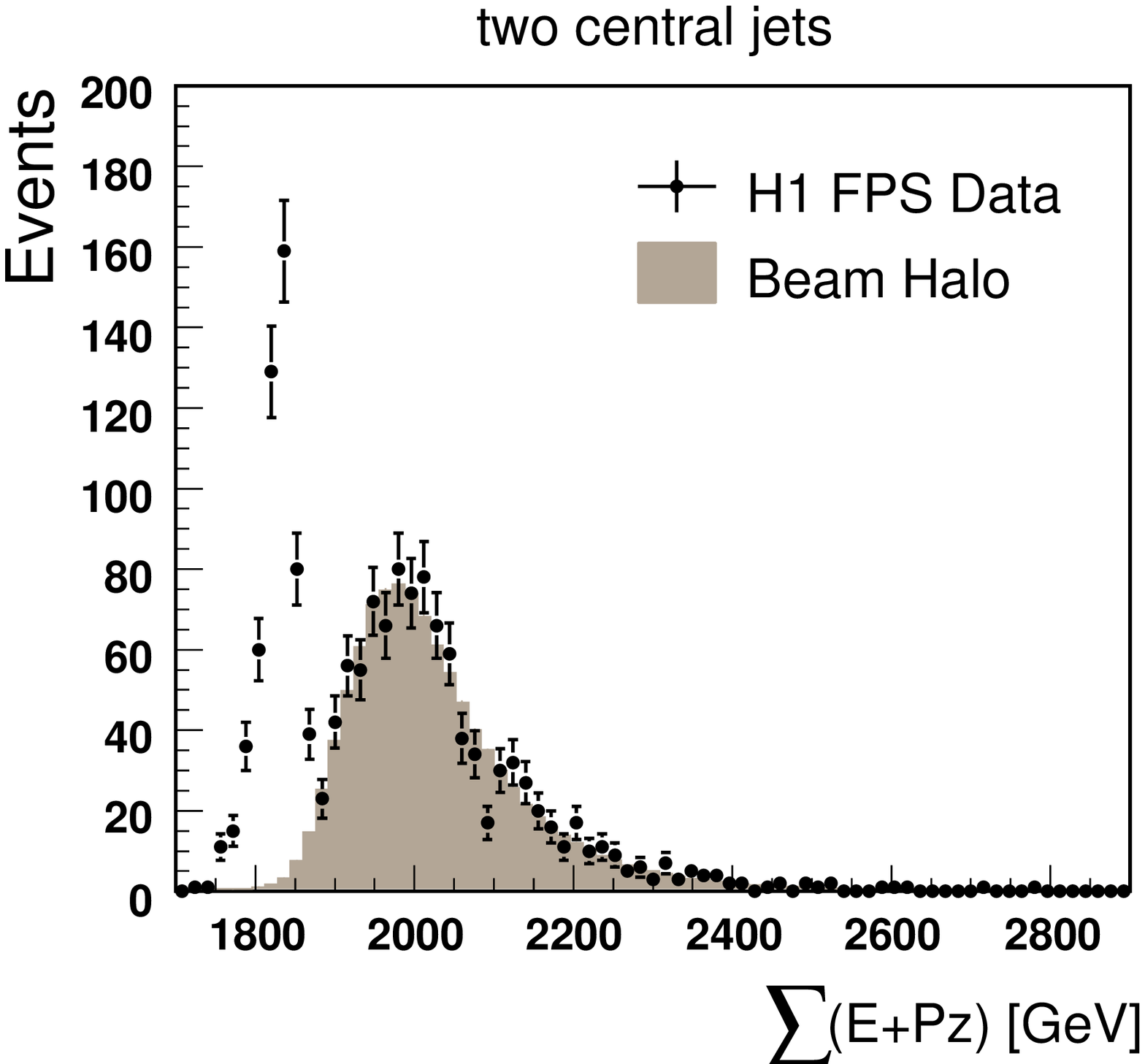 ,width=0.49\linewidth}
    \epsfig{file=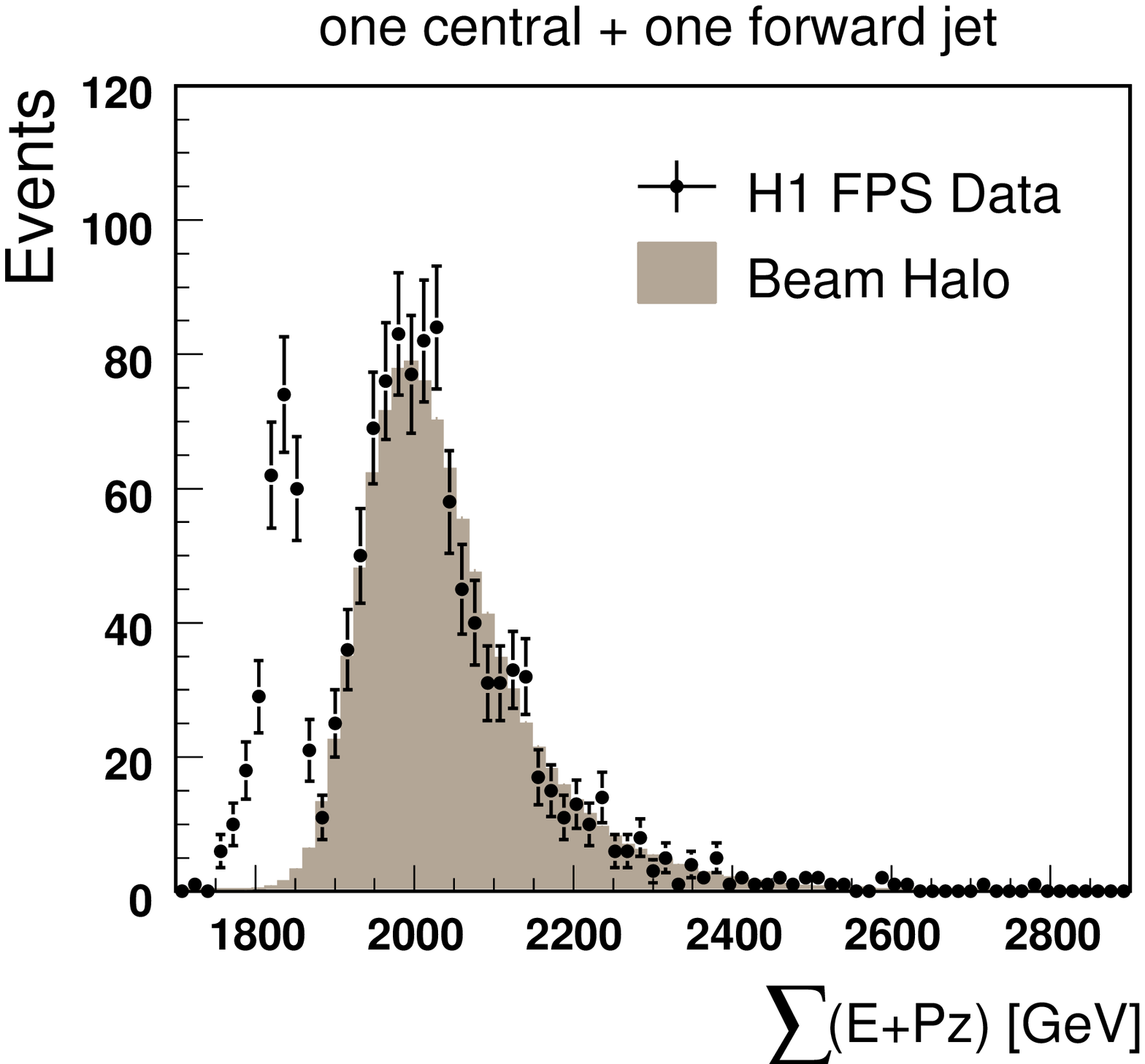 ,width=0.49\linewidth}
  \caption{The distribution of $\sum(E+P_z)$ for FPS DIS events (points) and for beam-halo DIS events 
  (shaded histogram).}
  \label{fig:epzplot}
\end{figure}

 The selected data samples contain background events arising from random
 coincidences of non-diffractive DIS events, 
 with off-momentum beam-halo protons producing a signal
 in the FPS. 
The beam-halo background contribution is estimated statistically by
 combining the quantity $\sum(E+P_z)$ summed over all reconstructed particles in the central 
 detector in DIS events (without the requirement of a track in the
 FPS) with the quantity $\sum(E+P_z)$ for beam-halo protons from randomly triggered events.
 The $\sum(E+P_z)$ spectra for leading proton and beam-halo 
 DIS events for both dijet event topologies are shown in figure~\ref{fig:epzplot}.
 The background distribution 
is normalised to the FPS DIS data distribution
 in the range $\sum(E+P_z)>1880$~GeV where the beam-halo background dominates. 
The ratio of signal to background depends on 
 the signal cross section and is found to be considerably larger than in the inclusive diffractive DIS 
 processes measured with the FPS detector~\cite{H1FPShera2}. 
 After the selection cut $\sum(E+P_z)<1880$~GeV the remaining background amounts on average to about $5\%$.
 The background is determined and subtracted bin-by-bin using this method.

\subsection{Detector simulation}
\label{mc}

 Monte Carlo simulations are used to correct the data for the effects of
 detector acceptance, inefficiencies, migrations
 between measurement intervals due to finite resolution and
 QED radiation.
 The response of the H1 detector is simulated in detail using the GEANT3 program~\cite{Geant3} and the events
 are passed through the same analysis chain as is used for the data.
 The reaction $ep \rightarrow eXp$
 is simulated with the RAPGAP program~\cite{RAPGAP}
 using the RP model and the DPDF set H1 2006 Fit B as described in section~\ref{models}.  
 QED radiative effects are simulated using the HERACLES~\cite{HERACLES} program within the
 RAPGAP event generator. %
In the \textquoteleft two central jets' analysis the $\eta_2^*$ distribution 
of the Monte Carlo simulation is reweighted in order to 
 describe the experimental data. 
A similar procedure is applied to the $\eta_f^*$ distribution in the \textquoteleft one central + one forward jet' sample.
More details of the analysis can be found elsewhere \cite{RPThesis}.

\begin{center}
   \begin{figure}
     \epsfig{file=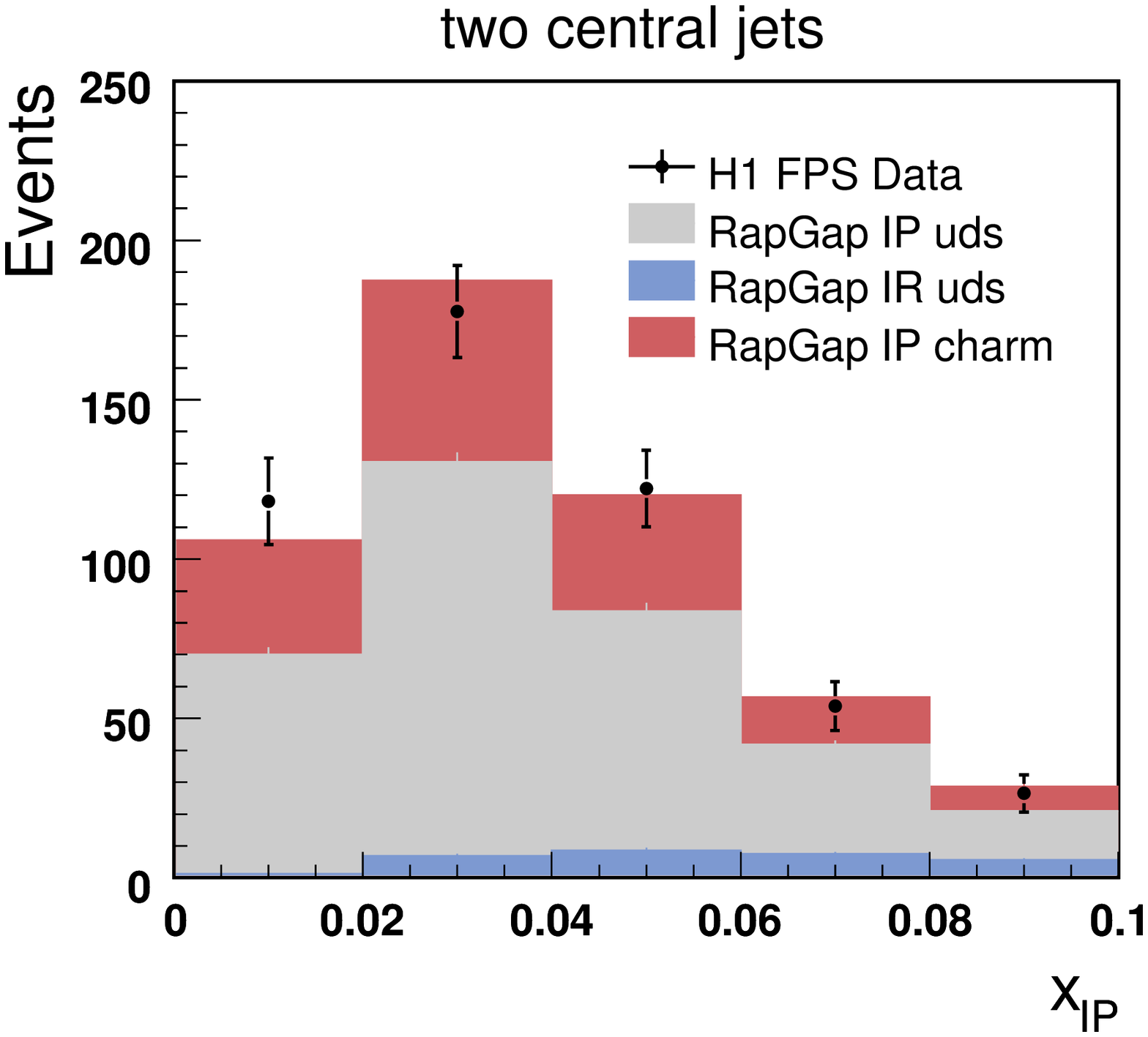 ,width=0.49\linewidth}
     \epsfig{file=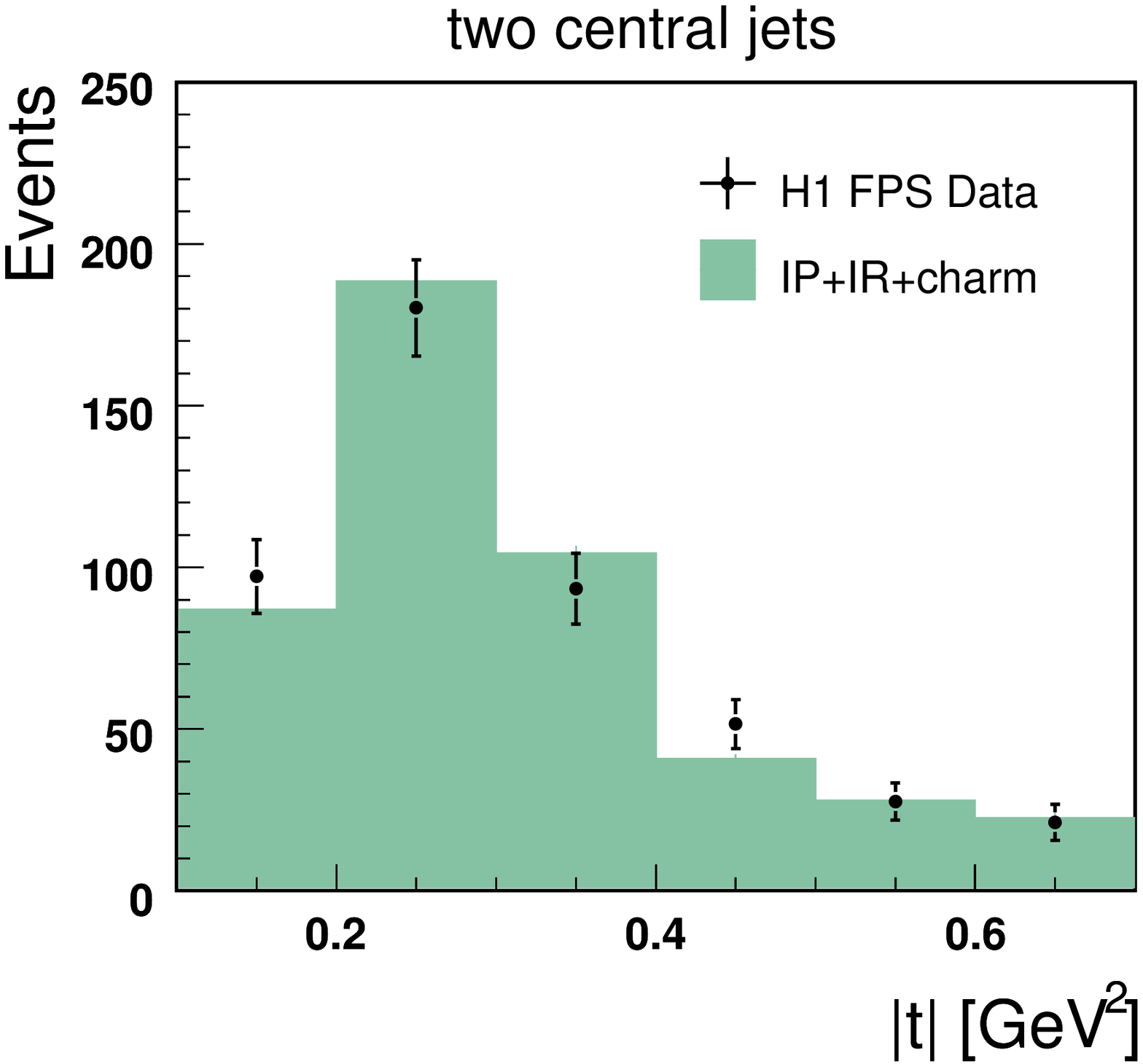 ,width=0.49\linewidth}
     \caption{ The distributions of the variables $\xpom$ (a) and $|t|$ (b) reconstructed using the 
 FPS (points) for events with two 
central jets. The beam-halo background is subtracted from the data. The RAPGAP Monte Carlo 
 simulation, reweighted to describe the $\eta_2^*$ distribution, 
 is shown as a histogram. Contributions from sub-processes are illustrated in the $\xpom$ distribution as areas
 filled with different colours.}
   \label{fig:fpsplots1}
 \end{figure}
 \end{center}

\begin{figure}
   \begin{center}
     \epsfig{file=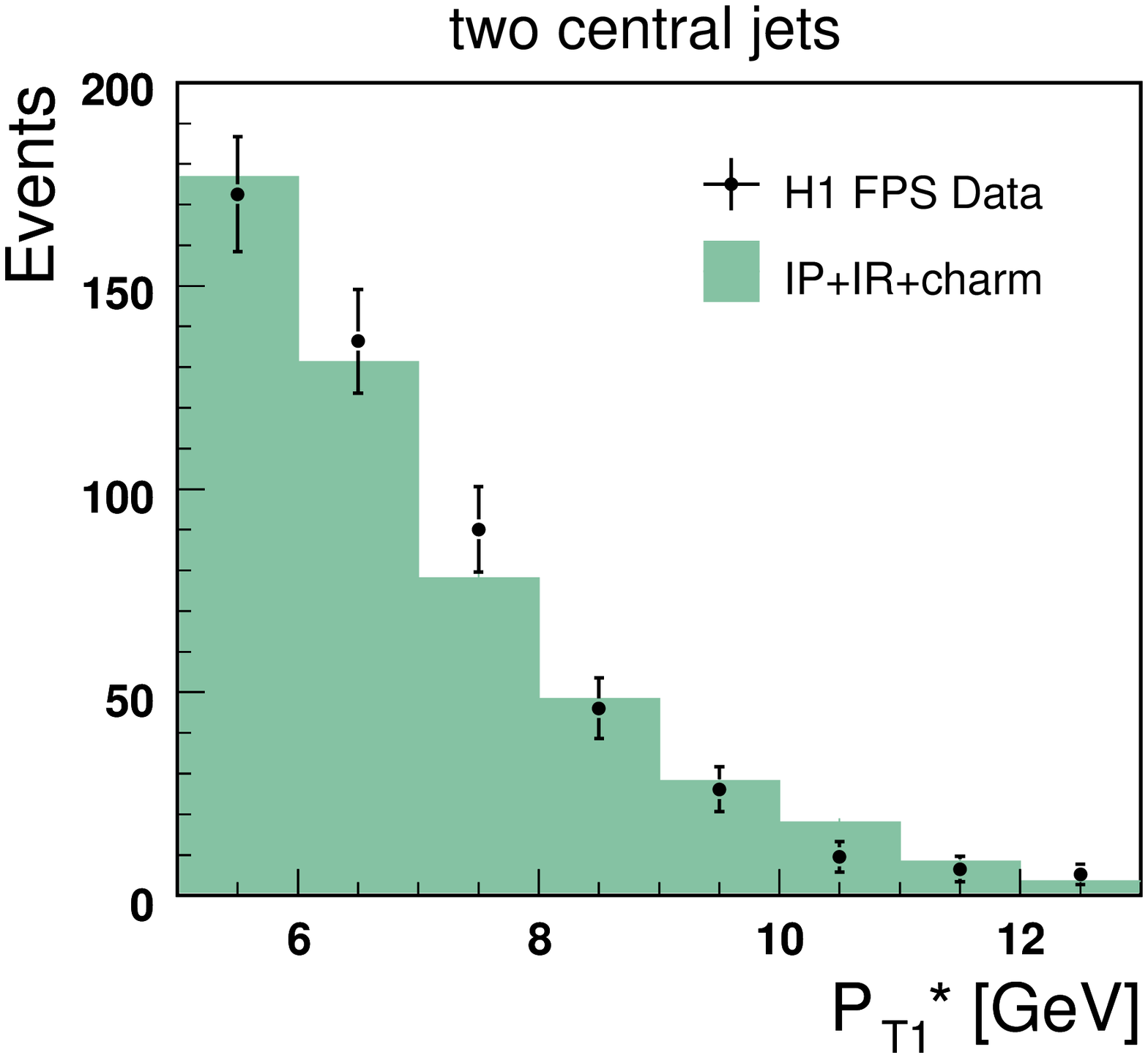,width=0.48\linewidth}
     \epsfig{file=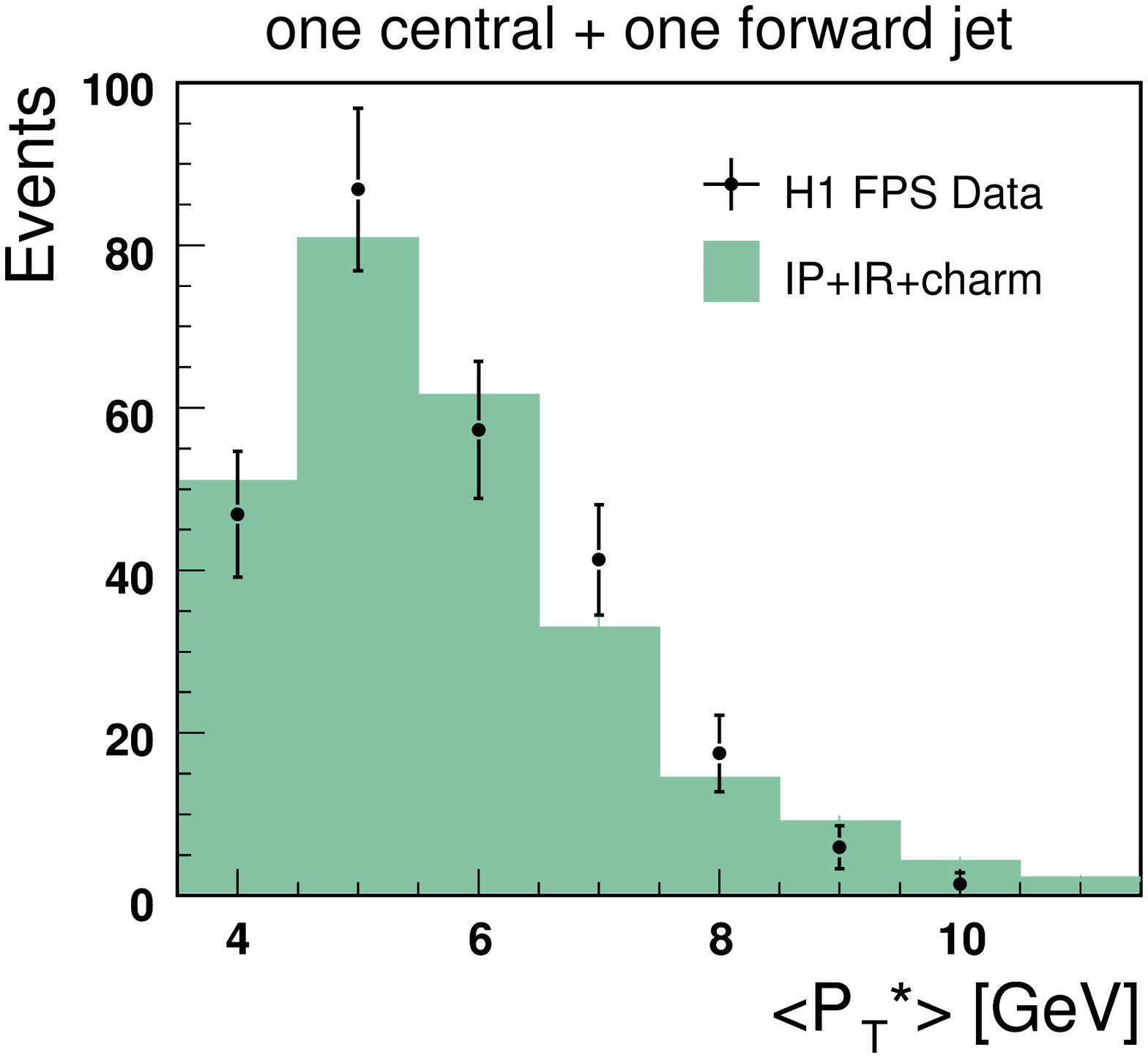,width=0.48\linewidth}
     \epsfig{file=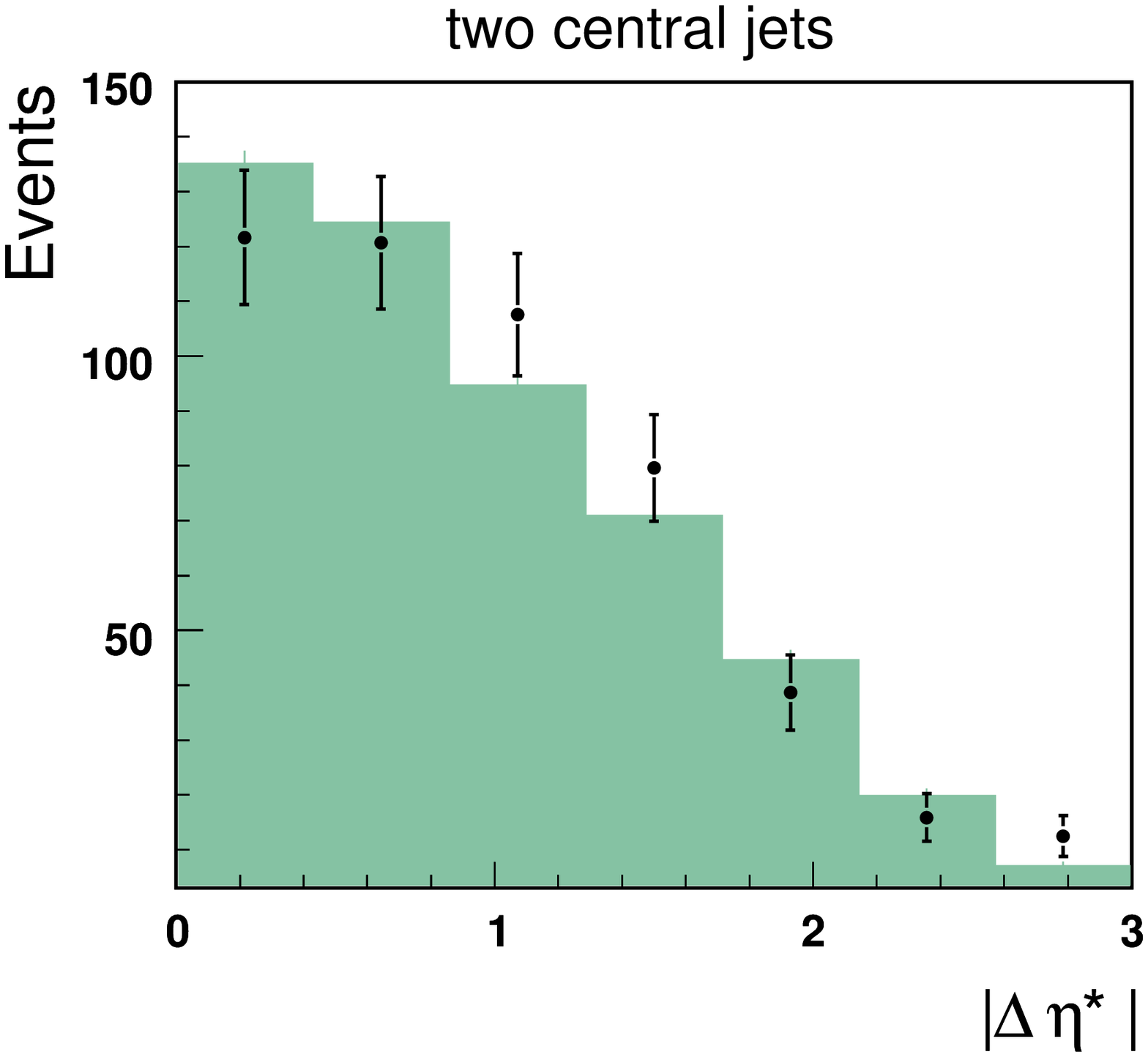,width=0.48\linewidth}
     \epsfig{file=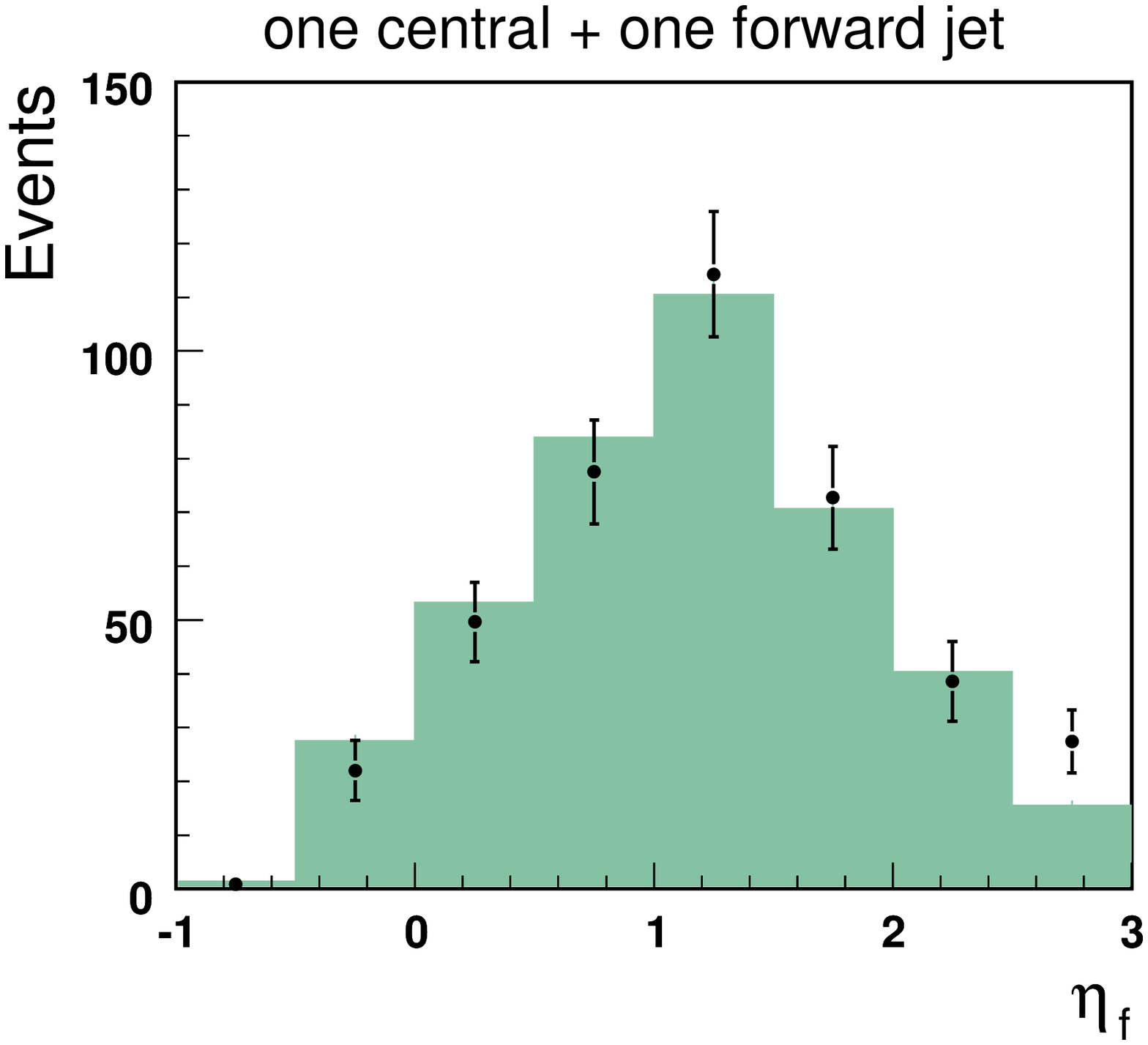,width=0.48\linewidth}
     \epsfig{file=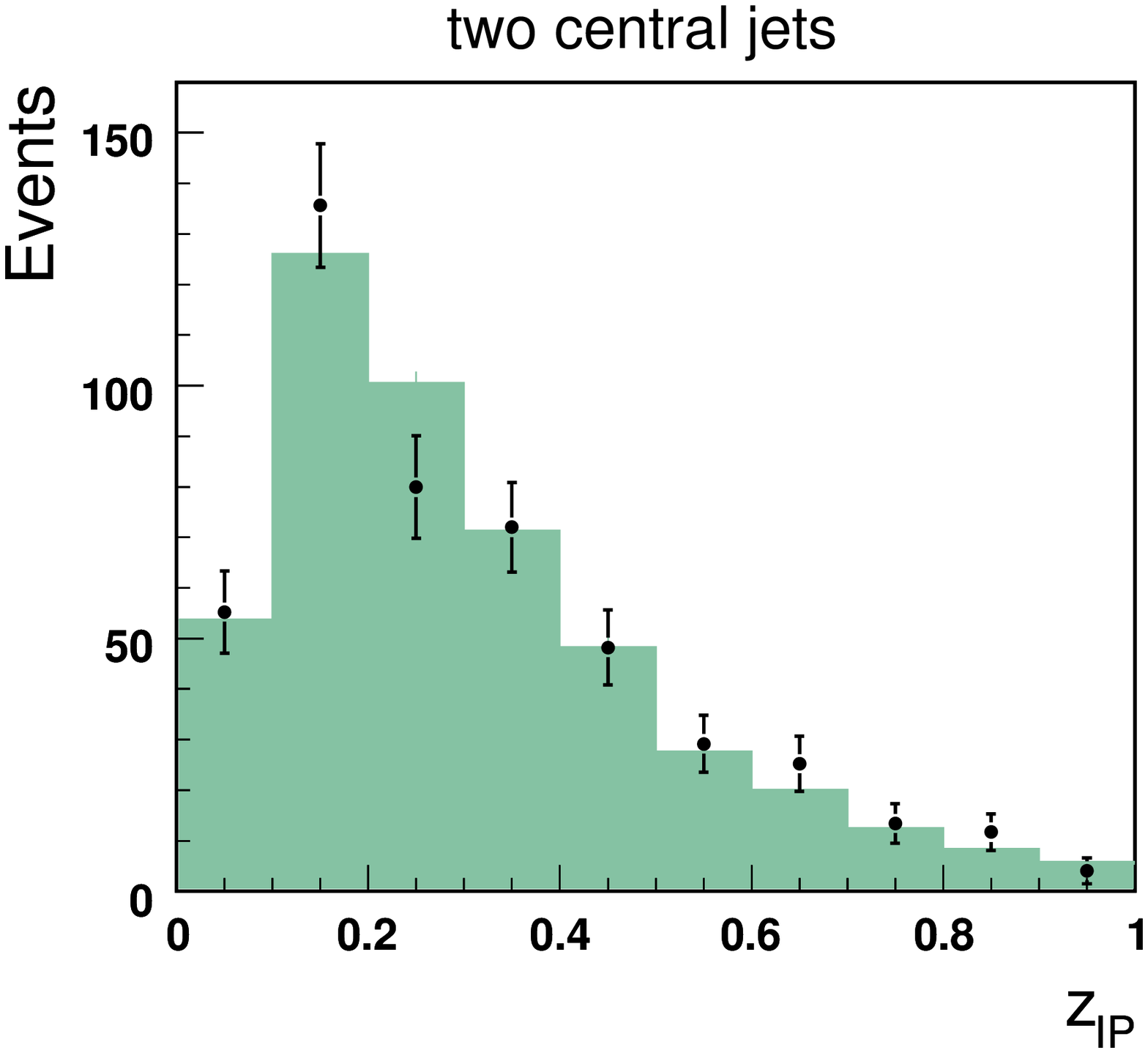,width=0.48\linewidth}
     \epsfig{file=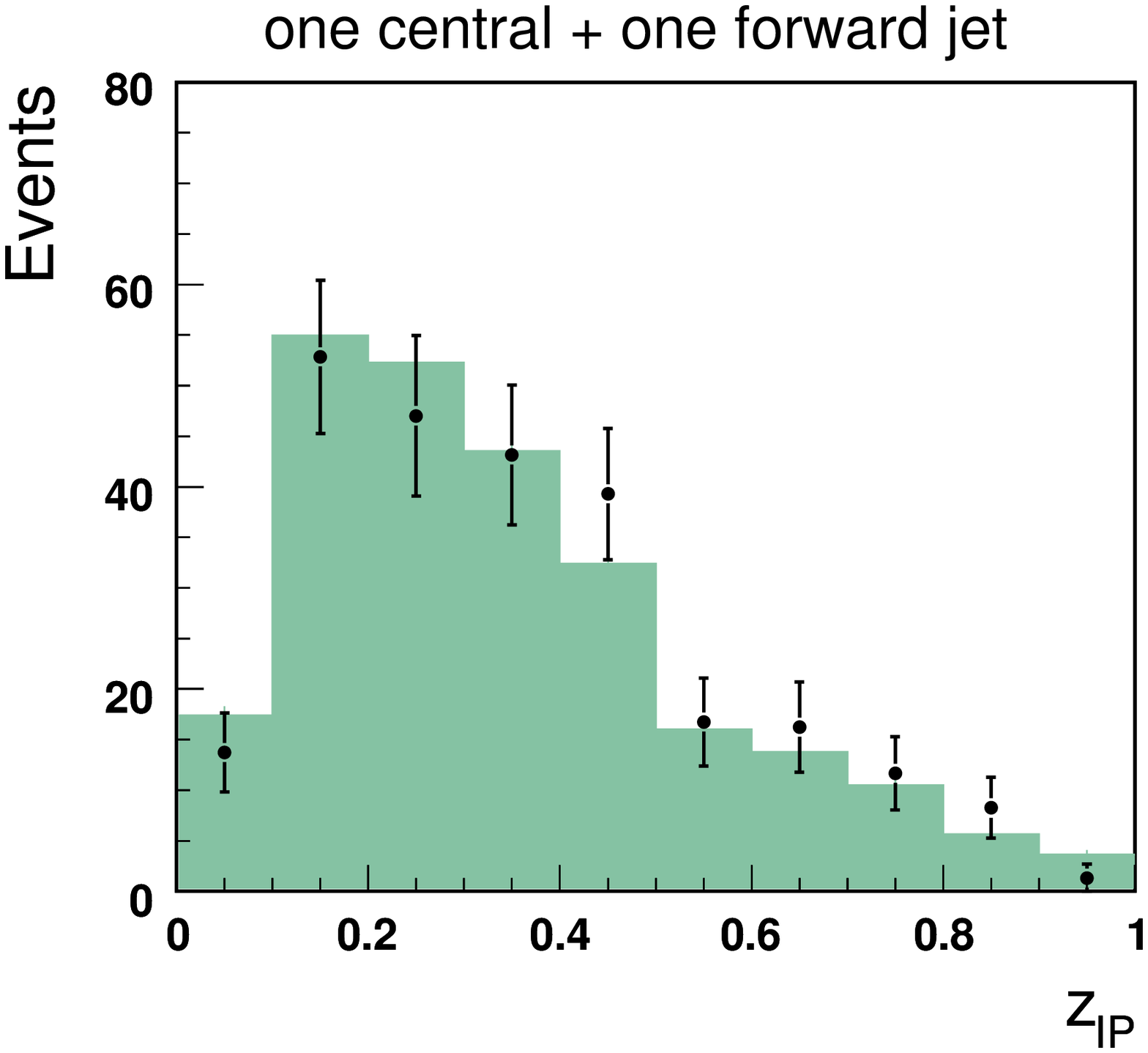,width=0.48\linewidth}
   \end{center}
   \caption{The distributions of the variables $P_{T,1}^{*}$, $|\Delta\eta^{*}|$ and $z_{I\!\!P}$ for events with two 
central jets and of the variables $\langle P_T^*\rangle$, $\eta_f$ and $z_{I\!\!P}$ for events with one central and one forward jet 
(histogram with the error bars). The beam-halo background is subtracted from 
the data.}
   \label{fig:fpsplots2}
 \end{figure}

 A comparison of the FPS data 
 and the RAPGAP simulation is presented in 
 figure~\ref{fig:fpsplots1} for the variables $\xpom$ and $|t|$ reconstructed with the FPS 
 detector. The contributions of light quarks (uds) to $\pom$ and $\reg$ exchanges and of charm quarks to $\pom$ exchange    
 are also shown in the $\log_{10}(\xpom)$ distribution.
 Figure~\ref{fig:fpsplots2} presents the data and the Monte Carlo distributions 
 of the variables $P_{T,1}^*, |\Delta\eta^*|$ and $\zpom$ for the \textquoteleft two central jets' sample and
 of the variables $\langle P_T^*\rangle, \eta_f$ and $\zpom$ for the 
 \textquoteleft one central + one forward jet' topology. 
For this comparison $\zpom$ is reconstructed from the scattered electron and the hadronic final state in the H1 detector.
 The MC simulation reproduces the data within the experimental systematic uncertainties.   

 \subsection{Cross section determination}
 \label{xsect}
 
In order to account for migration and smearing 
 effects and to evaluates the dijet cross sections at the level of stable hadrons, matrix unfolding of the reconstructed data is performed~\cite{blobel}.  The resolution and acceptance of the H1 detector is reflected 
 in the unfolding matrix {\bf A} which relates reconstructed variables $\vec{y}_{{\rm rec}}$ with  
 variables on the level of stable hadrons $\vec{x}_{{\rm true}}$ via the formula ${\bf A}\vec{x}_{{\rm true}} = 
 \vec{y}_{{\rm rec}}$. The matrix {\bf A}, obtained for each measured distribution using the RAPGAP simulation, is 
 constructed within an enlarged phase space in order to take into account possible migrations from outside of the 
 measured kinematic range. 
The following sources of migrations to the analysis phase space are considered: migrations from low $Q^2$, from low $y$, from large $\xpom$, from low $P_T$ jets, from the single jet topology, fulfilling the $P_T$ requirements for the leading jet as given table \ref{kinrange}, and in case of the \textquoteleft one central + one forward jet' analysis from large $\eta_f$. 
In order to treat 
the contamination of the measurement 
by these migrations correctly the analysis is performed in an extended phase space which includes side-bins in $\vec{y}_{rec}$ and $\vec{x}_{true}$ for each of the migration sources listed above. 


The unfolded true distribution on the level of stable hadrons is obtained from the measured one by minimising a $\chi^2$ function 
defined as

\begin{equation}
 \chi^2 = \chi^2_A + \tau^2\chi^2_L = 1/2(\vec{y}_{{\rm rec}} - {\bf A}\vec{x}_{{\rm true}})^T{\bf 
 V}^{-1}(\vec{y}_{{\rm rec}} - {\bf A}\vec{x}_{{\rm true}}) + \tau^2\chi^2_L
\end{equation}

\noindent
 where $\chi^2_A$ is a  measure of a deviation of ${\bf A}\vec{x}_{{\rm true}}$ from the data bins 
 $\vec{y}_{{\rm rec}}$. The matrix ${\bf V}$ is the covariance matrix of the data, based on the statistical 
 uncertainties. In order to 
 avoid statistical fluctuations, the regularisation term $\chi^2_L$ is implemented into the $\chi^2$ function and 
 defined as $\chi^2_L=(\vec{x}_{{\rm true}})^2$. The 
 regularisation parameter $\tau$  
 is tuned in  order to minimise the bin-to-bin correlations of the covariance matrix {\bf V}. Further details of the 
 unfolding method can be  found in~\cite{unf,unf2}


The 
Born level 
cross section is calculated in each bin $i$ according to the formula:

\begin{equation}
\sigma_i(ep \rightarrow ejjX^{\prime}p) = \frac{x_i}{\mathcal{L}}(1+\delta_{rad})
\end{equation}

\noindent
 where $x_i$ is the number of background subtracted events 
 as obtained with the unfolding procedure described 
 above, $\mathcal{L}$ is the total integrated luminosity 
 and $(1+\delta_{rad})$ are the QED 
 radiative corrections
which amount to about 5\% on average. 
The differential cross sections are obtained by dividing by the bin width.



\section{Systematic Uncertainties on the Measured Cross Sections}
\label{systs}

 The systematic uncertainties are implemented into the response matrix {\bf A} and propagated through the 
 unfolding  procedure. They are considered from the following sources listed below.

\begin{itemize}
\item The uncertainties on the leading proton
 energy and on the horizontal and vertical projections of the proton
 transverse momentum are $1$~GeV, $10$~MeV and $30$~MeV,
 respectively (section~\ref{detector}).
 The corresponding average uncertainties on the
 cross section  measurements are
 $0.5\%$, $5.3\%$ and $2.2\%$. 
 The dominant uncertainty originates from the FPS acceptance variation as a
 function of the leading proton transverse momentum in the horizontal
 projection.
 The above uncertainties result from the run-by-run variations of the
 incoming proton beam angle and of the FPS detector positions relative to the proton beam, as well as 
 from the imperfect knowledge of the HERA beam magnet optics.

\item The uncertainties of the measurements of the scattered electron energy $E'_e$ ($1\%$)
 and angle $\theta_e^\prime$ ($1$~mrad) on the SpaCal calorimeter
 lead to an average systematic uncertainty of the cross section of $1.5\%$ and $2.8\%$, 
 respectively.

\item The systematic uncertainty arising from
 the hadronic final state
 reconstruction is determined by varying the
 energy scale of the hadronic final state by $\pm 2\%$ as obtained using 
 a dedicated calibration~\cite{iter}. The $2\%$ uncertainty of the calibration is confirmed 
 by studies in the region of low jet transverse momenta and low photon virtuality. This source leads to an average uncertainty of the cross section measurements 
 of $6.2\%$ for production of two central jets and $9.5\%$ for production of one central and 
 one forward jet.

\item The model dependence of the acceptance and migration corrections is estimated by varying the shapes 
 of the distributions in the kinematic variables $\langle P_{T}^* \rangle$, $\eta_2^*$, $\eta_f^*$, $\xpom$, $\beta$ and $Q^2$  
 in the RAPGAP simulation within the constraints imposed on those distributions by the presented data. The $\eta_2^*$ and $\eta_f^*$  
 reweightings are varied within the errors of the parameters of the reweighting function, which amount up to a factor $4$. 
 The $\langle P_T^*\rangle$ distribution is reweighted by 
 $\langle P_{T}^* \rangle^{\pm 0.15}$, the $\xpom$ distribution by $(1/\xpom )^{\pm 0.05}$, the $\beta$ distribution 
 by $\beta^{\pm 0.05}$ and $(1- \beta )^{\mp 0.05}$ and the $Q^2$ distribution by $\log(Q^2)^{\pm 0.2}$. For the \textquoteleft two central jets' selection the largest 
 uncertainty is introduced by the $\eta_2^*$ reweighting ($4\%$), followed by $\beta$ ($2.7\%$), while the reweights 
 in $\xpom$, $\langle P_{T}^* \rangle$ and $Q^2$ result in an overall uncertainty of $2.3\%$. The uncertainties for the \textquoteleft one central + one forward jet' topology are $12.8\%$ for the $\eta_f^*$ reweighting, followed by $\langle P_{T}^* \rangle$ ($2.1\%$), while the reweights in $\xpom$, $\beta$ and $Q^2$ result in an overall uncertainty of $1.8\%$.

\item Reweighting the $t$ 
 distribution by $e^{\pm t}$ results in a normalisation uncertainty of $4.2\%$ for the extrapolation in  $t$ from the 
 measured range of $0.1 < |t|  < 0.7~\GeV^2$ to the region $|t_{\rm min}| < |t| < 1~{\GeV}^2$ covered by the LRG 
 data \cite{H1DiJets}. The uncertainty arising from the $t$ reweighting within the FPS acceptance range of 
 $0.1<|t|<0.7~\GeV^2$ is on average $1.4\%$.

\end{itemize}

The following uncertainties are considered to influence the normalisation of all measured cross sections in a correlated way:

\begin{itemize}

 \item 
Two sources of systematics related to the background subtraction are taken into account: the energy scale 
 uncertainty and the limited statistics in the data sample without the $\sum(E+p_z)$ cut. Firstly, 
 the beam-halo spectrum is
 shifted within the quoted uncertainties of the hadronic energy scale and proton energy scale. Secondly, the 
 normalisation of the background spectrum is shifted by $1\pm1/\sqrt{N_{{\rm bkg}}}$, where $N_{{\rm bkg}}$ is the number 
 of events in the FPS data sample in the range $\sum(E+P_z) > 1880$~GeV. The uncertainties from these two sources  
 are combined in quadrature. The uncertainty of the proton beam-halo background is considered as a normalisation error 
 and found to be $3.5\%$ for the production of two central jets and $1.5\%$ for the production of one central and one 
 forward jet. 

\item A normalisation uncertainty of $1\%$ is
 attributed to the trigger efficiencies, evaluated using event samples obtained with
 independent triggers.

\item The uncertainty in the FPS track reconstruction efficiency
 results in a normalisation uncertainty of $2\%$.

\item A normalisation uncertainty of $3.7\%$ arises from
 the luminosity measurement.

\end{itemize}
 
\noindent
 The systematic errors shown in the figures 
 are obtained by adding in quadrature
 all the contributions except for the normalisation uncertainties,
 leading to an average uncertainty of $11\%$ for \textquoteleft two central jets' and $17\%$ for \textquoteleft one 
 central + one forward jet'. The overall normalisation uncertainty of the cross section measurement
obtained by adding in quadrature all normalisation uncertainties is $7\%$ for 
 \textquoteleft two central jets' and $6.2\%$ for \textquoteleft one central + one forward jet'. 
The cross section measurement in $t$ has a normalisation uncertainty $4.6\%$. 

\section{Results}

 The $ep$ cross section for diffractive production of two central jets and one central + one forward jet,
 integrated over the full measured  kinematic range (table~\ref{kinrange}), is given in table~\ref{ccftable} 
together with the predictions obtained with NLO QCD calculations.

\begin{table}[ht]
\begin{center}
   \begin{tabular}{| c | c| c|}
\hline
 &              two central jets      &    one central + one forward jet                        \\
     & $\sigma$ [pb]                  & $\sigma$ [pb] \\
\hline


Data &  $254 \pm 20~ (\rm{stat.}) \pm 27~\rm{(syst.)}$   &  $150 \pm 19~ (\rm{stat.}) \pm 26~\rm{(syst.)}$ \\
\hline
NLO QCD &&\\
H1 2006 Fit B & $270~ ^{+134}_{-53}~\rm{(scale)}~ \pm 16~\rm{(hadr.)}$   &$148~ ^{+102}_{-33}~\rm{(scale)} \pm 6~\rm{(hadr.)}$ \\
H1 2007 Jets &  $257~ ^{+87}_{-51}~\rm{(scale)}~ \pm 22~\rm{(hadr.)}$ &$128~ ^{+66}_{-37}~\rm{(scale)} \pm 7~\rm{(hadr.)}$  \\
\hline
\end{tabular}
\end{center}
\caption{Total cross section for the \textquoteleft two central jets' and \textquoteleft one central + one forward jet' samples compared to the NLO QCD calculations.}
\label{ccftable}
\end{table}

Within the uncertainties, both cross sections are well described by the NLO QCD calculations. 

 The measured differential cross sections are presented in tables~\ref{dijet}-\ref{cfjet} and figures~\ref{dijnlo1}-\ref{cfmc2}. The tables also include the full covariance matrices of the experimental uncertainties. 
 The quoted differential cross sections  are averaged over the intervals specified in the
 tables~\ref{dijet}-\ref{cfjet}.

\subsection{Differential cross section for the production of two central jets}
\label{2cjxs}





\begin{figure}\unitlength 1mm
  \begin{center}
  \epsfig{file=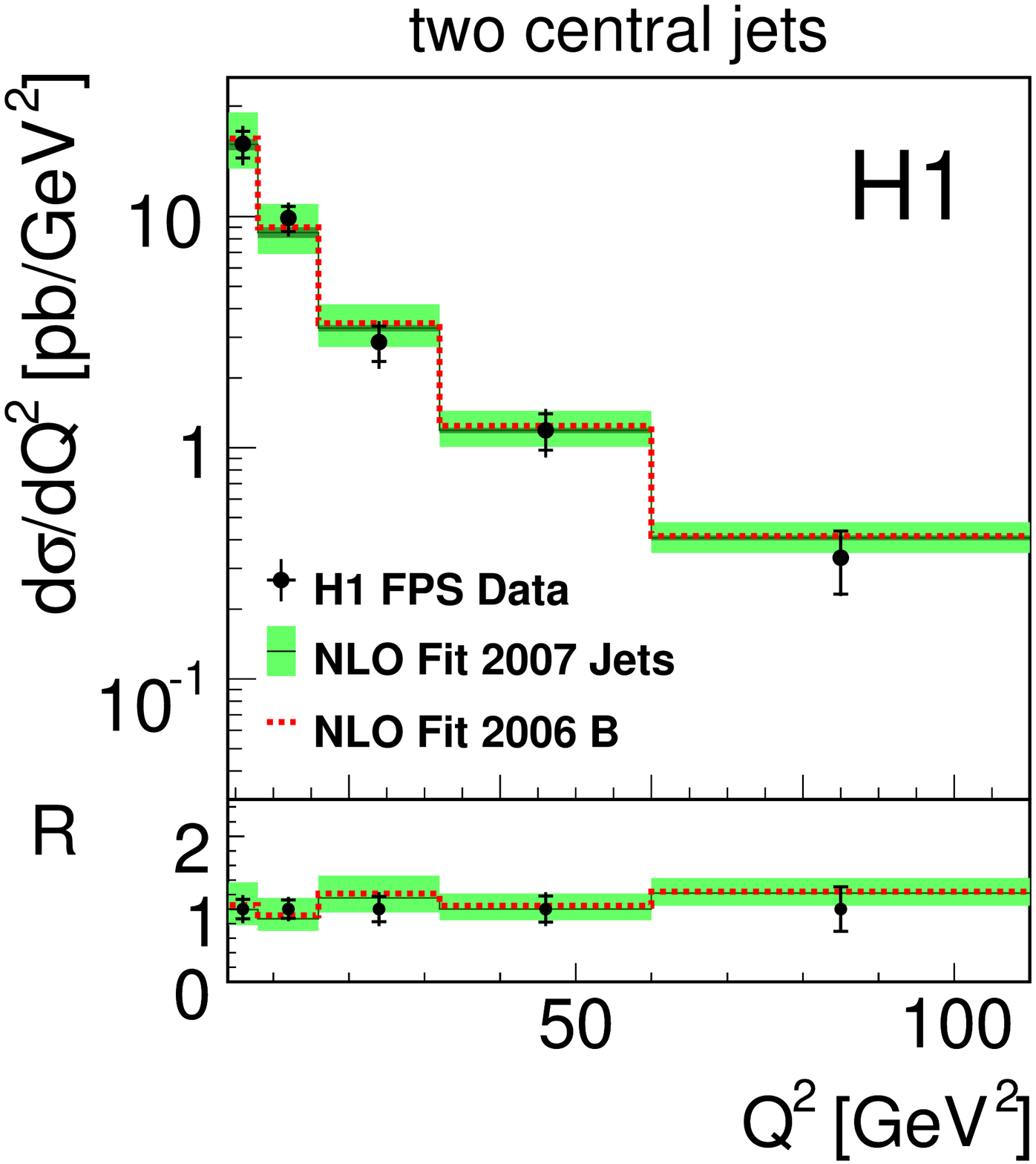,width=0.49\linewidth}
  \epsfig{file=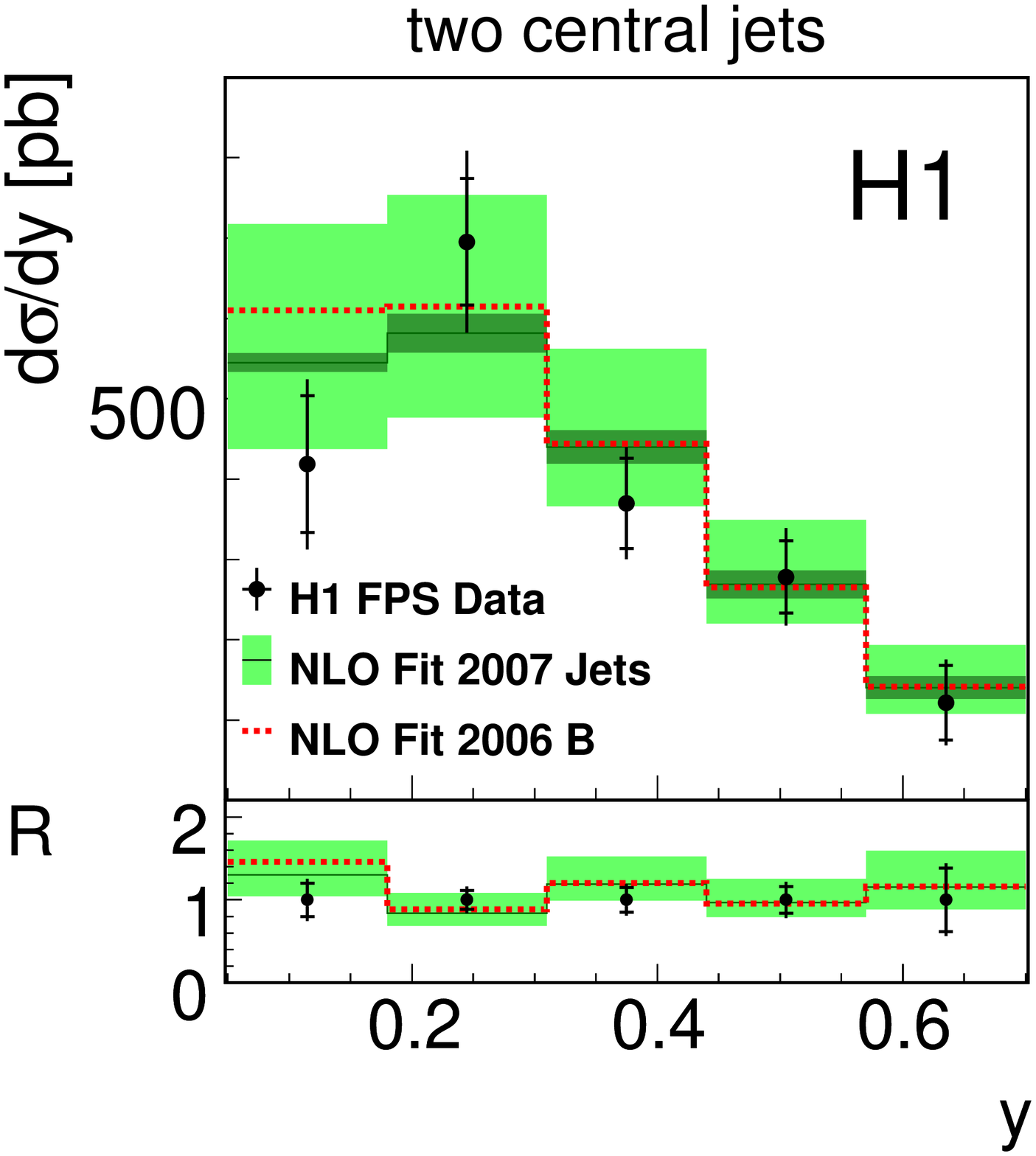,width=0.49\linewidth}
 \epsfig{file=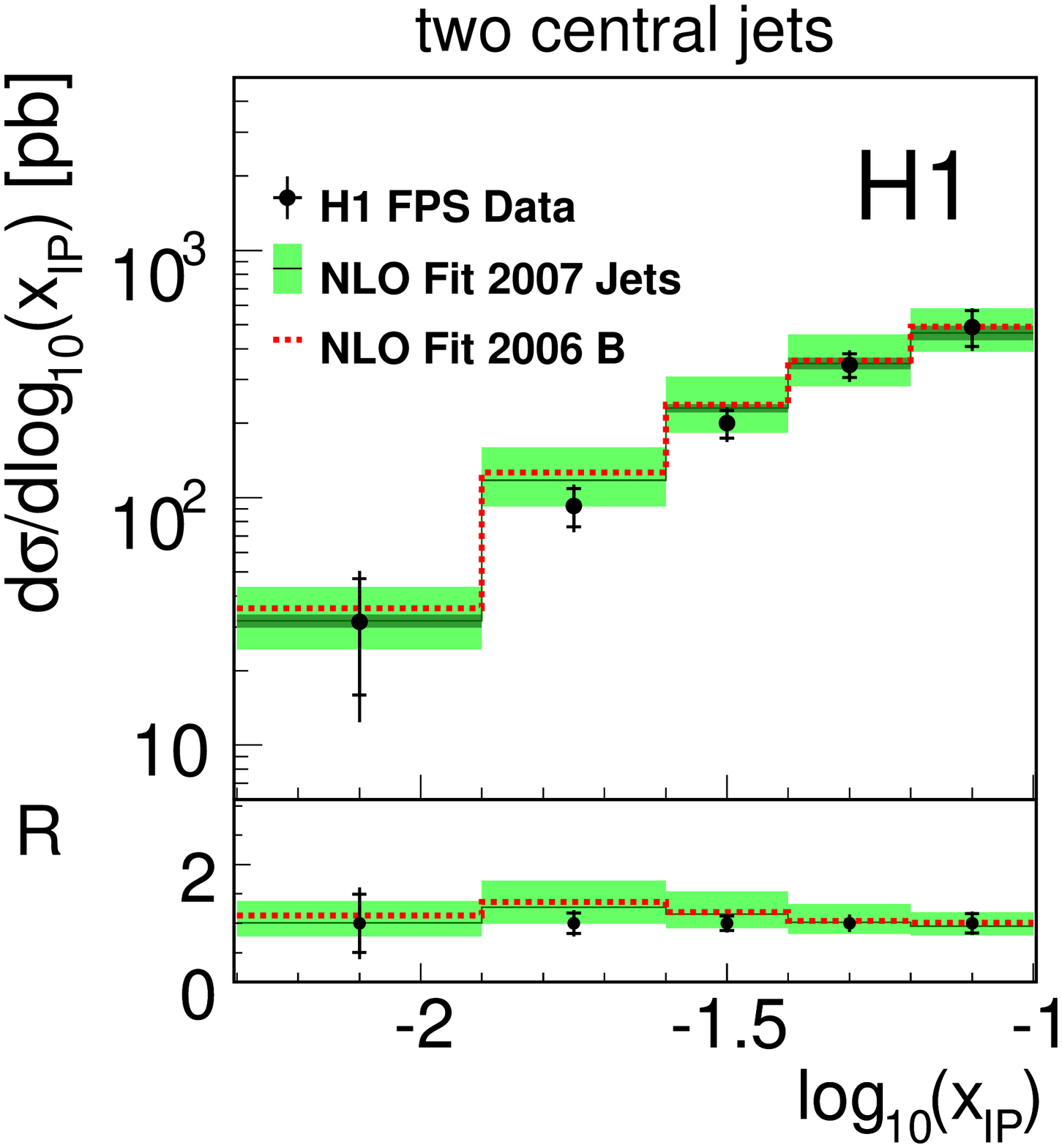,width=0.49\linewidth}
  \epsfig{file=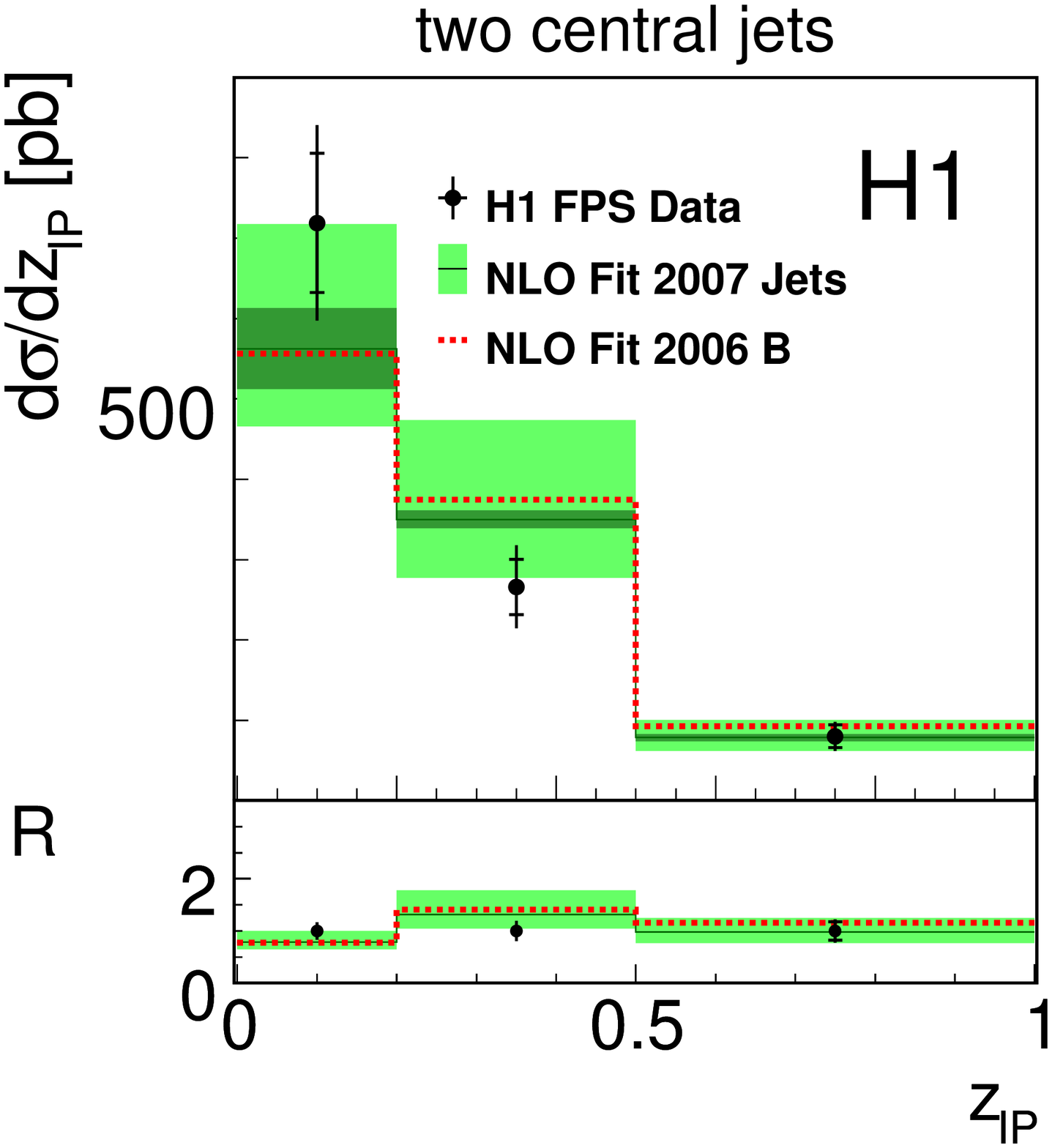,width=0.49\linewidth}


 \caption{The differential cross section for the production of two central
  jets shown as a function of $Q^2$, $y$, $\log_{10}(\xpom)$ and $z_{I\!\!P}$. The inner error bars represent the statistical errors. The outer
  error bars indicate the statistical and systematic errors added in quadrature. 
 NLO QCD predictions based on the 
 DPDF set  H1 2007 Jets, corrected to the level of stable hadrons, are shown as a solid line and
 a dark green band indicating the hadronisation uncertainties and light green band indicating the hadronisation and scale uncertainties added in quadrature.
 LO QCD predictions based on the same DPDF set are shown as a dotted line. 
 The NLO calculations 
 based on the DPDF set H1 2006 Fit B with applied hadronisation corrections are 
  shown as a thick line. R denotes the ratio of the measured cross sections and QCD predictions to the nominal values of
  the measured cross sections. The total normalisation error of $7.0\%$ is not shown.}
  \label{dijnlo1}
  \end{center}
 \end{figure}

 The measured differential cross sections are shown as a function of $Q^2$, $y$, $\log_{10}(x_{I\!\!P})$ and  $z_{I\!\!P}$ in 
 figure~\ref{dijnlo1}. 
 The calculations  obtained  with the DPDF sets H1 2006 Fit B  and H1 2007 Jets are presented as well as the ratio $R$ of the calculations to the data.
 Within the uncertainties, the normalisation and shape of the cross sections are reasonably well described by the NLO QCD 
 predictions. 
 Since dijet production is directly sensitive to the gluon DPDF, the measured 
cross sections  confirm the normalisation and shape of the gluon DPDFs 
 extracted from the NLO QCD fits to diffractive inclusive and dijet cross sections measured using 
 the LRG method~\cite{H1LRG,H1DiJets}.

\begin{figure}
  \begin{center}
    \epsfig{file=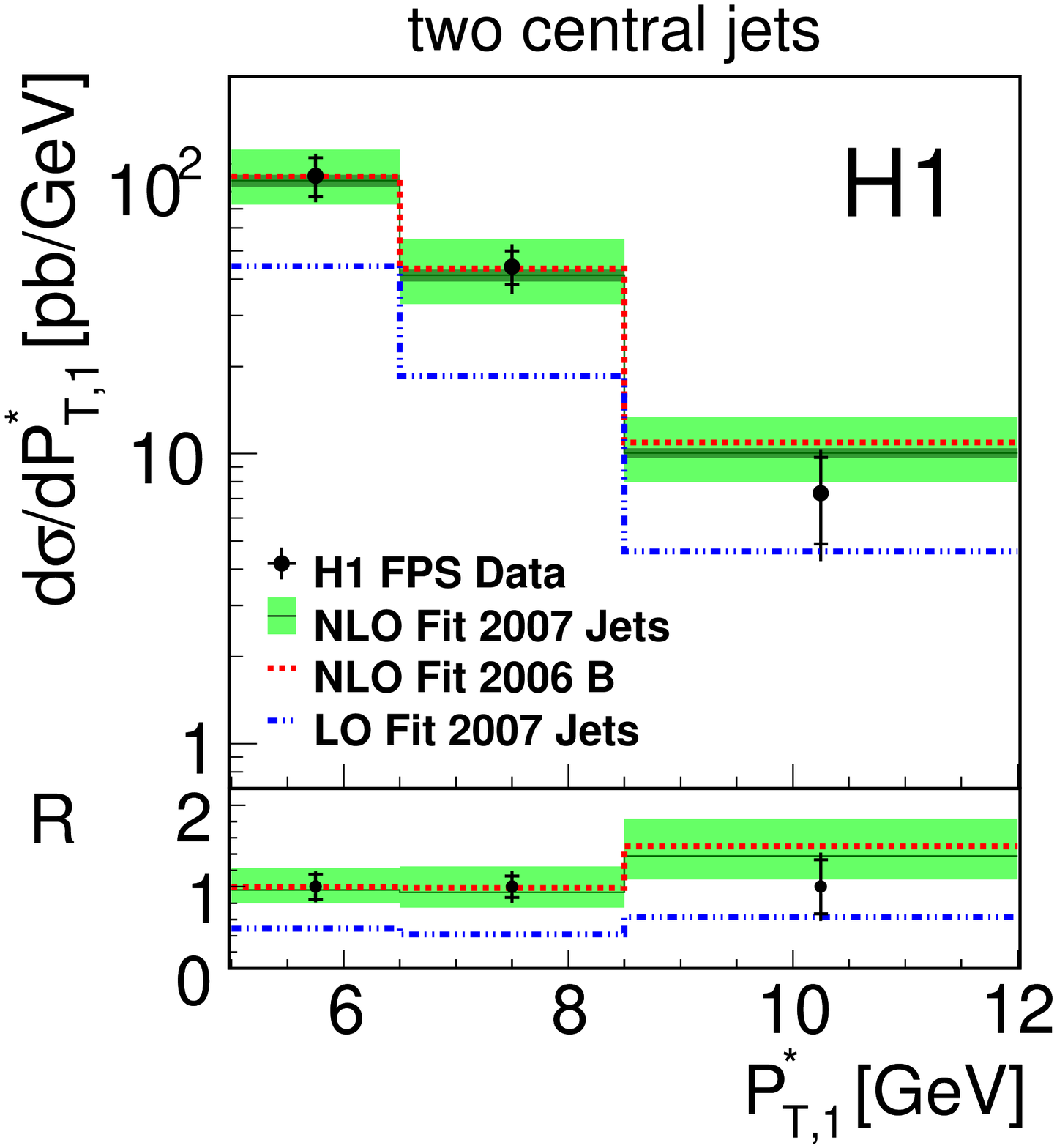,width=0.49\linewidth}
  \epsfig{file=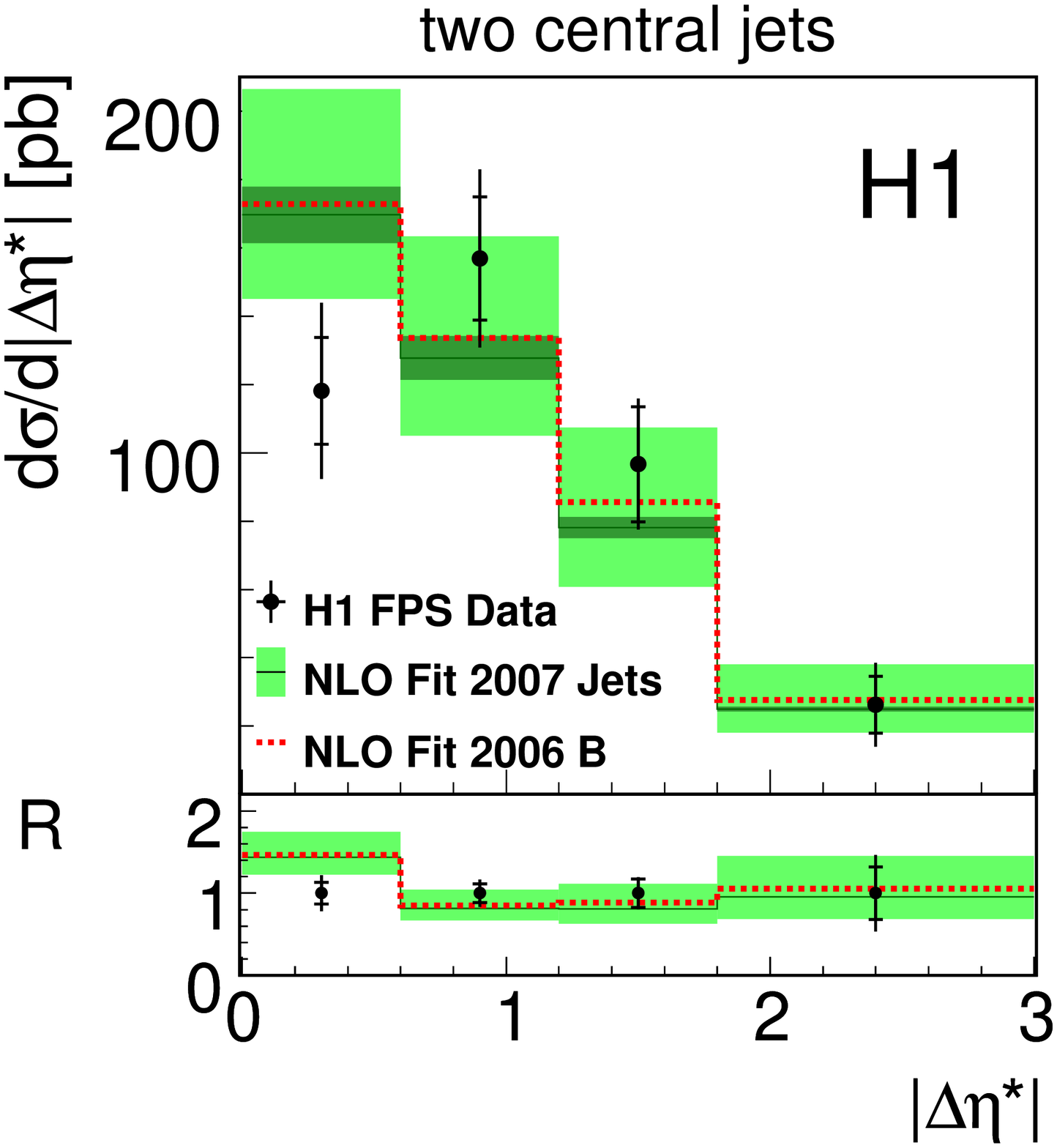,width=0.49\linewidth}
   \caption{The differential cross section for production of two central
  jets shown as a function of $P_{T,1}^*$ and $|\Delta\eta^*|$. For more details see figure \ref{dijnlo1}.}
  \label{dijnlo2}
  \end{center}
  \end{figure}

In figure~\ref{dijnlo2} the differential cross sections are shown as a function of $P^*_{T,1}$ and $|\Delta\eta^*|$. Within the errors, NLO QCD predictions describe the data. 
 A slight deviation of the theory from the data is observed for jets with a small separation in 
 pseudorapidity $|\Delta\eta^*|$.
 For the differential cross section in $P^*_{T,1}$, 
the LO QCD contribution is calculated as well using the DPDF set 
 H1 2007 Jets and is observed to underestimate the measured cross section by a factor of about $2$.

\begin{figure}
  \begin{center}
  \epsfig{file=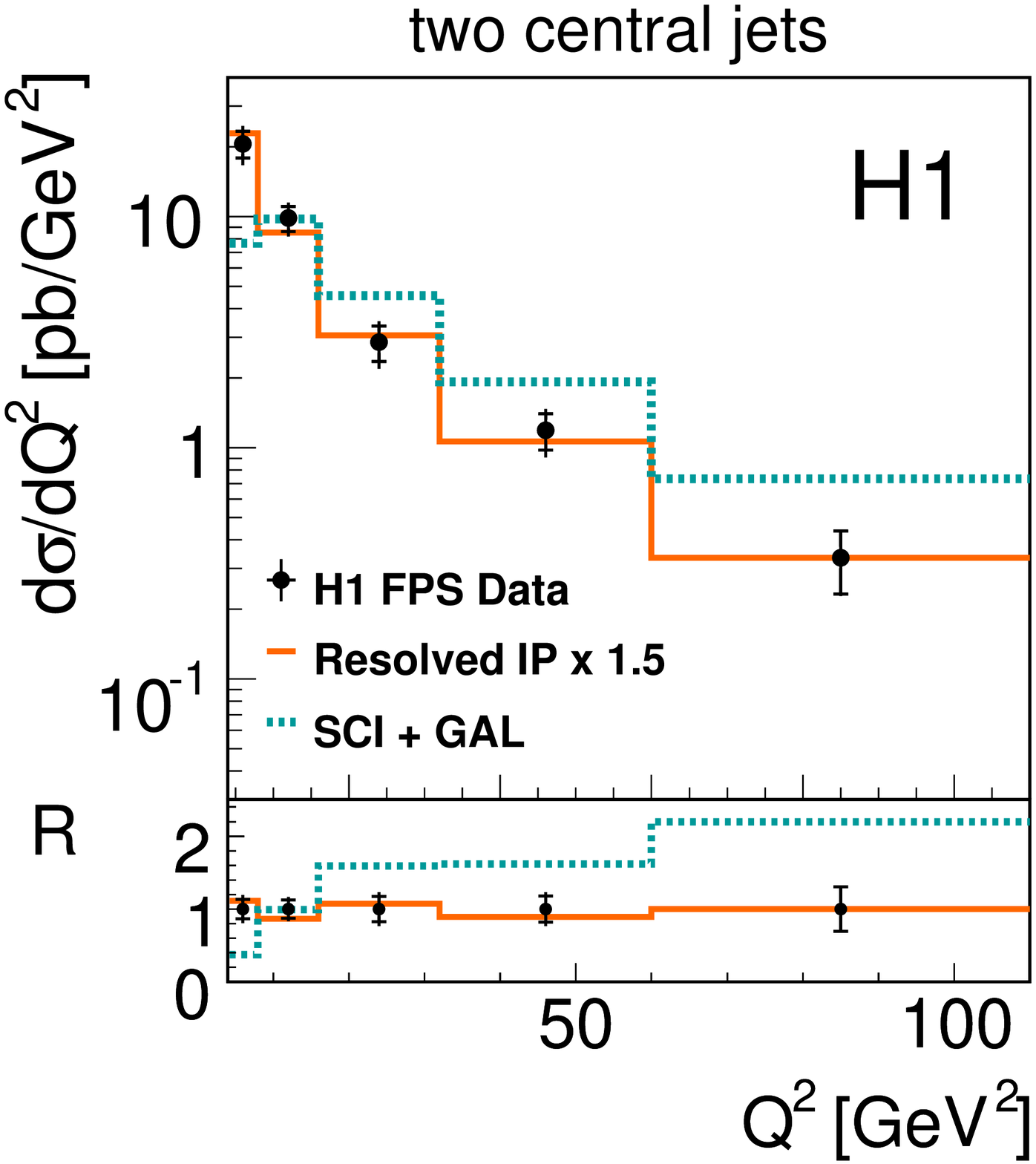,width=0.49\linewidth}
  \epsfig{file=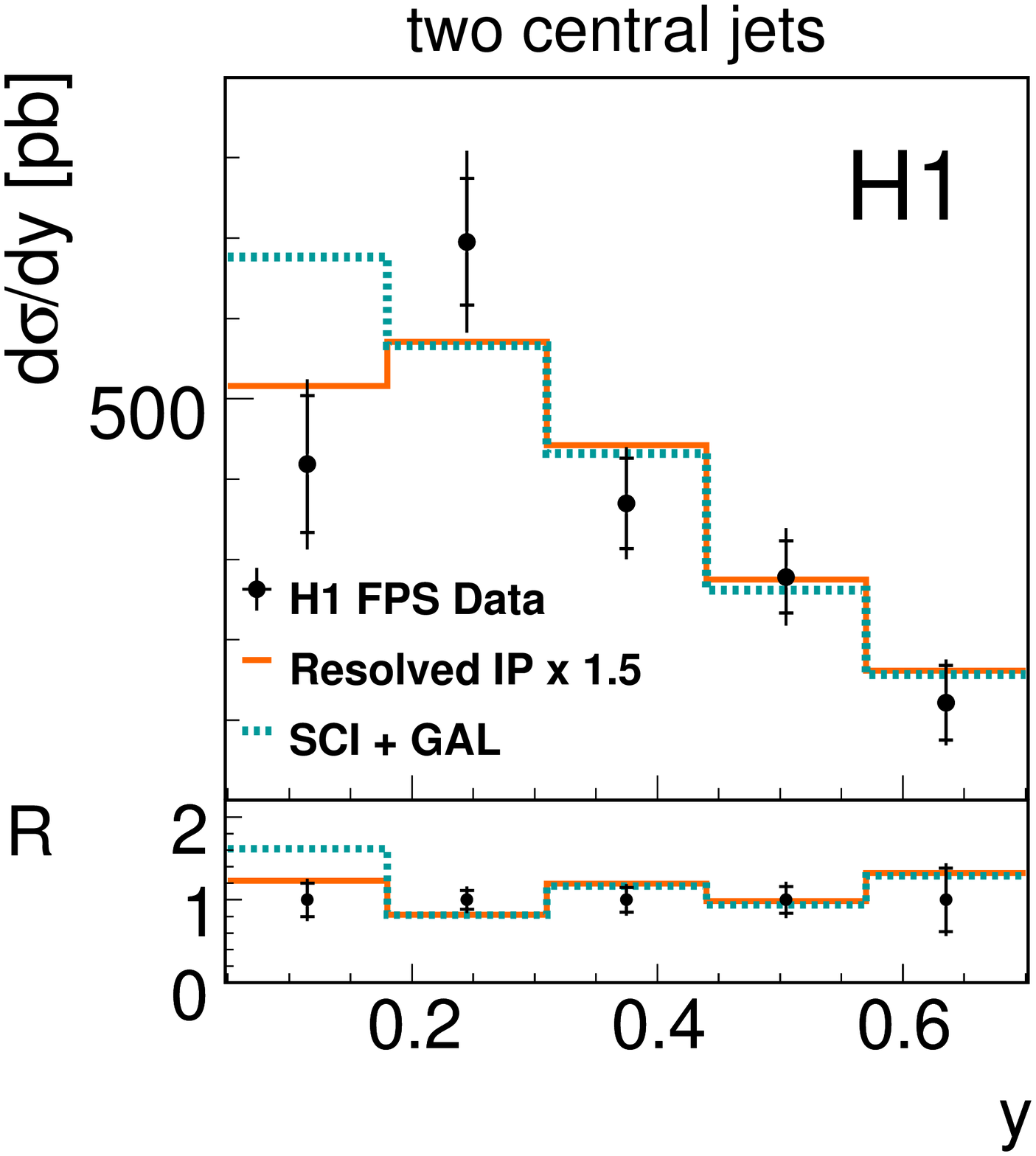,width=0.49\linewidth}
  \epsfig{file=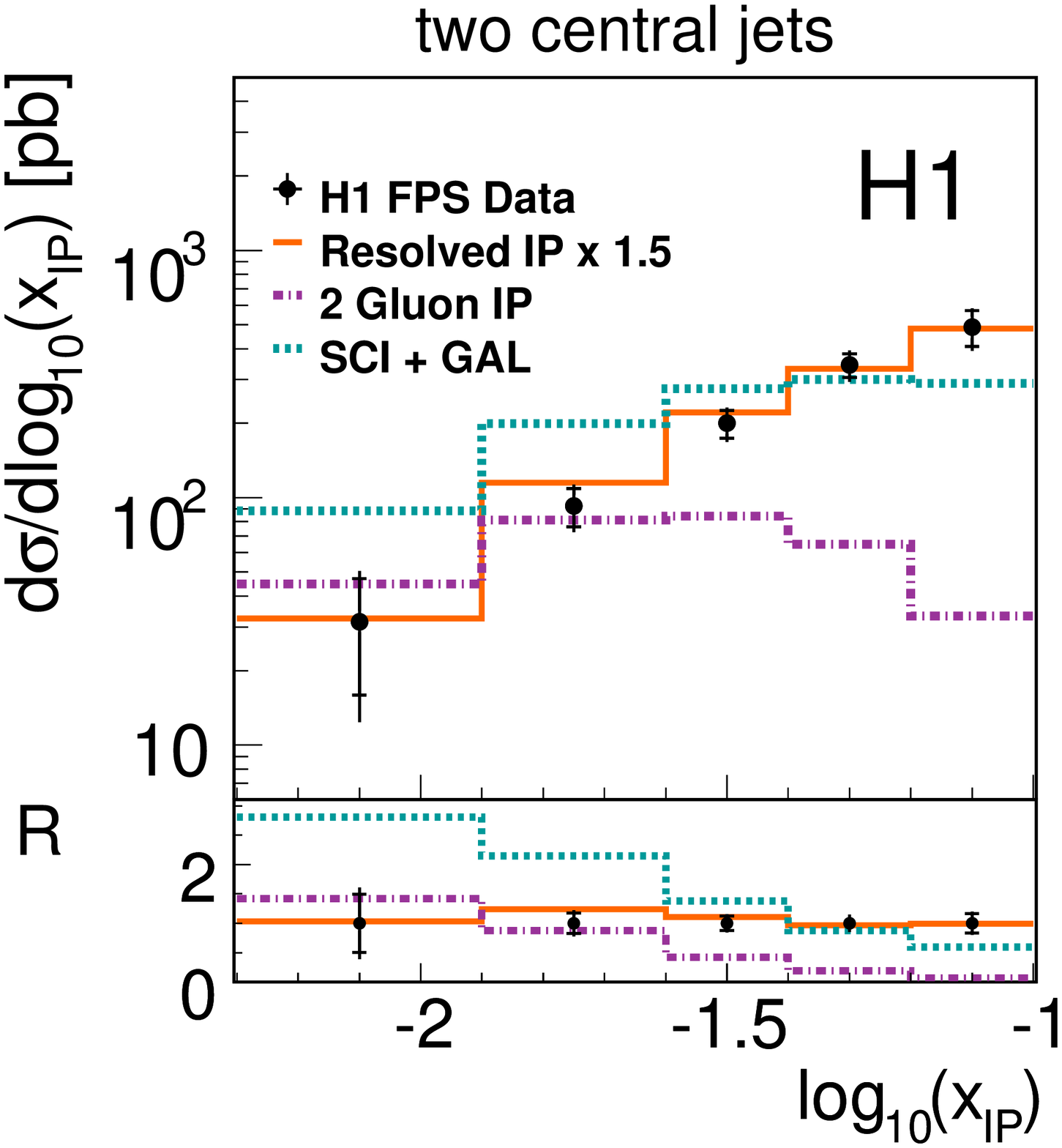,width=0.49\linewidth}
  \epsfig{file=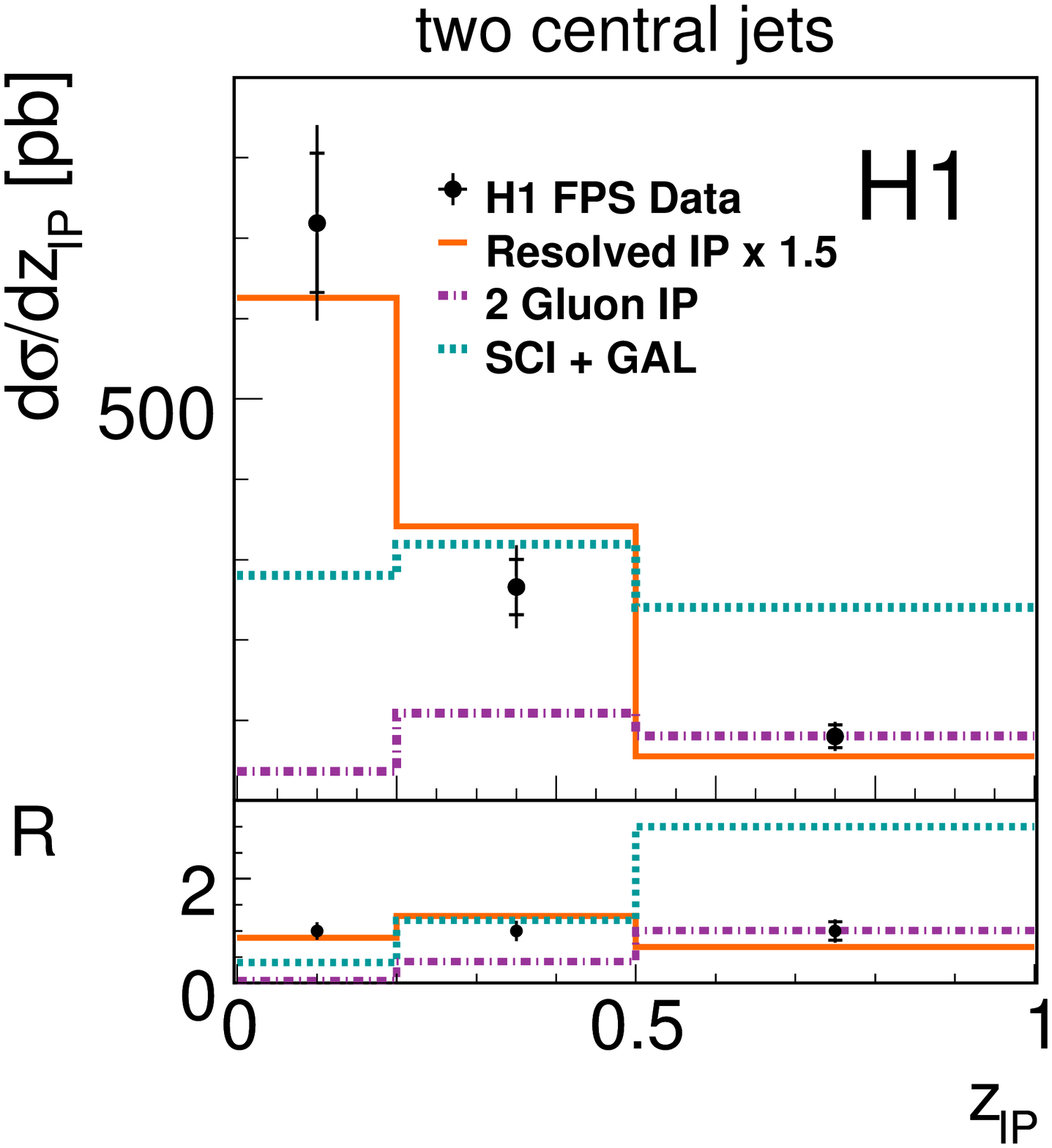,width=0.49\linewidth}
  \caption{The differential cross section for the production of two central
  jets shown as a function of $Q^2$, $y$, $\log_{10}(x_{I\!\!P})$ and $z_{I\!\!P}$. 
The inner error bars represent the statistical errors. The outer
  error bars indicate the statistical and systematic errors added in quadrature.
 The RP  
 and SCI+GAL models are shown as solid and dotted lines, respectively. 
 R denotes the ratio of the measured cross sections and MC model predictions to the nominal values of
  the measured cross sections. 
 The total normalisation error of $7.0\%$ is not 
 shown.}
  \label{dijmc1}
  \end{center}
  \end{figure}

 Figure~\ref{dijmc1} shows a comparison of the differential cross sections in $Q^2$, $y$, $\log_{10}(x_{I\!\!P})$ and  $z_{I\!\!P}$
with MC models based on the leading-logarithm approximation and parton 
 showers. 
The ratios of the measured cross sections to the MC predictions show
 that the RP model gives a good description of the shape, but underestimates the dijet cross section by a factor of $1.5$. 
 For this comparison the reweighting with respect to the $\eta_2^*$ distribution specified in section~\ref{mc} is not applied to the RP model.
Since the $\pom$ and $\reg$ fluxes which determine the $\xpom$ dependence in the RP model has been tuned to the inclusive 
 diffractive DIS LRG data~\cite{H1LRG} the good agreement in shape of the RP model with the dijet data supports the hypothesis of the proton vertex factorisation. Both the SCI+GAL and TGP models fail to describe the data.
The SCI+GAL model predicts harder spectra in $Q^2$ and $\zpom$ and a softer spectrum in $\log_{10}(x_{I\!\!P})$ than are seen in the data. 
It should be noted that 
 the probability of soft colour interactions and 
 hence the normalisation of diffractive processes in the SCI+GAL model is adjusted to the measured 
 dijet cross section.
The TGP model is in agreement with the data only 
at low $\xpom$ but underestimates the data significantly at larger $\xpom$ 
where the missing sub-leading contributions are expected to be large.

\begin{figure}
  \begin{center}
  \epsfig{file=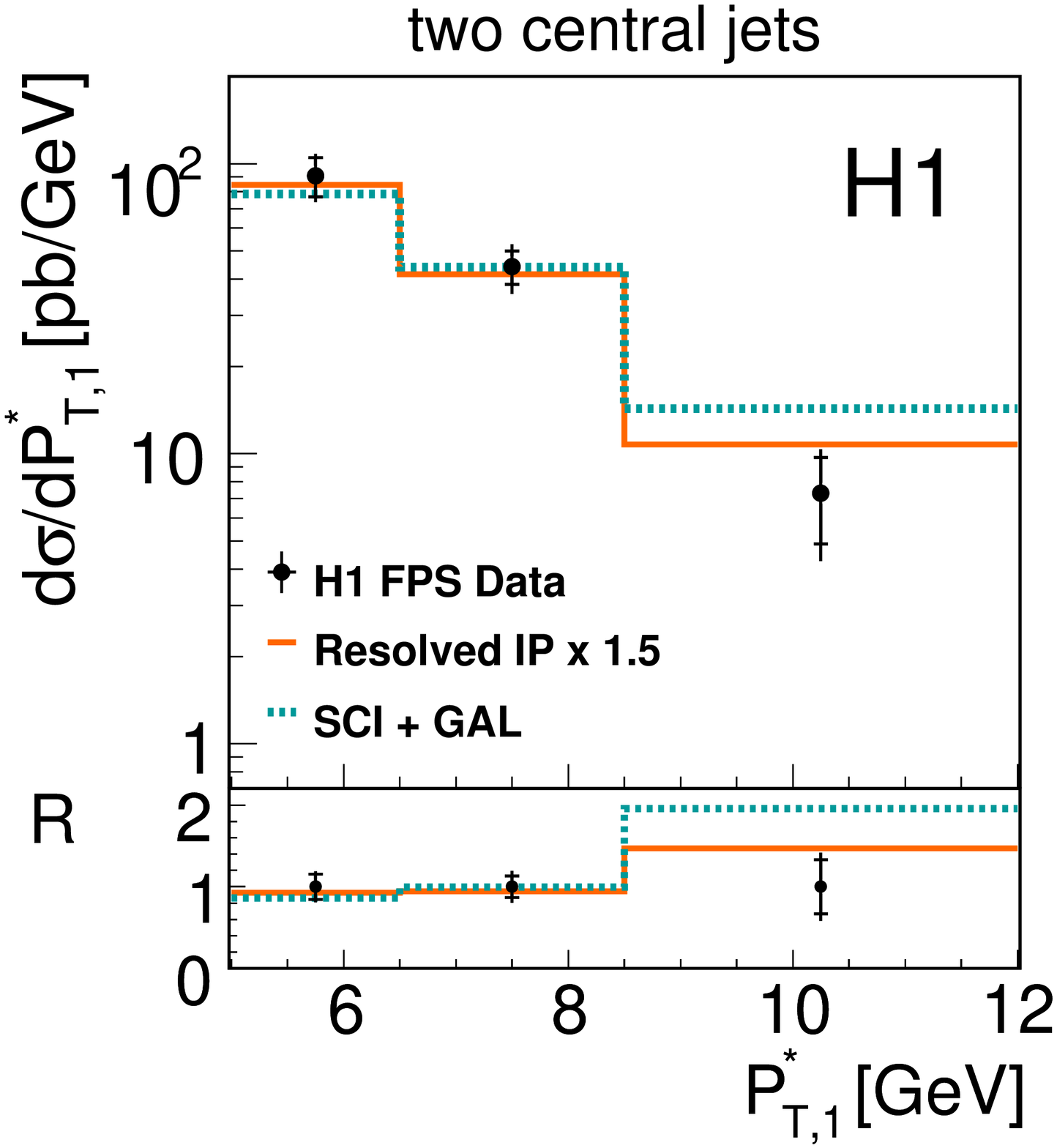,width=0.49\linewidth}
  \epsfig{file=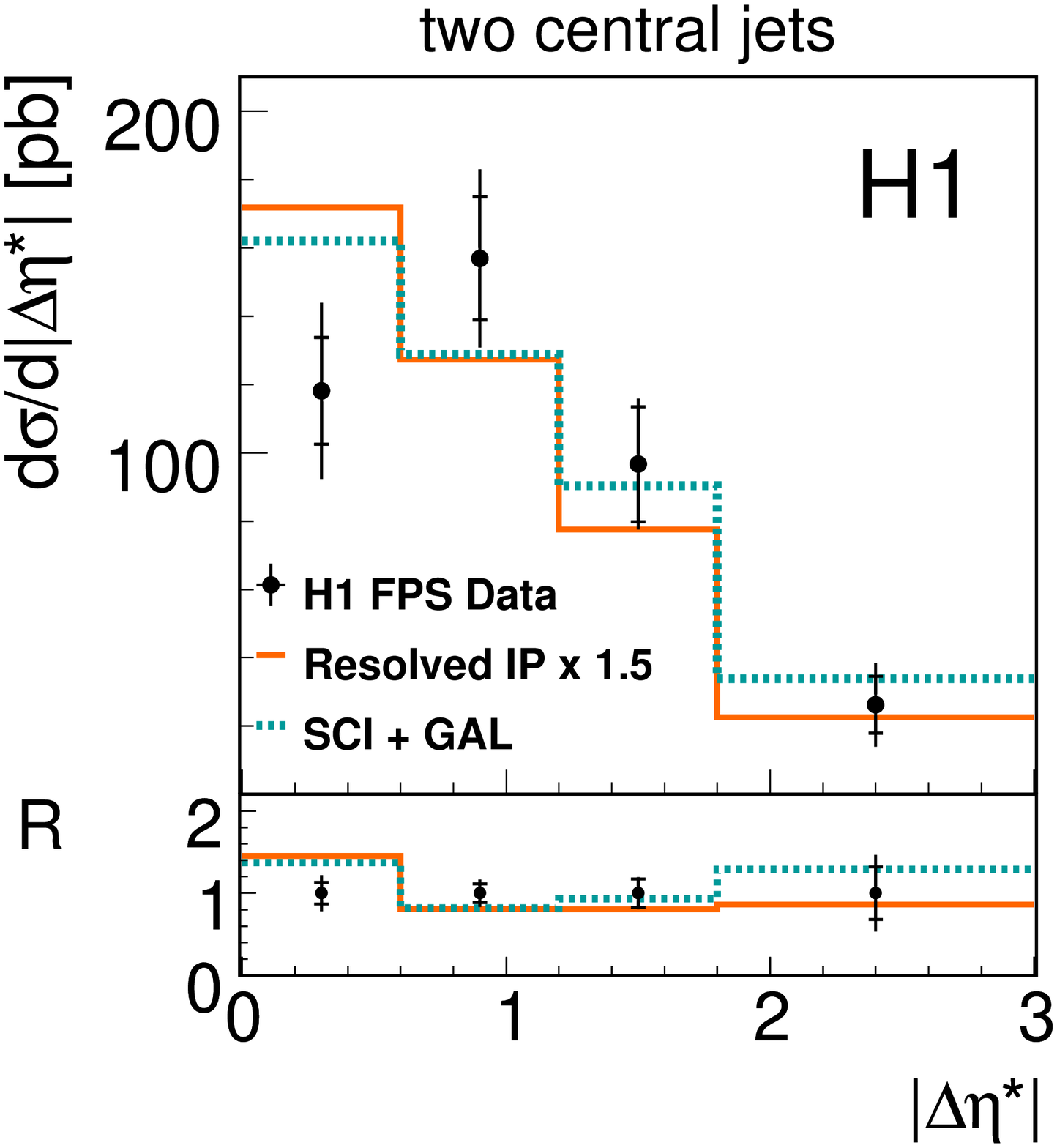,width=0.49\linewidth}
  \caption{The differential cross section for production of two central
  jets shown as a function of $P_{T,1}^{*}$ and $|\Delta\eta^*|$. For more details see figure \ref{dijmc1}.}
 \label{dijmc2}
  \end{center}
  \end{figure}

 Figure~\ref{dijmc2} shows the differential cross sections in $P^*_{T,1}$ and $|\Delta\eta^*|$
 for the data and the MC models. The shapes of these 
distributions are again well described by  the RP model. 
Although the SCI+GAL model is not able to describe the differential cross sections as a function of the diffractive kinematic variables $\xpom$ and $\zpom$ and of the DIS kinematic variable $Q^2$ this model reproduces reasonably well the measurements as a function of the jet variables $P^*_{T,1}$ and $|\Delta\eta^*|$.  

 None of the LO Monte Carlo models are able to describe all features of the measured differential 
 cross sections. The best shape description in all cases is provided by the RP model. 
 However, this model is a factor of $1.5$ below the data in normalisation.
 The TGP and SCI+GAL models fail to 
 describe the shape of the differential cross sections.

\begin{figure}
  \begin{center}
    \begin{picture}(160,90)
      \put(-3,5){\epsfig{file=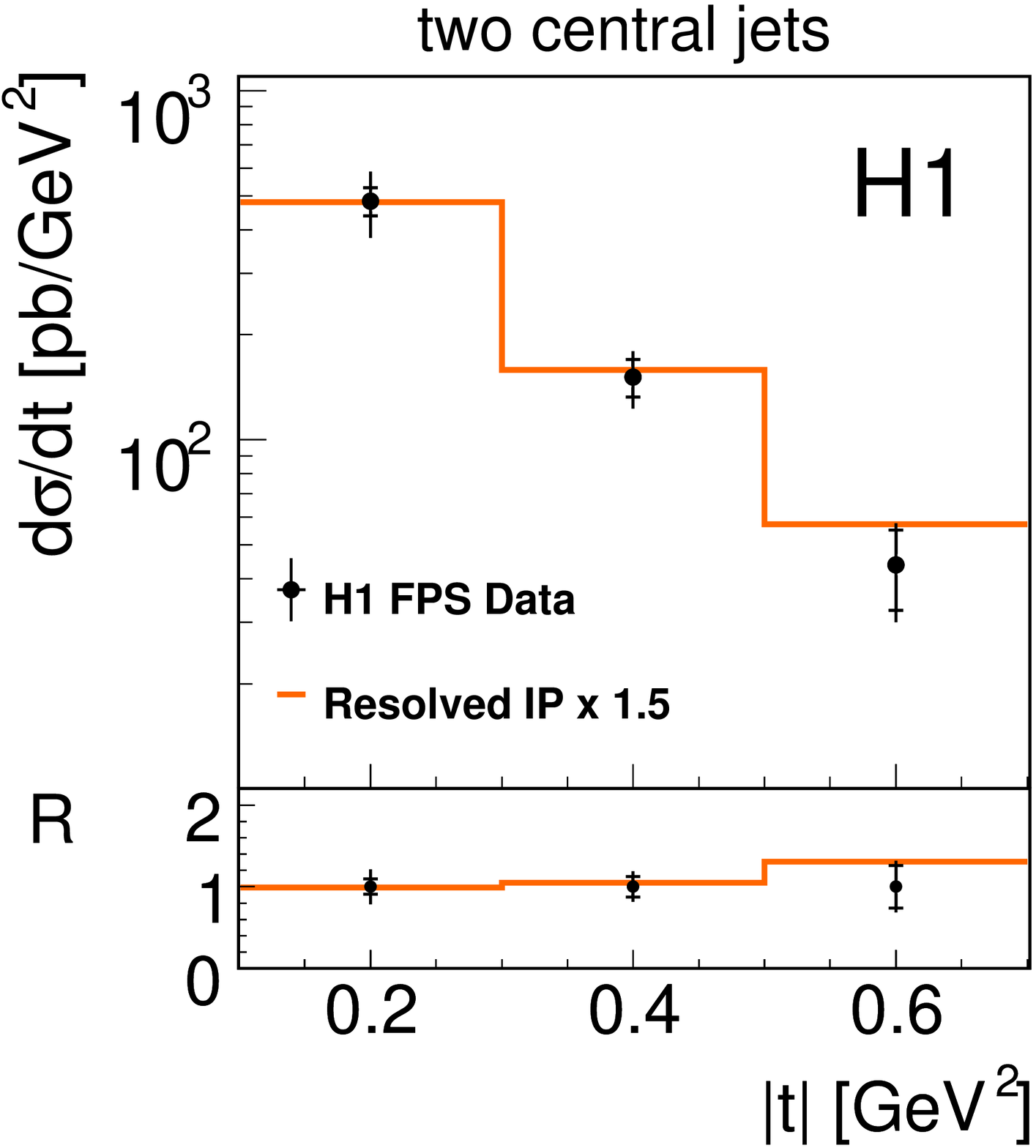,width=0.49\linewidth}}
      \put(80,5){\epsfig{file=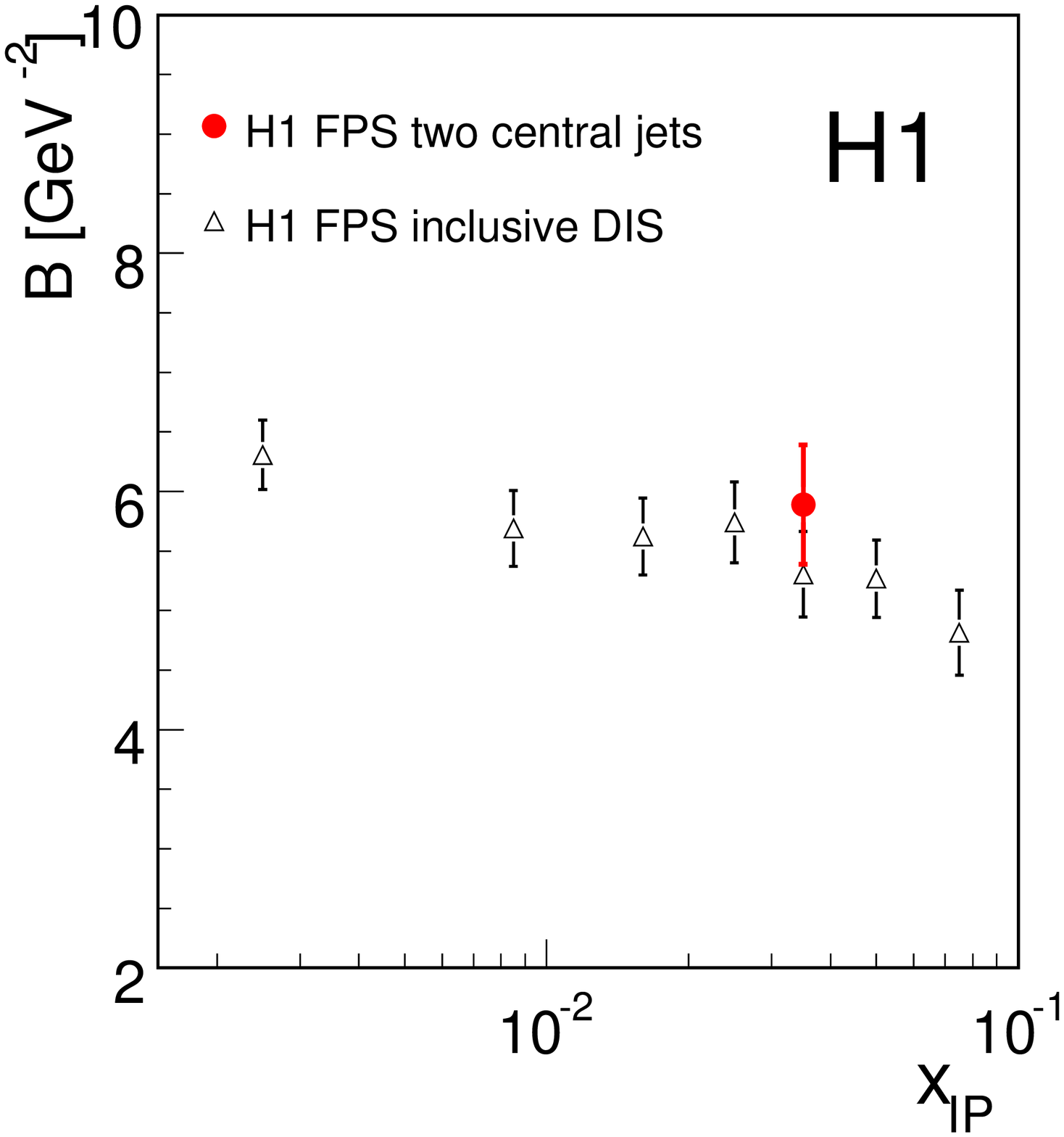,width=0.49\linewidth}}
      \put(61,60){\bf{\large{(a)}}}
      \put(140,60){\bf{\large{(b)}}}
    \end{picture}
  \end{center}
  \caption{The differential cross section for production of two central
  jets shown as a function of $t$ (a), the corresponding $t$-slope 
 (circle) shown as a function of $\xpom$ (b). The result is compared to the H1  
  inclusive  diffractive DIS data (triangles) \cite{H1FPShera2}. The 
  error bars indicate the statistical and systematic errors added in quadrature.}
  \label{tfit}
  \end{figure}

 The differential cross section in $|t|$ shown in figure~\ref{tfit}a is fit using an exponential
 form $\exp(Bt)$ motivated by Regge phenomenology. An iterative procedure is used to determine the slope parameter $B$, where bin centre corrections are applied to the differential cross section in $t$ using the value of $B$ extracted 
 from the previous fit iteration. The final fit results in $B = 5.89 \pm 
 0.50$~(exp.)~GeV$^{-2}$, where the experimental uncertainty is defined as the quadratic sum of the statistical 
 and  systematic 
 uncertainties and the full covariance matrix is taken into account in the fit. As shown in figure~\ref{tfit}b, this $t$-slope 
 parameter is consistent within the errors with the $t$-slope measured
 in inclusive diffractive DIS with a leading proton in the final state~\cite{H1FPShera2} at the same value of $\xpom$. 
 The consistency of the measured $t$ dependence 
 with that 
 for the  inclusive diffractive DIS cross sections supports the validity of the proton
 vertex factorisation hypothesis. 



\begin{figure}
   \begin{center}
   \epsfig{file=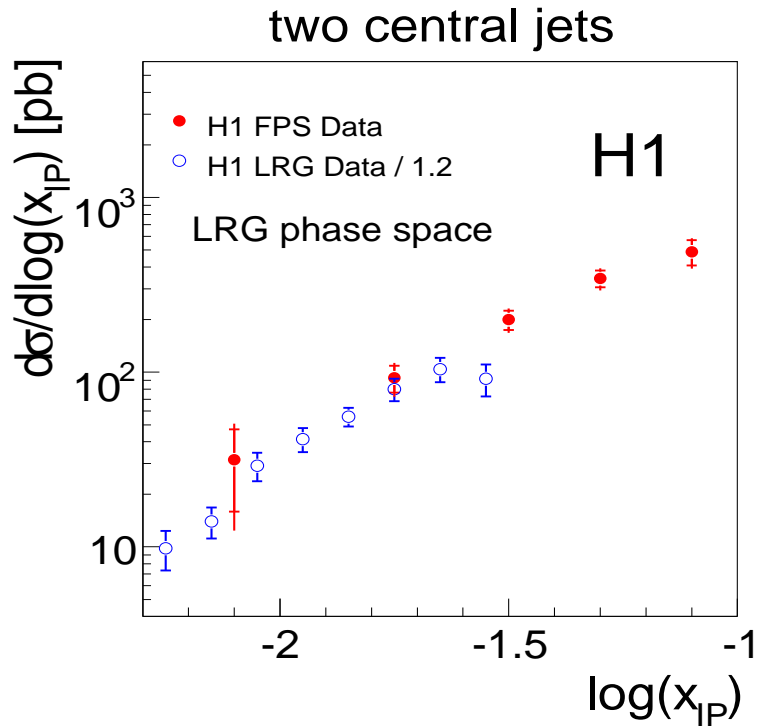,width=300pt,height=300pt}
    \caption{The differential cross section for the production of two central
    jets in the phase space of the LRG 
    measurement~\cite{H1DiJets} as described the text in section~\ref{2cjxs}. 
	The cross section is shown as a function of $\log_{10}(x_{I\!\!P})$.   
	The inner error bars represent the statistical errors. The outer
  error bars indicate the statistical and systematic errors added in quadrature.
    The published LRG dijet data are scaled down by a factor of $1.20$ to correct for the proton 
	dissociation contribution are shown as open circles with the 
	error bars indicating the statistical and systematic errors added in quadrature.} 
   \label{mozlog}
   \end{center}
   \end{figure}

 The cross section for the production of two central jets can be compared with the diffractive dijet measurement 
obtained 
 using the LRG technique~\cite{H1DiJets}. The LRG measurement includes proton dissociation 
 to states $Y$ with masses 
 $M_Y<1.6$~GeV. To correct for the contributions of proton dissociation processes, the LRG dijet data are scaled down by 
 a factor of $1.20$, taken from the diffractive inclusive DIS measurement~\cite{H1FPShera2}.
 To compare to the results of the LRG method, dijet events are selected in the same kinematic range.
 The DIS and jet variables $Q^2$, $y$, $P_{T,1}^*$ and $\eta_{1,2}$ are restricted to the ranges $4<Q^2<80$~GeV$^2$, 
 $0.1<y<0.7$, $P_{T,1}^* > 5.5$~GeV, and $-1 < \eta_{1,2} < 2$, respectively. 
 The results are presented in figure \ref{mozlog}. The comparison shows 
 consistency of the results within the experimental errors. 
 Compared to the LRG measurement, the phase space of the present analysis extends to $\xpom$ values that are
 a factor of three larger.

\subsection{Differential cross section for the production of one central + one forward jet}





\begin{figure}
  \begin{center}
  \epsfig{file=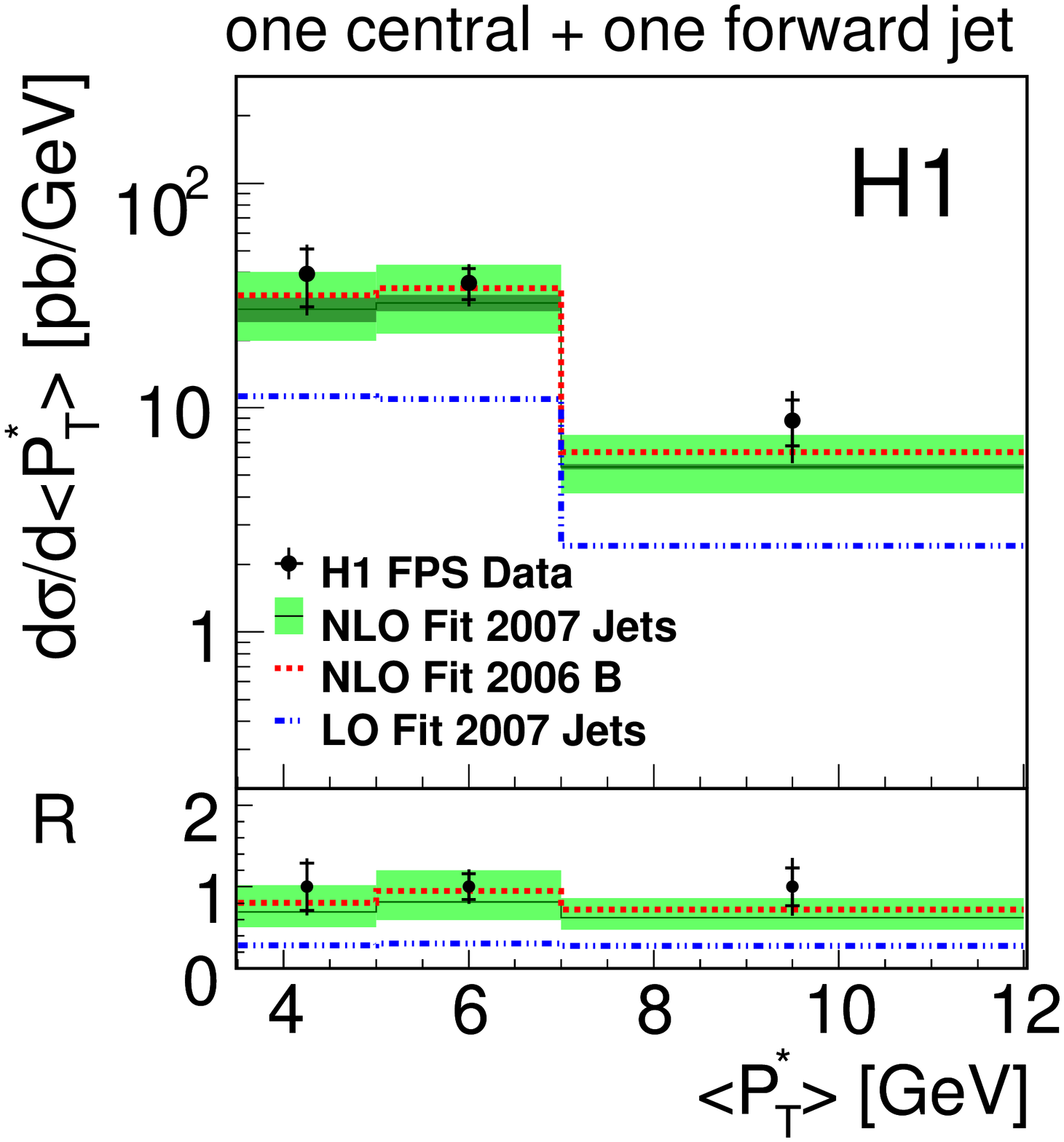,width=0.49\linewidth}
  \epsfig{file=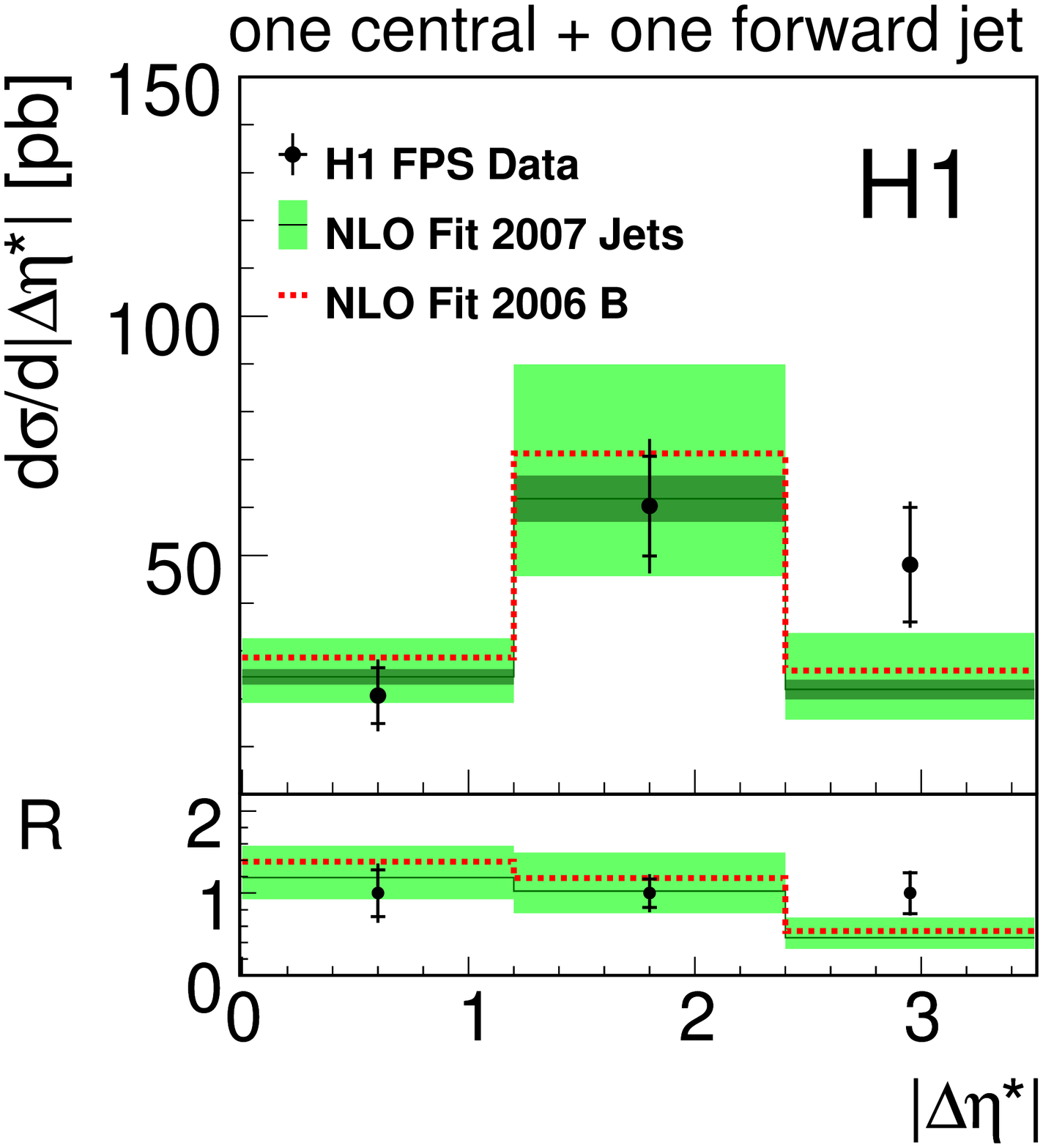,width=0.49\linewidth}
  \epsfig{file=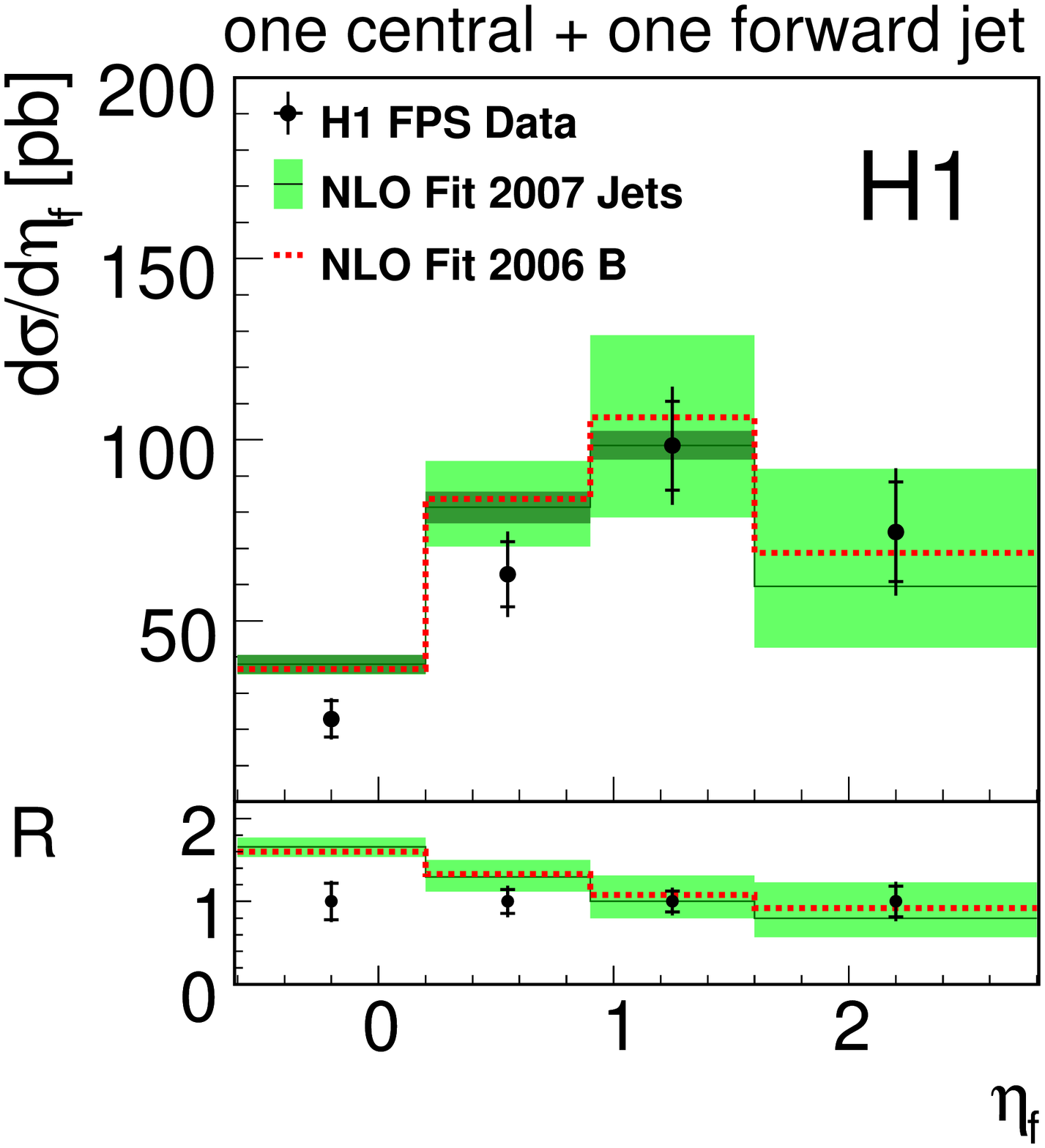,width=0.49\linewidth}
   \caption{The differential cross section for the production of one central and one forward
    jet shown as a function of the mean transverse momentum of two jets $\langle P_T^*\rangle$, $|\Delta\eta^*|$ and $\eta_f$. 
   The inner error bars represent the statistical errors. The outer
  error bars indicate the statistical and systematic errors added in quadrature.
  NLO QCD
 predictions based on the DPDF set H1 2007 Jets, 
 corrected to the level of stable hadrons, are shown as a line with a dark green band indicating the hadronisation error and light green band indicating the hadronisation and scale errors added in quadrature.
 LO predictions based on the same DPDF set are shown as a dotted line. 
 The NLO calculations based on the DPDF set H1 2006 Fit B with applied hadronisation 
 corrections is shown as a thick line. 
 R denotes the ratio of the measured cross sections and QCD predictions to the nominal values of
  the measured cross sections. 
 The total normalisation error of $6.2\%$ is not shown.}
  \label{cfnlo1}
  \end{center}
 \end{figure}

 Figure \ref{cfnlo1} shows the differential cross sections 
for the production of \textquoteleft one central + one forward jet'
as a function of $|\Delta\eta^*|$, $\eta_f$ and the mean transverse momentum of the forward and  central  jets $\langle P_T^*\rangle$ together with the expectations from the NLO QCD. Within the errors, the measured data 
 are  described 
 by NLO QCD  predictions. 
In order to test the predictions in a wider 
 kinematic range, the 
 $\eta_f$  distribution of the forward jet shown in figure \ref{cfnlo1} is extended down to a minimum value of 
 $-0.6$
 where the prediction overshoots the data. LO QCD calculations, performed using the DPDF set H1 2007 Jets 
 underestimate the measured cross section by a factor of about $2.5$.

\begin{figure}
     \begin{center}
      \epsfig{file=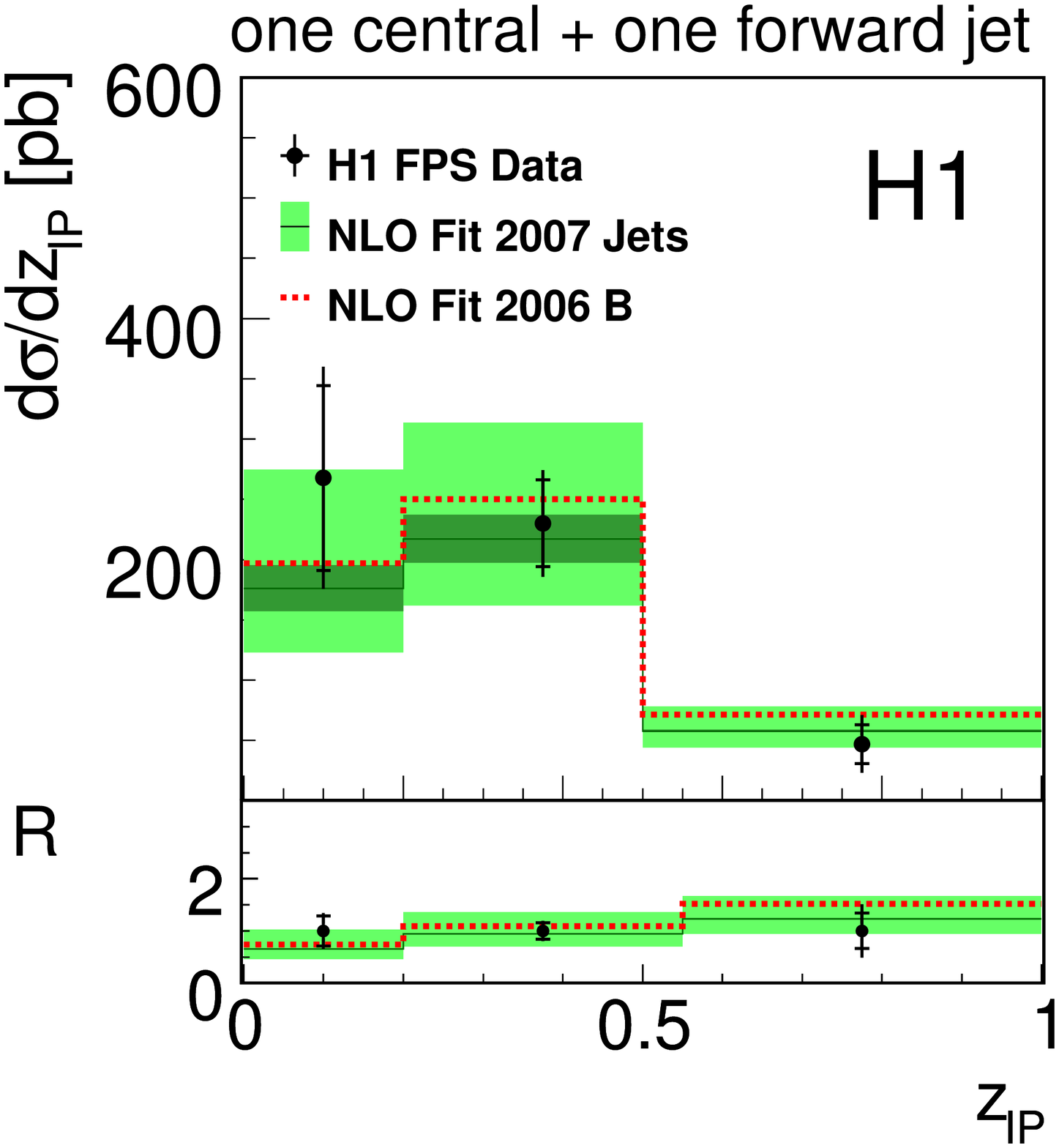,width=0.49\linewidth}
     \epsfig{file=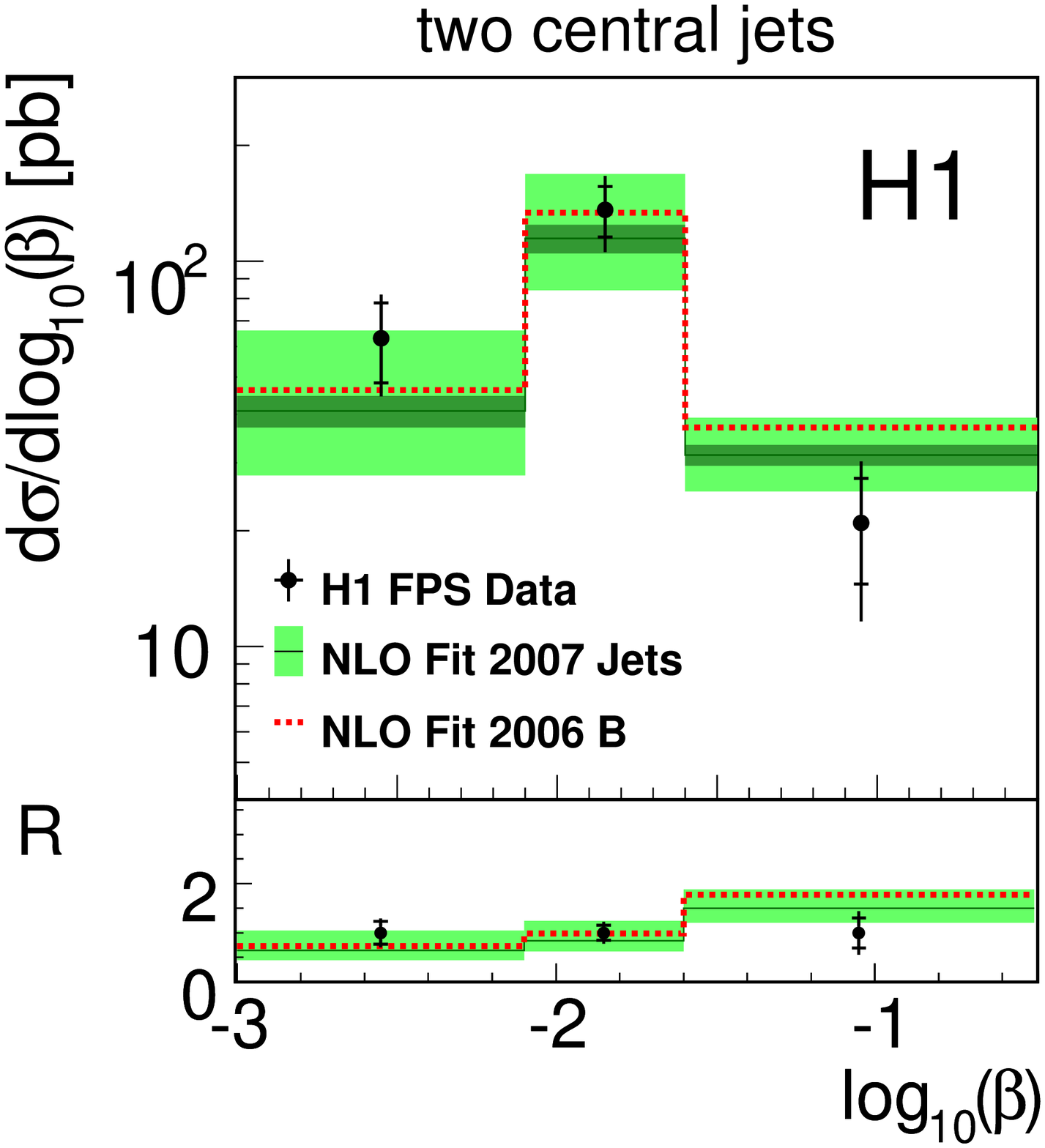,width=0.49\linewidth}
     \epsfig{file=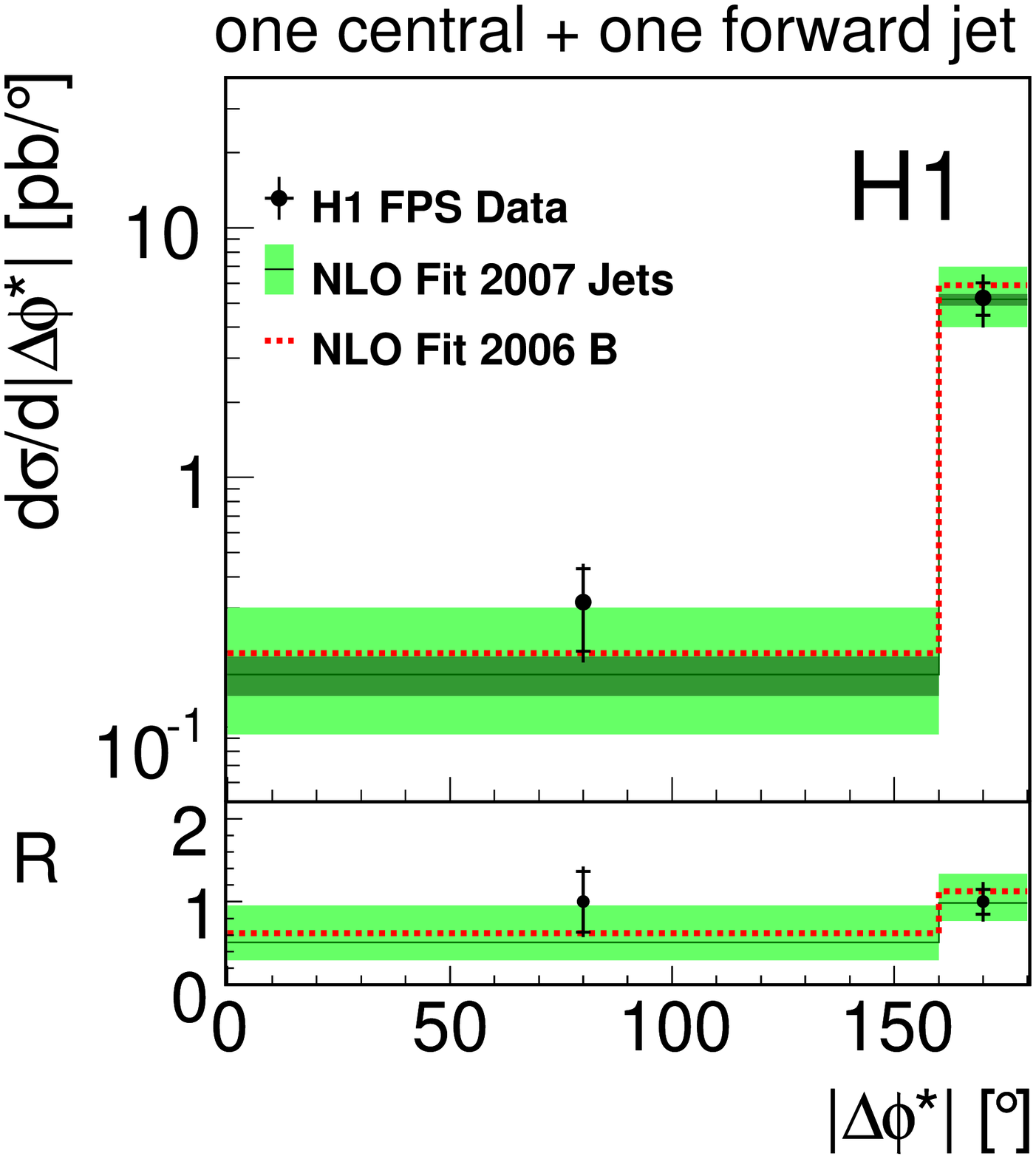,width=0.49\linewidth}
     \caption{The differential cross section for production of one central and one forward
    jet shown as a function of  $z_{I\!\!P}$, $\log_{10}(\beta)$ and $|\Delta\phi^*|$. For more details see figure \ref{cfnlo1}.}
     \label{cfnlo2}
    \end{center}
 \end{figure}

 The differential cross sections measured as a function of $z_{I\!\!P}$, $\log_{10}(\beta)$ and $|\Delta\phi^*|$ 
 are presented in figure~\ref{cfnlo2}.   The data are well described by the NLO QCD predictions.
In the BFKL approach~\cite{BFKL,BFKL2,BFKL3}, additional gluons can be emitted in the gap between the
two jets, leading to a de-correlation in azimuthal angle $|\Delta\phi^*|$.
The observed agreement between the measured cross sections and NLO DGLAP predictions in
this distribution shows no evidence for such an effect in the kinematic region accessible in this analysis. 

\begin{figure}
  \begin{center}
   \epsfig{file=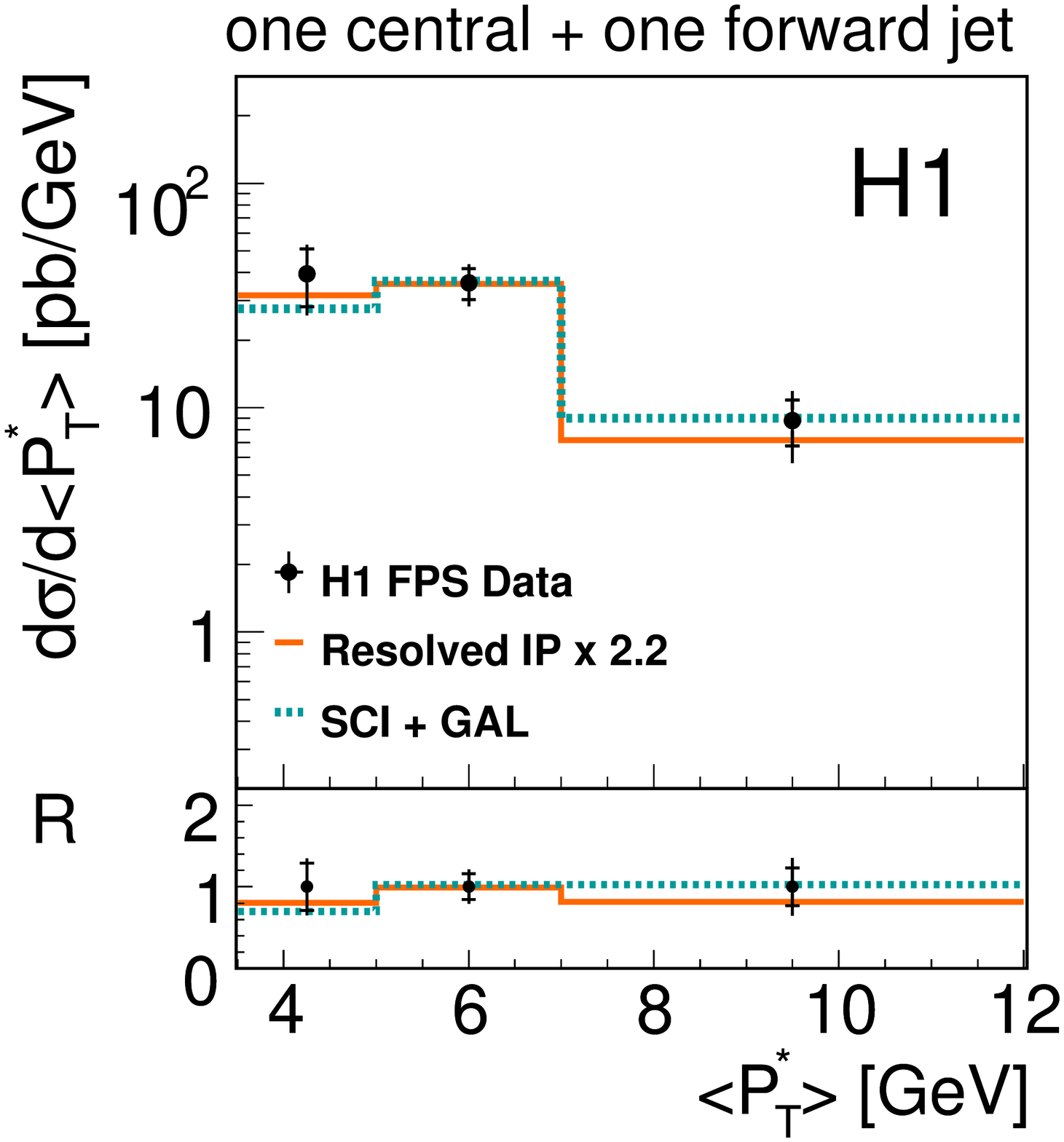,width=0.49\linewidth}
   \epsfig{file=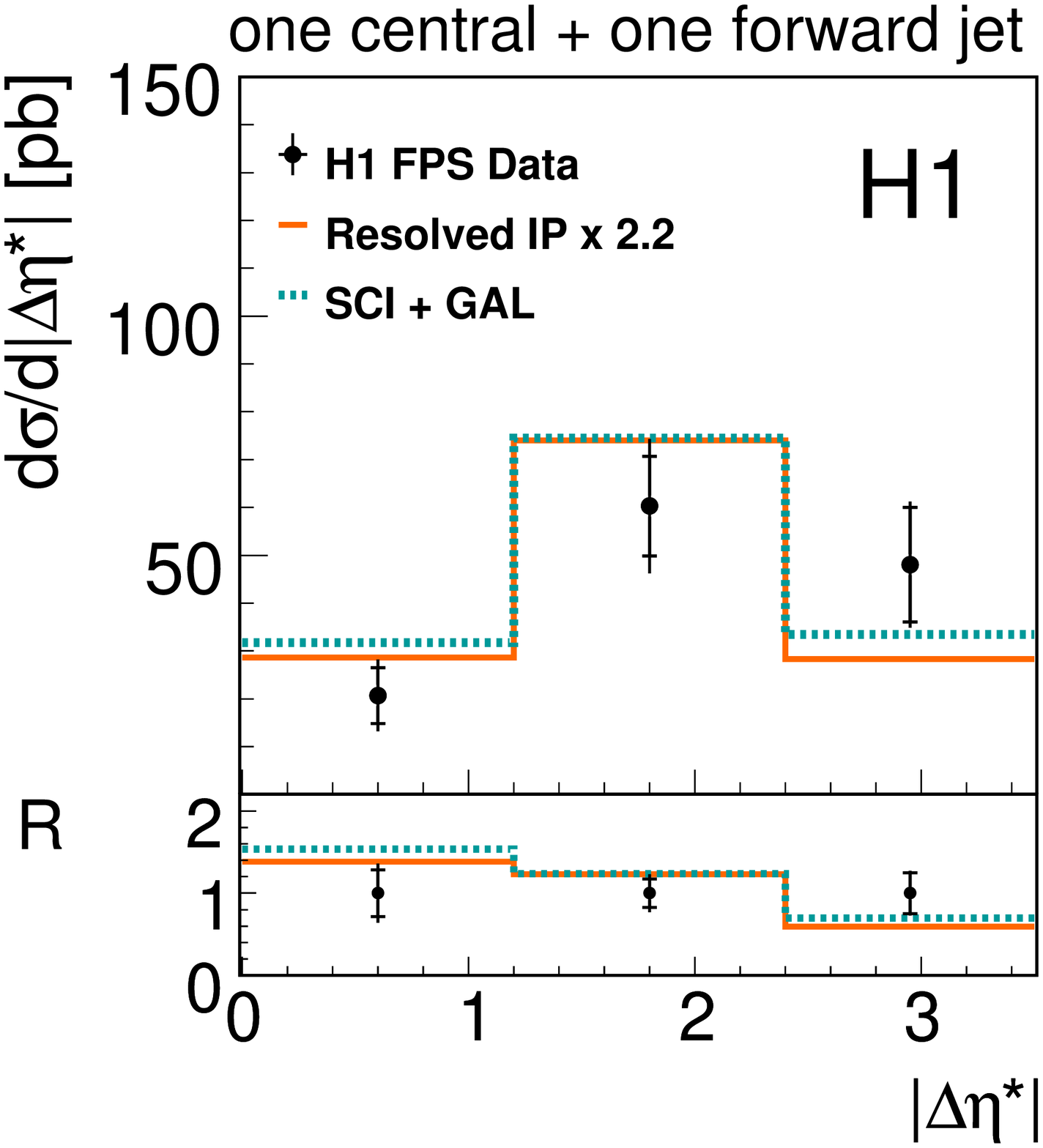,width=0.49\linewidth}
   \epsfig{file=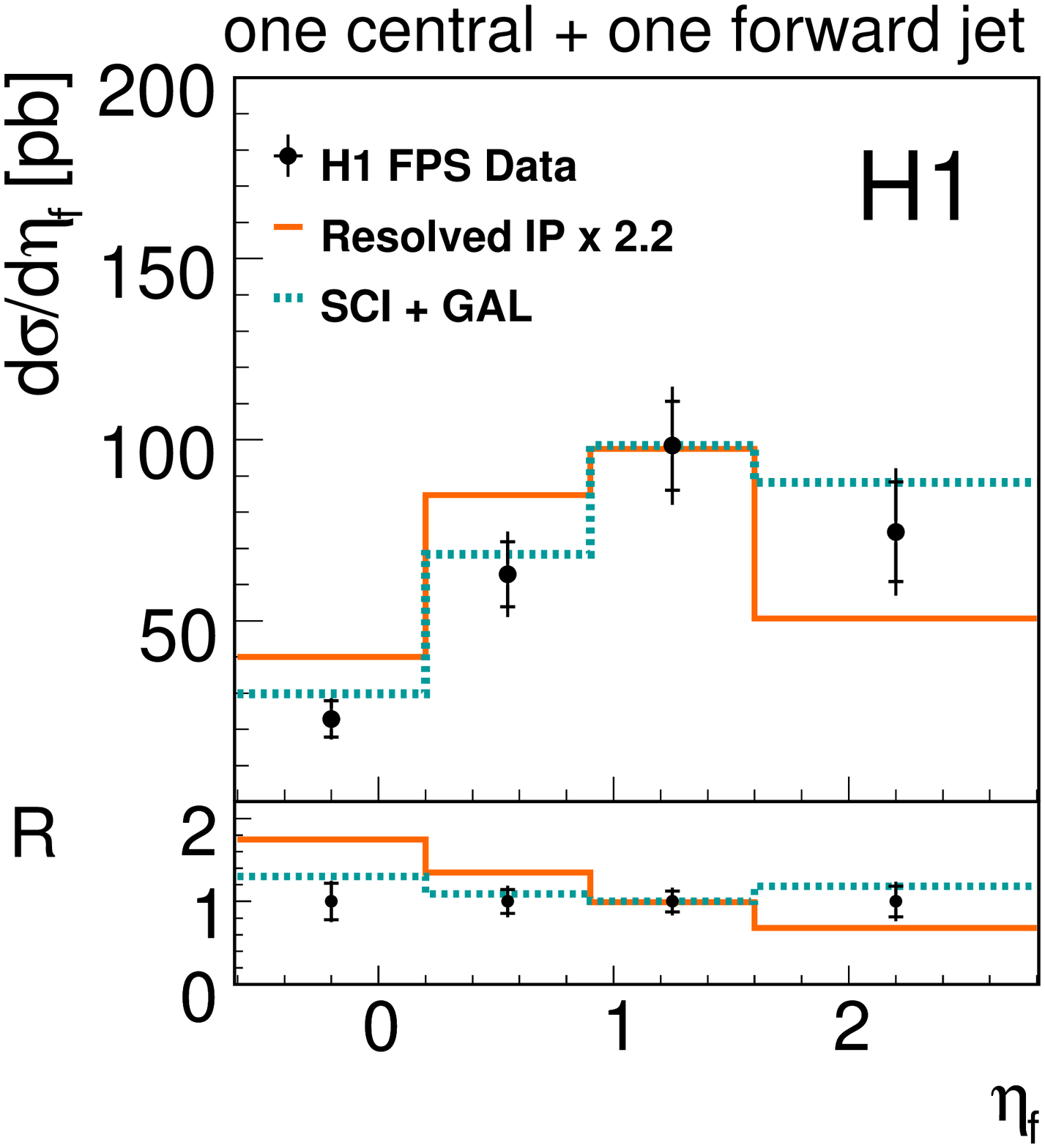,width=0.49\linewidth}
   \caption{The differential cross section for production of one central and one forward
    jet shown as a function of  the mean transverse momentum of two jets $\langle P_T^*\rangle$, $|\Delta\eta^*|$ and $\eta_f$. 
   The inner error bars represent the statistical errors. The outer
  error bars indicate the statistical and systematic errors added in quadrature.
 The RP 
 and the SCI+GAL models are shown as solid and dotted lines, respectively. 
 R denotes the ratio of the measured cross sections and MC model predictions to the nominal values of
  the measured cross sections. 
 The total normalisation error of $6.2\%$ is not 
 shown.}
   \label{cfmc1}
  \end{center}
   \end{figure}

 Figure~\ref{cfmc1} presents the differential cross sections for the production of \textquoteleft one central + one forward jet' as a 
 function of the variables $\langle P_T^*\rangle$, $|\Delta\eta^*|$ and $\eta_f$. 
 The RP model is a factor of $2.2$ below the data which is a larger discrepancy in normalisation than that observed in the \textquoteleft two central jets' sample. A similar trend is seen for the LO QCD contributions in the two samples.
 The normalisation of the SCI+GAL model, tuned to \textquoteleft two central jets', agrees with the cross section 
for \textquoteleft one central + one forward jet'.
 The shapes of the distributions are reasonably well described by both the RP and SCI+GAL models.

\begin{figure}
  \begin{center}
  \epsfig{file=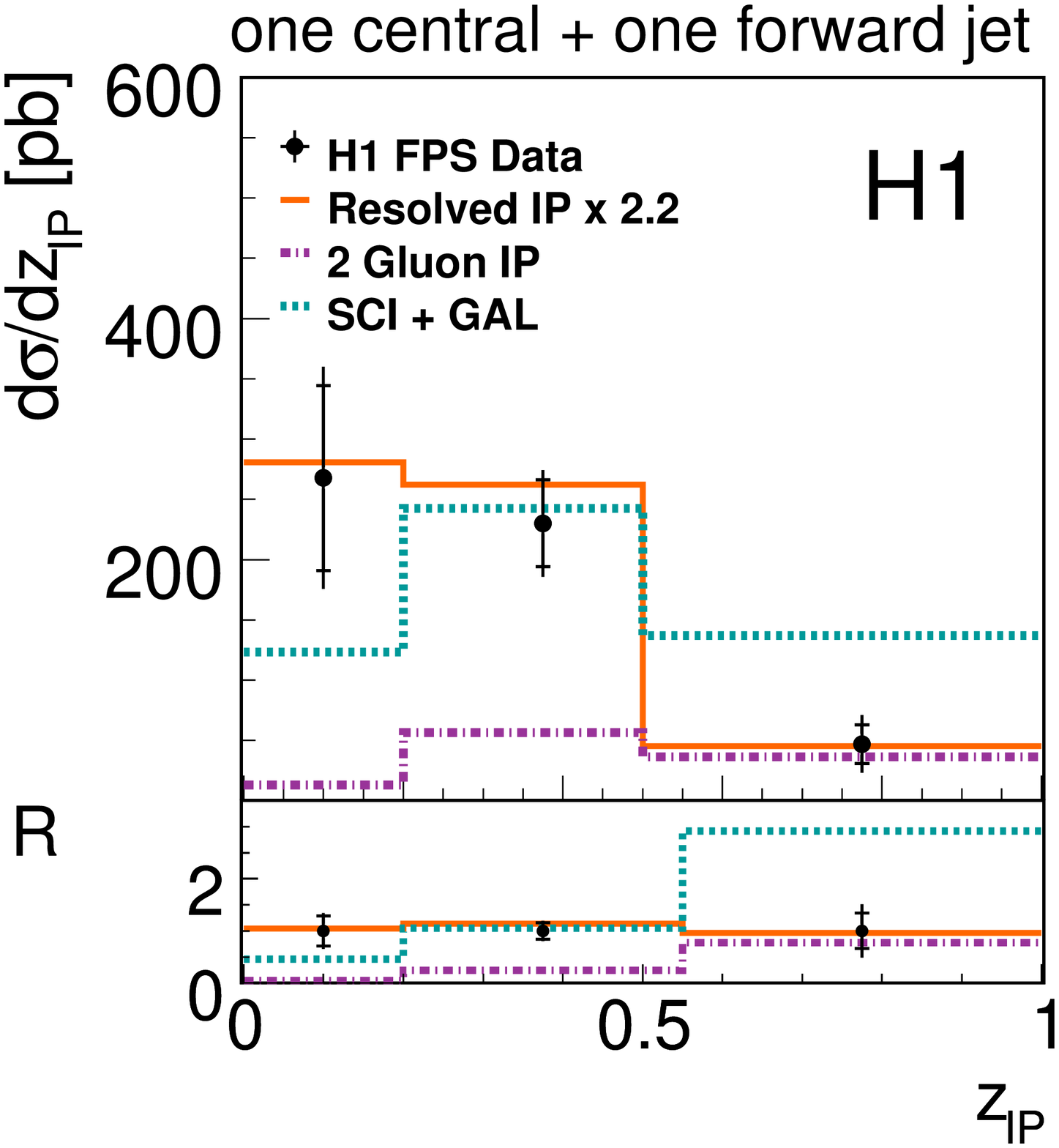,width=0.49\linewidth}
  \epsfig{file=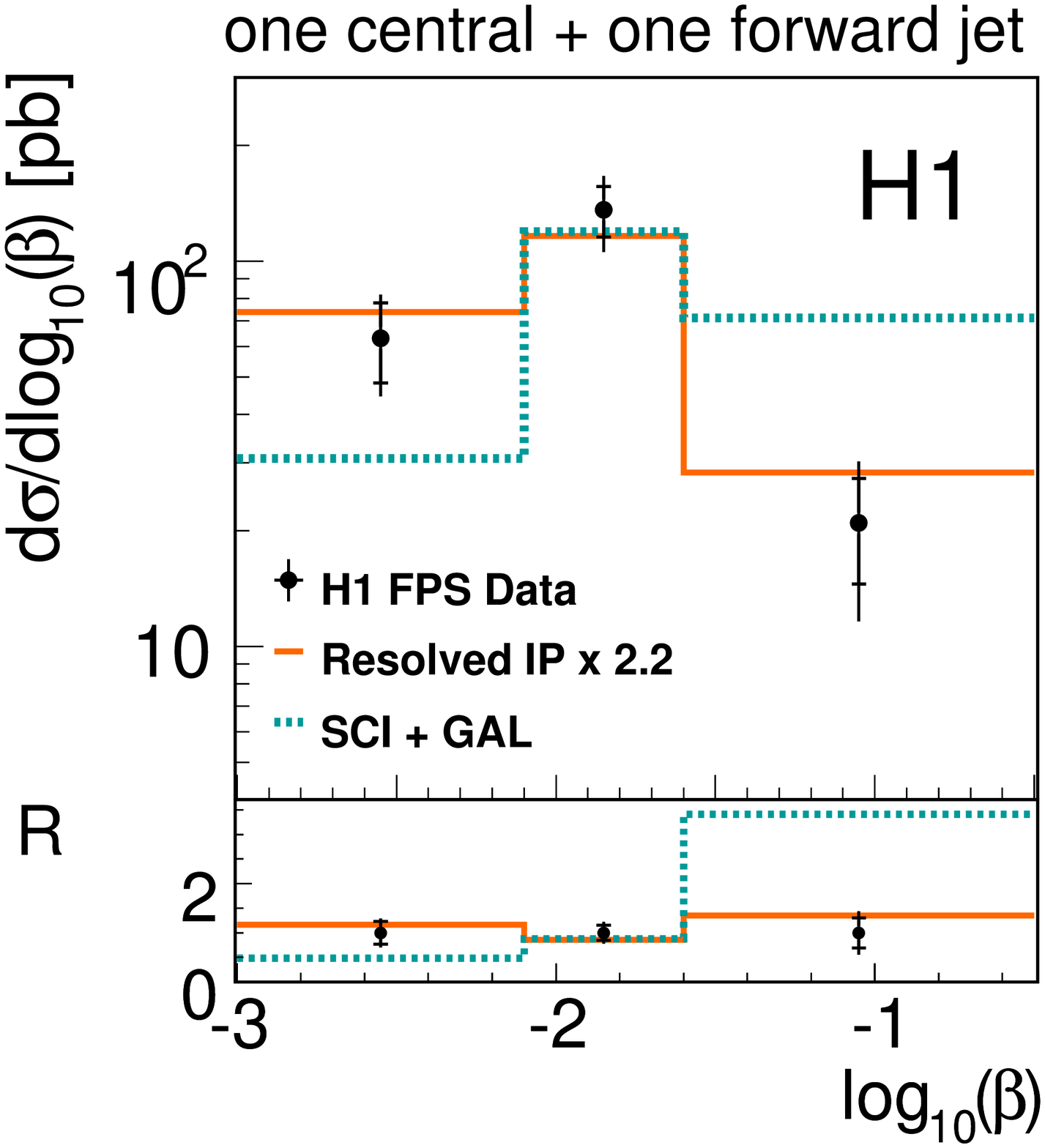,width=0.49\linewidth}
  \epsfig{file=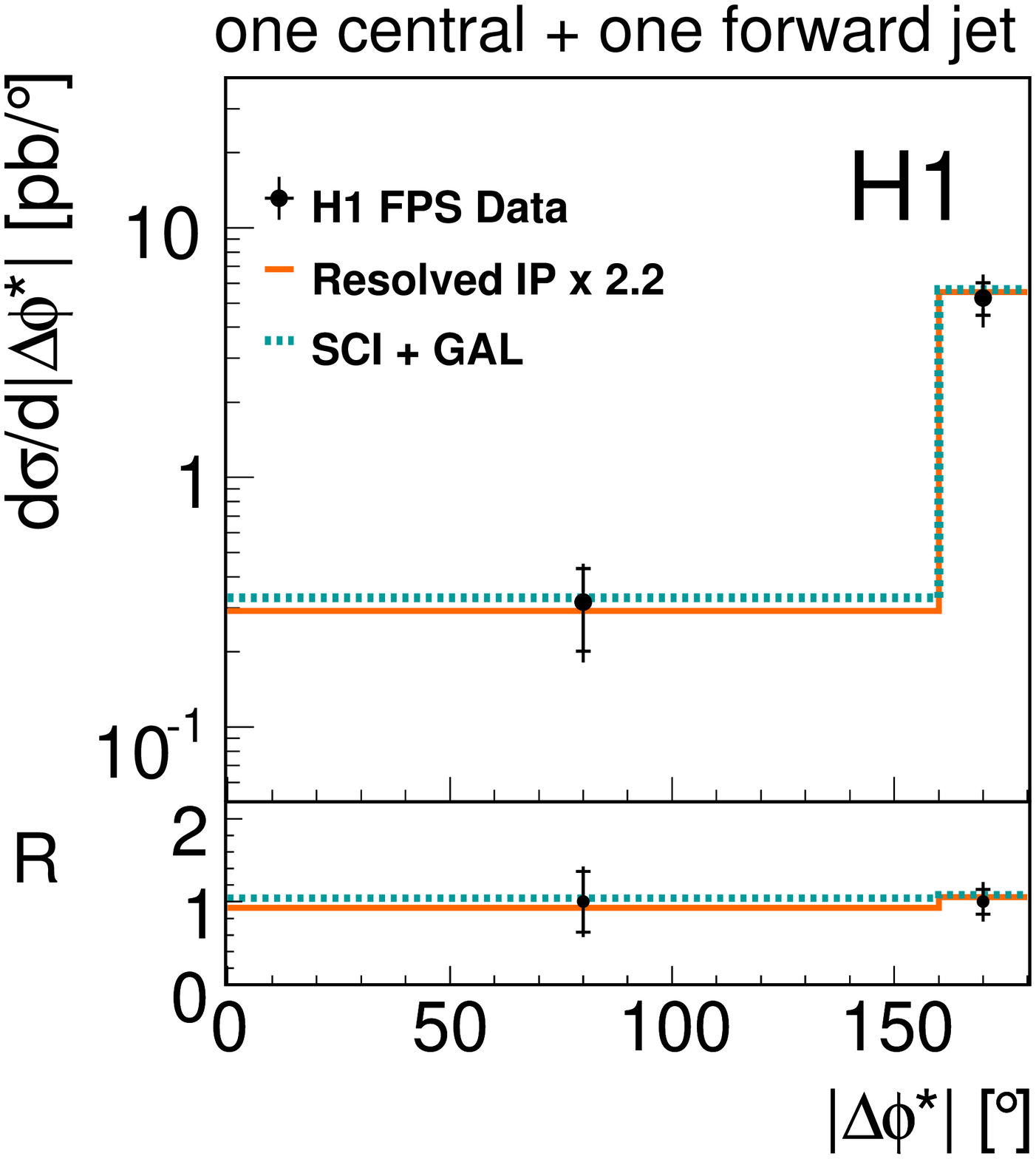,width=0.49\linewidth}
  \caption{The differential cross section for production of one central and one forward
    jet shown as a function of   $z_{I\!\!P}$, $\log_{10}(\beta)$ and $|\Delta\phi^*|$. For more details see figure \ref{cfmc1}.}
  \label{cfmc2}
  \end{center}
  \end{figure}

 The differential cross sections in $z_{I\!\!P}$, $\log_{10}(\beta)$ and $|\Delta\phi^*|$ are 
 shown in figure~\ref{cfmc2}. The shapes of all distributions are well described only by 
 the RP model. 
As for the case of the \textquoteleft two central jets' the SCI+GAL model is not able to describe the distributions of the diffractive kinematic variables but it well reproducing the shape of the $|\Delta\phi^*|$ distribution. 
The TGP model completely fails again to describe the $z_{I\!\!P}$ spectrum.

\section{Summary}

 Integrated and differential cross sections are measured for dijet production in the diffractive DIS process $ep 
\rightarrow e jjX^{\prime}p$. 
 In the process studied, the scattered proton carries at least $90\%$ of the incoming proton momentum and is measured in 
 the H1 Forward Proton Spectrometer.
The presented results are compatible with the previous measurements based on the LRG method and explore a new domain at large $\xpom$.

 Dijet cross sections are measured for an event topology with two jets produced in the central 
 pseudorapidity region, where DGLAP parton evolution mechanism is expected to dominates, and for a topology with one jet in the central region 
 and one 
 jet in the forward region, where effects of non-DGLAP parton evolution may be observed. 
 NLO QCD predictions 
based on the DGLAP approach and using DPDFs 
extracted from inclusive diffraction measurements
describe the dijet cross sections within the errors for both event 
topologies, supporting  the universality of DPDFs.
 The measured $t$-slope of the dijet cross section is consistent within uncertainties with the value 
 measured  in inclusive diffractive DIS with a leading proton in the final state.
 This confirms the validity of the proton
 vertex factorisation hypothesis for dijet production in diffractive DIS. 

 The measured cross sections are compared with predictions from 
Monte Carlo models based on leading order matrix elements and parton 
 showers. The Resolved Pomeron model describes the
 shape of the cross sections well, 
but is too low in normalisation.
 This suggests that contributions from higher
 order processes are expected to be sizable in this approach.
 The SCI+GAL model is able to reproduce the normalisation of the cross section for both dijet topologies
 presented after tuning the model to the \textquoteleft two central jets' data. 
The dependence of the diffractive dijet cross section on $\xpom$ and
 $\zpom$ is able to distinguish between the models. 
 The SCI+GAL and Two Gluon Pomeron models fail to describe the shape of the distributions of the diffractive variables, 
 while the Resolved Pomeron model describes the shape of these distributions well.

\section*{Acknowledgements}

We are grateful to the HERA machine group whose outstanding
efforts have made this experiment possible.
We thank the engineers and technicians for their work in constructing and
maintaining the H1 detector, our funding agencies for
financial support, the DESY technical staff for continual assistance
and the DESY directorate for support and for the
hospitality which they extend to the non-DESY members of the
collaboration.




\renewcommand{\arraystretch}{1.35}


\begin{sidewaystable}
\begin{center}

\begin{tiny}  
  \begin{tabular}{|c|c|c|c|c|c|c|c|c|c|c|c|c|c|c|c|c|c|c|c|c| }
\hline
$Q^2$ & $d\sigma/dQ^2$ & $\delta_{tot}$ & $\delta_{stat}$ & $\delta_{syst}$ & $\rho_{i,i+1}$ & $\rho_{i,i+2}$ & 
$\rho_{i,i+3}$ & $\rho_{i,i+4}$ & $\delta_{E_{e}}$ & $\delta_{\theta_{e}}$ & $\delta_{E_p}$ & $\delta_{P_x}$ & $\delta_{P_y}$ & $\delta_{\eta^*_2}$ & $\delta_{x_{I\!\!P}}$ & $\delta_{E_{had}}$ & $\delta_{\beta}$ & $\delta_{Q^2}$ & $\delta_{P^*_T}$ & $1+\delta_{had}$\\
$[GeV^2]$ & $[pb/GeV^2]$ & $[\%]$ & $[\%]$ & $[\%]$ & $ $ & $ $ & $ $ & $ $ & $[\%]$ & $[\%]$ & $[\%]$ & $[\%]$ & $[\%]$ & $[\%]$ & $[\%]$ & $[\%]$ & $[\%]$ & $[\%]$ & $[\%]$ & $$ \\
\hline
$4.0 - 8.0$ & $21$ & $17.0$ & $13.2$ & $10.6$ & $0.628$ & $0.643$ & $0.596$ & $0.268$  &$-0.5$ & $-3.4$ & $0.2$ & $5.6$ & $-1.8$ & $9.3$ & $1.7$ & $-7.4$ & $-0.2$ & $1.4$ & $2.7$ & $0.87 \pm 0.05$\\
$8.0 - 16.0$ & $9.8$ & $14.8$ & $12.5$ & $7.9$ & $0.646$ & $0.588$ & $0.272$ & $ - $ & $-1.4$ & $-3.0$ & $0.3$ & $4.6$ & $-2.2$ & $7.6$ & $1.6$ & $4.3$ & $1.0$ & $0.5$ & $1.8$ & $0.88 \pm 0.05$\\
$16.0 - 32.0$ & $2.9$ & $20.2$ & $17.3$ & $10.5$ & $0.605$ & $0.256$ & $ - $ & $ - $ &$0.9$ & $-5.0$ & $-0.2$ & $-5.3$ & $-2.3$ & $11.3$ & $2.0$ & $6.4$ & $-0.9$ & $1.3$ & $2.1$ & $0.89 \pm 0.03$\\
$32.0 - 60.0$ & $1.2$ & $20.1$ & $18.1$ & $8.9$ & $0.221$ & $ - $ & $ - $ & $ - $ & $1.0$ & $-1.6$ & $0.1$ & $-5.7$ & $-0.9$ & $-12.2$ & $2.1$ & $5.9$ & $0.7$ & $1.4$ & $1.1$  & $0.89 \pm 0.02$\\
$60.0 - 110.0$ & $0.3$ & $31.6$ & $30.6$ & $8.2$ & $ - $ & $ - $ & $ - $ & $ - $ & $0.6$ & $-3.1$ & $0.0$ & $4.2$ & $-2.5$ & $2.5$ & $1.4$ & $5.3$ & $-1.4$ & $0.3$ & $1.2$  & $0.89 \pm 0.02$\\
\hline
\hline
$y$ & $d\sigma/dy$ & $\delta_{tot}$ & $\delta_{stat}$ & $\delta_{syst}$ & $\rho_{i,i+1}$ & $\rho_{i,i+2}$ & 
$\rho_{i,i+3}$ & $\rho_{i,i+4}$ & $\delta_{E_{e}}$ & $\delta_{\theta_{e}}$ & $\delta_{E_p}$ & $\delta_{P_x}$ & $\delta_{P_y}$ & $\delta_{\eta^*_2}$ & $\delta_{x_{I\!\!P}}$ & $\delta_{E_{had}}$ & $\delta_{\beta}$ & $\delta_{Q^2}$ & $\delta_{P^*_T}$ & $1+\delta_{had}$\\
$ $ & $[pb]$ & $[\%]$ & $[\%]$ & $[\%]$ & $ $ & $ $ & $ $ & $ $ & $[\%]$ & $[\%]$ & $[\%]$ & $[\%]$ & $[\%]$ & $[\%]$ & $[\%]$ & $[\%]$ & $[\%]$ & $[\%]$ & $[\%]$ & $$ \\
\hline
$0.05 - 0.18$ & $419$ & $24.6$ & $20.4$ & $13.7$ & $0.506$ & $0.427$ & $0.366$ & $0.224$ & $7.0$ & $-3.0$ & $-0.1$ & $5.9$ & $-2.3$ & $6.2$ & $0.4$ & $7.8$ & $4.7$ & $-0.2$ & $2.9$ & $0.83 \pm 0.02$\\
$0.18 - 0.31$ & $696$ & $13.8$ & $11.3$ & $8.0$ & $0.557$ & $0.579$ & $0.449$ & $ - $ & $0.6$ & $-1.5$ & $0.1$ & $-4.6$ & $-1.3$ & $8.6$ & $1.4$ & $5.3$ & $2.4$ & $0.1$ & $1.4$ & $0.86 \pm 0.04$\\
$0.31 - 0.44$ & $370$ & $17.8$ & $15.2$ & $9.3$ & $0.439$ & $0.335$ & $ - $ & $ - $ & $-2.8$ & $1.6$ & $0.1$ & $6.8$ & $-0.7$ & $5.5$ & $2.0$ & $5.2$ & $2.4$ & $-0.2$ & $1.1$ & $0.90 \pm 0.04$\\
$0.44 - 0.57$ & $279$ & $18.6$ & $16.3$ & $9.1$ & $0.366$ & $ - $ & $ - $ & $ - $ & $2.5$ & $-2.4$ & $0.2$ & $-6.4$ & $3.6$ & $-11.4$ & $2.0$ & $3.1$ & $-0.6$ & $-0.6$ & $1.2$ & $0.97 \pm 0.06$\\
$0.57 - 0.70$ & $122$ & $39.7$ & $38.4$ & $10.0$ & $ - $ & $ - $ & $ - $ & $ - $ & $-6.3$ & $-1.1$ & $0.1$ & $-5.7$ & $-0.8$ & $18.9$ & $2.9$ & $-2.4$ & $2.5$ & $-0.1$ & $1.7$ & $0.98 \pm 0.10$\\
\hline
\hline
$\log_{10}(\xpom)$ & $d\sigma/d\log_{10}(\xpom)$ & $\delta_{tot}$ & $\delta_{stat}$ & $\delta_{syst}$ & $\rho_{i,i+1}$ 
& 
$\rho_{i,i+2}$ & $\rho_{i,i+3}$ & $\rho_{i,i+4}$ & $\delta_{E_{e}}$ & $\delta_{\theta_{e}}$ & $\delta_{E_p}$ & $\delta_{P_x}$ & $\delta_{P_y}$ & $\delta_{\eta^*_2}$ & $\delta_{x_{I\!\!P}}$ & $\delta_{E_{had}}$ & $\delta_{\beta}$ & $\delta_{Q^2}$ & $\delta_{P^*_T}$ & $1+\delta_{had}$\\
$ $ & $[pb]$ & $[\%]$ & $[\%]$ & $[\%]$ & $ $ & $ $ & $ $ & $ $ & $[\%]$ & $[\%]$ & $[\%]$ & $[\%]$ & $[\%]$ & $[\%]$ & $[\%]$ & $[\%]$ & $[\%]$ & $[\%]$ & $[\%]$ & $$ \\
\hline
$-2.3 - -1.9$ & $32$ & $57.0$ & $49.5$ & $28.2$ & $-0.240$ & $0.179$ & $0.161$ & $0.056$ & $5.9$ & $3.3$ & $24.0$ & $6.2$ & $13.3$ & $-5.9$ & $1.7$ & $-18.9$ & $0.9$ & $-0.7$ & $0.4$ & $0.96 \pm 0.06$\\
$-1.9 - -1.6$ & $93$ & $20.2$ & $17.6$ & $10.0$ & $0.136$ & $-0.027$ & $-0.047$ & $ - $ & $1.3$ & $-1.1$ & $-8.4$ & $-6.0$ & $-0.6$ & $-5.4$ & $-0.6$ & $5.0$ & $0.4$ & $0.7$ & $2.0$ & $0.91 \pm 0.01$\\
$-1.6 - -1.4$ & $200$ & $15.9$ & $12.8$ & $9.5$ & $0.579$ & $0.334$ & $ - $ & $ - $ & $1.1$ & $-1.9$ & $-3.8$ & $6.0$ & $-1.2$ & $0.4$ & $0.1$ & $5.5$ & $-0.4$ & $0.2$ & $1.8$ & $0.89 \pm 0.03$\\
$-1.4 - -1.2$ & $344$ & $13.9$ & $11.0$ & $8.6$ & $0.709$ & $ - $ & $ - $ & $ - $ & $0.4$ & $-1.7$ & $-4.3$ & $-4.9$ & $-2.5$ & $4.4$ & $-0.7$ & $4.2$ & $-0.2$ & $0.2$ & $1.9$ & $0.87 \pm 0.05$\\
$-1.2 - -1.0$ & $488$ & $18.8$ & $16.5$ & $9.0$ & $ - $ & $ - $ & $ - $ & $ - $ & $-0.3$ & $-2.1$ & $-5.6$ & $-4.4$ & $-0.8$ & $5.8$ & $-1.7$ & $3.8$ & $0.7$ & $0.1$ & $2.6$ & $0.87 \pm 0.06$\\
\hline
\hline
$z_{I\!\!P}$ & $d\sigma/dz_{I\!\!P}$ & $\delta_{tot}$ & $\delta_{stat}$ & $\delta_{syst}$ & $\rho_{i,i+1}$ & 
$\rho_{i,i+2}$ &&& $\delta_{E_{e}}$ & $\delta_{\theta_{e}}$ & $\delta_{E_p}$ & $\delta_{P_x}$ & $\delta_{P_y}$ & $\delta_{\eta^*_2}$ & $\delta_{x_{I\!\!P}}$ & $\delta_{E_{had}}$ & $\delta_{\beta}$ & $\delta_{Q^2}$ & $\delta_{P^*_T}$ & $1+\delta_{had}$ \\
$ $ & $[pb]$ & $[\%]$ & $[\%]$ & $[\%]$ & $ $ & $ $ &&& $[\%]$ & $[\%]$ & $[\%]$ & $[\%]$ & $[\%]$ & $[\%]$ & $[\%]$ & $[\%]$ & $[\%]$ & $[\%]$ & $[\%]$ & $$ \\
\hline
$0.0 - 0.2$ & $719$ & $14.7$ & $12.0$ & $8.5$ & $0.601$ & $0.127$ & &&$-1.5$ & $-2.4$ & $5.6$ & $-4.5$ & $2.3$ & $-8.4$ & $0.2$ & $-1.5$ & $0.7$ & $0.6$ & $1.9$ & $0.88 \pm 0.08$\\
$0.2 - 0.5$ & $266$ & $16.5$ & $12.9$ & $10.3$ & $0.336$ & $ - $ & &&$-1.4$ & $-1.6$ & $2.0$ & $6.7$ & $1.9$ & $10.1$ & $1.6$ & $6.3$ & $2.3$ & $0.1$ & $1.4$ & $0.88 \pm 0.03$ \\
$0.5 - 1.0$ & $80$ & $22.3$ & $17.8$ & $13.4$ & $ - $ & $ - $ & &&$-3.2$ & $-1.8$ & $6.4$ & $4.7$ & $5.0$ & $4.2$ & $-1.5$ & $8.3$ & $2.1$ & $0.0$ & $0.5$& $0.90 \pm 0.05$ \\
\hline
\end{tabular}
\end{tiny}
\caption{Bin averaged hadron level differential cross sections for the 
production of two central jets in diffractive DIS as a function of $Q^2$, $y$, $\log_{10}(\xpom)$ and $z_{I\!\!P}$.
The total ($\delta_{tot}$), statistical ($\delta_{stat}$) and systematic ($\delta_{sys}$) 
uncertainties and  the correlation coefficients $\rho$ of the cross section covariance matrix 
defined in section~\ref{xsect} are given together with the changes of the cross sections
due to a $+ 1 \sigma$ variation of the various systematic error
sources described in section~\ref{systs}:
the electromagnetic energy scale ($\delta_{ele}$),
the scattering angle of the electron ($\delta_{\theta}$),
the leading proton energy $E_p$ ($\delta_{E_p}$),
the proton transverse momentum components $P_x$ ($\delta_{P_x}$) and $P_y$ ($\delta_{P_y}$),
the reweighting of the simulation in $\eta_2$ ($\delta_{\eta_2}$) and $\xpom$ ($\delta_{\xpom}$)
the hadronic energy scale ($\delta E_{had}$),
the reweighting of the simulation in $\beta$ ($\delta_{\beta}$),
 $Q^2$ ($\delta_{Q^2}$) and $P_T^*$ ($\delta_{P_T^*}$).
 All uncertainties are given in per cent.
 The normalisation uncertainty of $7\%$ is not included. 
 The hadronisation correction factors $(1+\delta_{had})$ applied to the NLO calculations and the associated uncertainty are shown in the last column.}
\label{dijet}
\end{center}
\end{sidewaystable}

\begin{sidewaystable}
\begin{center}

\begin{tiny}  
  \begin{tabular}{|c|c|c|c|c|c|c|c|c|c|c|c|c|c|c|c|c|c|c|c|c| }
\hline
$p^*_{T1}$ & $d\sigma/dp^*_{T1}$ & $\delta_{tot}$ & $\delta_{stat}$ & $\delta_{syst}$ & $\rho_{i,i+1}$ & 
$\rho_{i,i+2}$ &&& $\delta_{E_{e}}$ & $\delta_{\theta_{e}}$ & $\delta_{E_p}$ & $\delta_{P_x}$ & $\delta_{P_y}$ & $\delta_{\eta^*_2}$ & $\delta_{x_{I\!\!P}}$ & $\delta_{E_{had}}$ & $\delta_{\beta}$ & $\delta_{Q^2}$ & $\delta_{P^*_T}$ & $1+\delta_{had}$ \\
$[GeV]$ & $[pb/GeV]$ & $[\%]$ & $[\%]$ & $[\%]$ & $ $ & $ $ &&& $[\%]$ & $[\%]$ & $[\%]$ & $[\%]$ & $[\%]$ & $[\%]$ & $[\%]$ & $[\%]$ & $[\%]$ & $[\%]$ & $[\%]$ & $$ \\
\hline
$5.0 - 6.5$ & $91$ & $17.6$ & $15.3$ & $8.6$ & $0.402$ & $0.180$ &&& $1.7$ & $-3.2$ & $0.2$ & $-5.4$ & $2.4$ & $-6.8$ & $2.1$ & $-2.2$ & $3.5$ & $-0.5$ & $2.2$ & $0.81 \pm 0.04$\\
$6.5 - 8.5$ & $44$ & $17.0$ & $13.2$ & $10.8$ & $0.395$ & $ - $ &&& $-0.7$ & $0.6$ & $-0.3$ & $6.6$ & $3.2$ & $9.6$ & $1.6$ & $7.3$ & $2.1$ & $0.2$ & $0.7$ & $0.96 \pm 0.05$ \\
$8.5 - 12.0$ & $7.3$ & $39.1$ & $33.0$ & $20.9$ & $ - $ & $ - $ &&& $3.2$ & $-5.8$ & $-2.0$ & $1.7$ & $4.0$ & $13.9$ & $2.4$ & $19.4$ & $-0.5$ & $0.1$ & $0.5$& $0.99 \pm 0.04$ \\
\hline
\hline
$|\Delta\eta^{*}|$ & $d\sigma/d|\Delta\eta^{*}|$ & $\delta_{tot}$ & $\delta_{stat}$ & $\delta_{syst}$ & $\rho_{i,i+1}$ 
& $\rho_{i,i+2}$ & $\rho_{i,i+3}$ && $\delta_{E_{e}}$ & $\delta_{\theta_{e}}$ & $\delta_{E_p}$ & $\delta_{P_x}$ & $\delta_{P_y}$ & $\delta_{\eta^*_2}$ & $\delta_{x_{I\!\!P}}$ & $\delta_{E_{had}}$ & $\delta_{\beta}$ & $\delta_{Q^2}$ & $\delta_{P^*_T}$ & $1+\delta_{had}$ \\
$ $ & $[pb]$ & $[\%]$ & $[\%]$ & $[\%]$ & $ $ & $ $ & $ $ && $[\%]$ & $[\%]$ & $[\%]$ & $[\%]$ & $[\%]$ & $[\%]$ & $[\%]$ & $[\%]$ & $[\%]$ & $[\%]$ & $[\%]$ & $$ \\
\hline
$0.0 - 0.6$ & $118$ & $16.1$ & $13.2$ & $9.2$ & $0.682$ & $0.276$ & $-0.479$ && $-0.6$ & $-2.3$ & $0.1$ & $-5.7$ & $-1.6$ & $-14.9$ & $1.8$ & $5.7$ & $1.9$ & $-0.3$ & $1.5$ & $0.88 \pm 0.04$\\
$0.6 - 1.2$ & $157$ & $14.2$ & $11.5$ & $8.3$ & $0.343$ & $-0.342$ & $ - $ && $1.2$ & $-1.7$ & $0.0$ & $5.0$ & $2.8$ & $-8.7$ & $1.6$ & $4.8$ & $2.0$ & $-0.1$ & $1.3$ & $0.89 \pm 0.04$\\
$1.2 - 1.8$ & $97$ & $19.8$ & $17.4$ & $9.4$ & $0.089$ & $ - $ & $ - $ && $1.1$ & $-2.3$ & $0.1$ & $-4.8$ & $-2.1$ & $2.2$ & $1.7$ & $6.1$ & $-2.6$ & $0.0$ & $1.7$& $0.90 \pm 0.04$ \\
$1.8 - 3.0$ & $26$ & $33.3$ & $31.8$ & $9.8$ & $ - $ & $ - $ & $ - $ && $2.0$ & $2.8$ & $0.3$ & $-5.1$ & $-0.7$ & $-32.9$ & $0.5$ & $4.3$ & $4.7$ & $0.6$ & $3.1$ & $0.84 \pm 0.02$\\
\hline  
\hline
$|t|$ & $d\sigma/d|t|$ & $\delta_{tot}$ & $\delta_{stat}$ & $\delta_{syst}$ & $\rho_{i,i+1}$ & $\rho_{i,i+2}$ &&& 
$\delta_{E_{e}}$ & $\delta_{\theta_{e}}$ & $\delta_{E_p}$ & $\delta_{P_x}$ & $\delta_{P_y}$ & $\delta_{\eta^*_2}$ & $\delta_{x_{I\!\!P}}$ & $\delta_{E_{had}}$ & $\delta_{\beta}$ & $\delta_{Q^2}$ & $\delta_{P^*_T}$ & \\
$[GeV^2]$ & $[pb/GeV^2]$ & $[\%]$ & $[\%]$ & $[\%]$ & $ $ & $ $ &&& $[\%]$ & $[\%]$ & $[\%]$ & $[\%]$ & $[\%]$ & $[\%]$ & $[\%]$ & $[\%]$ & $[\%]$ & $[\%]$ & $[\%]$ &\\
\hline
$0.1 - 0.3$ & $483$ & $17.3$ & $9.3$ & $14.6$ & $0.495$ & $0.420$ &&& $-0.5$ & $-1.5$ & $-0.3$ & $10.3$ & $3.8$ & $12.8$ & $1.9$ & $9.2$ & $1.0$ & $0.3$ & $0.6$ &\\
$0.3 - 0.5$ & $151$ & $16.6$ & $12.5$ & $11.0$ & $0.288$ & $ - $ &&& $0.5$ & $1.9$ & $-0.1$ & $-4.6$ & $3.1$ & $8.6$ & $1.6$ & $9.0$ & $0.6$ & $-0.3$ & $0.8$ &\\
$0.5 - 0.7$ & $44$ & $29.5$ & $25.9$ & $14.0$ & $ - $ & $ - $ &&& $2.3$ & $-1.8$ & $1.1$ & $-3.6$ & $10.4$ & $-11.4$ & $-1.0$ & $-8.0$ & $0.0$ & $0.2$ & $1.0$ &\\
\hline
\end{tabular}
\end{tiny}
\caption{Bin averaged hadron level differential cross sections for the 
production of two central jets in diffractive DIS as a function of $P^*_{T,1}$, $|\Delta\eta^*|$ and $|t|$.  The normalisation uncertainties of $4.6\%$ for the differential cross section in $|t|$ and $7\%$ for other cross sections are not included. For details see table \ref{dijet}.}
\end{center}
\end{sidewaystable}

\begin{sidewaystable}
\begin{center}

\begin{tiny}  
  \begin{tabular}{|c|c|c|c|c|c|c|c|c|c|c|c|c|c|c|c|c|c|c|c| }
\hline
$\langle p^*_{T}\rangle$ & $d\sigma/d\langle p^*_{T}\rangle$ & $\delta_{tot}$ & $\delta_{stat}$ & $\delta_{syst}$ & 
$\rho_{i,i+1}$ & $\rho_{i,i+2}$ && $\delta_{E_{e}}$ & $\delta_{\theta_{e}}$ & $\delta_{E_p}$ & $\delta_{P_x}$ & $\delta_{P_y}$ & $\delta_{\eta^*_f}$ & $\delta_{x_{I\!\!P}}$ & $\delta_{E_{had}}$ & $\delta_{\beta}$ & $\delta_{Q^2}$ & $\delta_{P^*_T}$ & $1+\delta_{had}$ \\
$[GeV]$ & $[pb/GeV]$ & $[\%]$ & $[\%]$ & $[\%]$ & $ $ & $ $ && $[\%]$ & $[\%]$ & $[\%]$ & $[\%]$ & $[\%]$ & $[\%]$ & $[\%]$ & $[\%]$ & $[\%]$ & $[\%]$ & $[\%]$ & $$ \\
\hline
$3.5 - 5.0$ & $40$ & $33.0$ & $28.8$ & $16.2$ & $0.433$ & $0.403$ && $-3.5$ & $-0.9$ & $-0.7$ & $4.5$ & $-1.8$ & $11.1$ & $0.6$ & $-14.1$ & $3.6$ & $1.8$ & $1.8$ & $0.7 \pm 0.09$\\
$5.0 - 7.0$ & $36$ & $17.3$ & $15.8$ & $7.2$ & $0.577$ & $ - $ && $-0.8$ & $-3.9$ & $-0.3$ & $3.2$ & $2.2$ & $12.1$ & $1.0$ & $3.4$ & $-0.9$ & $1.1$ & $1.9$ & $0.93 \pm 0.08$\\
$7.0 - 12.0$ & $8.8$ & $26.0$ & $22.9$ & $12.2$ & $ - $ & $ - $ && $1.3$ & $-2.0$ & $0.4$ & $4.7$ & $-2.2$ & $24.4$ & $1.6$ & $9.9$ & $-1.5$ & $-0.1$ & $0.2$ & $1.05 \pm 0.03$\\
\hline
\hline
$|\Delta\eta^*|$ & $d\sigma/d|\Delta\eta^*|$ & $\delta_{tot}$ & $\delta_{stat}$ & $\delta_{syst}$ & $\rho_{i,i+1}$ & 
$\rho_{i,i+2}$ && $\delta_{E_{e}}$ & $\delta_{\theta_{e}}$ & $\delta_{E_p}$ & $\delta_{P_x}$ & $\delta_{P_y}$ & $\delta_{\eta^*_f}$ & $\delta_{x_{I\!\!P}}$ & $\delta_{E_{had}}$ & $\delta_{\beta}$ & $\delta_{Q^2}$ & $\delta_{P^*_T}$ & $1+\delta_{had}$ \\
$ $ & $[pb]$ & $[\%]$ & $[\%]$ & $[\%]$ & $ $ & $ $ && $[\%]$ & $[\%]$ & $[\%]$ & $[\%]$ & $[\%]$ & $[\%]$ & $[\%]$ & $[\%]$ & $[\%]$ & $[\%]$ & $[\%]$ & $$ \\
\hline
$0.0 - 1.2$ & $21$ & $30.0$ & $28.3$ & $10.2$ & $0.489$ & $0.321$ && $0.6$ & $-3.9$ & $-0.3$ & $4.4$ & $-3.6$ & $20.1$ & $0.0$ & $6.0$ & $0.2$ & $0.7$ & $4.1$ & $1.04 \pm 0.07$\\
$1.2 - 2.4$ & $60$ & $20.9$ & $17.3$ & $11.7$ & $0.329$ & $ - $ && $-1.8$ & $3.0$ & $0.2$ & $4.0$ & $2.3$ & $10.2$ & $1.3$ & $9.5$ & $2.1$ & $1.6$ & $1.9$ & $0.88 \pm 0.07$\\
$2.4 - 3.5$ & $48$ & $26.5$ & $25.0$ & $8.8$ & $ - $ & $ - $ && $-2.0$ & $-0.8$ & $1.0$ & $4.8$ & $1.6$ & $7.0$ & $0.0$ & $6.7$ & $-0.9$ & $0.4$ & $1.1$ & $0.69 \pm 0.06$\\
\hline
\hline
$\eta_{f}$ & $d\sigma/d\eta_{f}$ & $\delta_{tot}$ & $\delta_{stat}$ & $\delta_{syst}$ & $\rho_{i,i+1}$ & 
$\rho_{i,i+2}$& $\rho_{i,i+3}$ & $\delta_{E_{e}}$ & $\delta_{\theta_{e}}$ & $\delta_{E_p}$ & $\delta_{P_x}$ & $\delta_{P_y}$ & $\delta_{\eta^*_f}$ & $\delta_{x_{I\!\!P}}$ & $\delta_{E_{had}}$ & $\delta_{\beta}$ & $\delta_{Q^2}$ & $\delta_{P^*_T}$ & $1+\delta_{had}$ \\
$ $ & $[pb]$ & $[\%]$ & $[\%]$ & $[\%]$ & $ $ & $ $ & $$ & $[\%]$ & $[\%]$ & $[\%]$ & $[\%]$ & $[\%]$ & $[\%]$ & $[\%]$ & $[\%]$ & $[\%]$ & $[\%]$ & $[\%]$ & $$\\
\hline
$-0.6 - 0.2$ & $23$ & $24.1$ & $21.9$ & $10.0$ & $0.391$ & $0.437$ & $0.360$ & $3.5$ & $-2.3$ & $-0.1$ & $5.5$ & $2.3$ & $-6.4$ & $-1.9$ & $6.5$ & $0.0$ & $0.7$ & $0.5$ & $0.89 \pm 0.06$\\
$0.2 - 0.9$ & $63$ & $17.0$ & $14.3$ & $9.2$ & $0.567$ & $0.427$ & $ - $ & $-0.5$ & $1.4$ & $-0.1$ & $-6.4$ & $-1.9$ & $8.0$ & $-2.0$ & $5.3$ & $-1.5$ & $0.2$ & $-1.7$ & $0.93 \pm 0.05$\\
$0.9 - 1.6$ & $98$ & $15.4$ & $12.5$ & $9.1$ & $0.549$ & $ - $ & $ - $ & $-0.2$ & $-1.7$ & $0.1$ & $-4.9$ & $-1.7$ & $6.2$ & $1.5$ & $6.9$ & $-0.4$ & $0.7$ & $0.3$ & $0.89 \pm 0.04$\\
$1.6 - 2.8$ & $75$ & $21.9$ & $18.5$ & $11.7$ & $ - $ & $ - $ & $ - $ & $-0.7$ & $2.9$ & $-0.4$ & $2.9$ & $-1.0$ & $9.0$ & $-0.2$ & $10.1$ & $-0.5$ & $0.2$ & $3.4$ & $0.86 \pm 0.01$\\
\hline
\hline
$z_{I\!\!P}$ & $d\sigma/dz_{I\!\!P}$ & $\delta_{tot}$ & $\delta_{stat}$ & $\delta_{syst}$ & $\rho_{i,i+1}$ & 
$\rho_{i,i+2}$ && $\delta_{E_{e}}$ & $\delta_{\theta_{e}}$ & $\delta_{E_p}$ & $\delta_{P_x}$ & $\delta_{P_y}$ & $\delta_{\eta^*_f}$ & $\delta_{x_{I\!\!P}}$ & $\delta_{E_{had}}$ & $\delta_{\beta}$ & $\delta_{Q^2}$ & $\delta_{P^*_T}$ & $1+\delta_{had}$ \\
$ $ & $[pb]$ & $[\%]$ & $[\%]$ & $[\%]$ & $ $ & $ $ && $[\%]$ & $[\%]$ & $[\%]$ & $[\%]$ & $[\%]$ & $[\%]$ & $[\%]$ & $[\%]$ & $[\%]$ & $[\%]$ & $[\%]$ & $$ \\
\hline
$0.0 - 0.2$ & $268$ & $33.5$ & $28.6$ & $17.4$ & $0.134$ & $-0.138$ && $-4.4$ & $-3.7$ & $8.4$ & $8.0$ & $5.2$ & $5.4$ & $-2.0$ & $7.7$ & $8.4$ & $2.6$ & $1.5$ & $0.93 \pm 0.10$\\
$0.2 - 0.6$ & $230$ & $17.2$ & $15.6$ & $7.0$ & $0.308$ & $ - $ && $-1.2$ & $1.5$ & $3.0$ & $-3.2$ & $3.4$ & $8.3$ & $0.0$ & $-4.3$ & $-1.8$ & $0.5$ & $1.5$ & $0.85 \pm 0.08$ \\
$0.6 - 1.0$ & $47$ & $39.0$ & $34.1$ & $18.9$ & $ - $ & $ - $ && $3.1$ & $-4.5$ & $8.4$ & $2.6$ & $1.5$ & $33.6$ & $-3.2$ & $14.3$ & $-1.2$ & $0.6$ & $1.3$ & $0.82 \pm 0.02$\\
\hline
\hline
$\log_{10}(\beta)$ & $d\sigma/d\log_{10}(\beta)$ & $\delta_{tot}$ & $\delta_{stat}$ & $\delta_{syst}$ & $\rho_{i,i+1}$ & 
$\rho_{i,i+2}$ && $\delta_{E_{e}}$ & $\delta_{\theta_{e}}$ & $\delta_{E_p}$ & $\delta_{p_x}$ & $\delta_{p_y}$ & $\delta_{\eta^*_f}$ & $\delta_{x_{I\!\!P}}$ & $\delta_{E_{had}}$ & $\delta_{\beta}$ & $\delta_{Q^2}$ & $\delta_{P^*_T}$ & $1+\delta_{had}$ \\
$ $ & $[pb]$ & $[\%]$ & $[\%]$ & $[\%]$ & $ $ & $ $ && $[\%]$ & $[\%]$ & $[\%]$ & $[\%]$ & $[\%]$ & $[\%]$ & $[\%]$ & $[\%]$ & $[\%]$ & $[\%]$ & $[\%]$ & $$ \\
\hline
$-3.0 - -2.1$ & $63$ & $29.1$ & $23.4$ & $17.2$ & $0.419$ & $0.115$ && $3.3$ & $-5.2$ & $4.8$ & $-9.6$ & $-3.0$ & $5.2$ & $-1.1$ & $6.0$ & $8.8$ & $3.4$ & $3.0$ & $0.89 \pm 0.08$\\
$-2.1 - -1.6$ & $136$ & $18.6$ & $14.9$ & $11.0$ & $0.396$ & $ - $ && $-1.4$ & $-1.7$ & $-1.1$ & $3.3$ & $-1.5$ & $12.5$ & $1.2$ & $9.4$ & $1.7$ & $1.7$ & $1.6$ & $0.85 \pm 0.07$\\
$-1.6 - -0.5$ & $21$ & $38.4$ & $30.6$ & $23.2$ & $ - $ & $ - $ && $-5.3$ & $-2.9$ & $-19.8$ & $3.6$ & $5.6$ & $21.9$ & $1.6$ & $-8.4$ & $-1.1$ & $0.1$ & $3.2$ & $0.84 \pm 0.05$\\
\hline
\hline
$|\Delta\phi^*|$ & $d\sigma/d|\Delta\phi^*|$ & $\delta_{tot}$ & $\delta_{stat}$ & $\delta_{syst}$ & $\rho_{i,i+1}$ &&&
$\delta_{E_{e}}$ & $\delta_{\theta_{e}}$ & $\delta_{E_p}$ & $\delta_{P_x}$ & $\delta_{P_y}$ & $\delta_{\eta^*_f}$ & $\delta_{x_{I\!\!P}}$ & $\delta_{E_{had}}$ & $\delta_{\beta}$ & $\delta_{Q^2}$ & $\delta_{P^*_T}$ & $1+\delta_{had}$ \\
$ [degree] $ & $[pb/degree]$ & $[\%]$ & $[\%]$ & $[\%]$ & $ $ &&& $[\%]$ & $[\%]$ & $[\%]$ & $[\%]$ & $[\%]$ & $[\%]$ & $[\%]$ & $[\%]$ & $[\%]$ & $[\%]$ & $[\%]$ & $$ \\
\hline
$0.0 - 160.0$ & $0.3$ & $41.9$ & $36.4$ & $20.9$ & $0.261$ &&& $-1.2$ & $-1.8$ & $-2.5$ & $-2.9$ & $-2.9$ & $6.7$ & $-1.0$ & $19.6$ & $3.8$ & $0.7$ & $4.1$ & $0.91 \pm 0.16$\\
$160.0 - 180.0$ & $5.2$ & $17.2$ & $14.9$ & $8.7$ & $ - $ &&& $0.2$ & $-1.6$ & $-0.4$ & $4.5$ & $-1.7$ & $16.6$ & $1.3$ & $6.0$ & $1.0$ & $0.8$ & $2.2$ & $0.85 \pm 0.05$\\
\hline
\end{tabular}
\end{tiny}
\caption{Bin averaged hadron level differential cross sections for the 
production of one central and one forward jet in diffractive DIS as a function of $\langle P^*_{T}\rangle$, $|\Delta\eta^*|$, $\eta_{f}$,
$z_{I\!\!P}$, $\log_{10}(\beta)$ and $|\Delta\phi^*|$.
 The normalisation uncertainty of $6.2\%$ is not included. 
For more details see table \ref{dijet}. }
\label{cfjet}
\end{center}
\end{sidewaystable}

\end{document}